\documentclass[twocolumn]{aastex62}

\pdfoutput=1 
\usepackage{amsmath,amssymb,amstext}
\usepackage[all]{hypcap} 
\usepackage{natbib}

\newcommand\Change[1]{}

\received{March 21, 2018}
\revised{April 18, 2018}
\accepted{April 20, 2018}
\submitjournal{ApJ}

\shorttitle{CR propagation from the Galactic Wind Termination Shock}
\shortauthors{Merten et al.}

\begin{document}
\title{The Propagation of Cosmic Rays From the Galactic Wind Termination Shock: Back to the Galaxy?}

\correspondingauthor{Lukas Merten}
\email{lukas.merten@rub.de, bustard@wisc.edu\\ zweibel@astro.wisc.edu, julia.tjus@rub.de}

\author[0000-0003-1332-9895]{Lukas Merten}
\affil{Theoretische Physik IV, Ruhr-Universit\"at Bochum, Universit\"atsstrasse 150, 45801 Bochum, Germany}

\author{Chad Bustard}
\affiliation{Physics Department, University of Wisconsin-Madison, 1150 University Avenue, Madison, WI 53706}

\author{Ellen G.\ Zweibel}
\affiliation{Physics Department, University of Wisconsin-Madison, 1150 University Avenue, Madison, WI 53706}
\affiliation{Department of Astronomy, University of Wisconsin-Madison, 2535 Sterling Hall, 475 N. Charter Street, Madison, WI 53706}

\author{Julia Becker Tjus}
\affiliation{Theoretische Physik IV, Ruhr-Universit\"at Bochum, Universit\"atsstrasse 150, 45801 Bochum, Germany}

\begin{abstract}

Although several theories for the origin of cosmic rays in the region between the spectral `knee' and `ankle' exist, this problem is still unsolved. A variety of observations suggest that the transition from Galactic to extragalactic sources occurs in this energy range. In this work we examine whether a Galactic wind which eventually forms a termination shock far outside the Galactic plane can contribute as a possible source to the observed flux in the region of interest. Previous work by \cite{2017ApJ...835...72B} estimated that particles can be accelerated up to energies above the `knee' up to $R_\mathrm{max} = 10^{16}$~eV for parameters drawn from a model of a Milky Way wind \citep{2010ApJ...711...13E}. A remaining question is whether the accelerated cosmic rays can propagate back into the Galaxy. To answer this crucial question, we simulate the propagation of the cosmic rays using the low energy extension of the CRPropa framework, based on the solution of the transport equation via stochastic differential equations. The setup includes all relevant processes, including three-dimensional anisotropic spatial diffusion, advection, and corresponding adiabatic cooling. We find that, assuming realistic parameters for the shock evolution, a possible Galactic termination shock can contribute significantly to the energy budget in the `knee' region and above. We estimate the resulting produced neutrino fluxes and find them to be below measurements from IceCube and limits by KM3NeT. 

\end{abstract}


\section{Introduction}
\label{sec:Intro}

There is a growing amount of evidence that galactic cosmic rays (CRs) are accelerated at supernova remnant (SNR) shock fronts. The dominant acceleration mechanism, revealed by both 
theoretical \citep{1949PhRv...75.1169F, 1977ICRC...11..132A, 1978MNRAS.182..147B, 1978ApJ...221L..29B, 1989ApJ...336..243S,1989ApJ...336..264S,  2013A&ARv..21...70B, 2015ICRC...34....8C} and observational \citep{1994A&A...287..959D, 
2011ApJ...742L..30G, 2013Sci...339..807A, 2016arXiv160908671H} studies, is diffusive shock acceleration, a first order Fermi acceleration process in which about 10\% of the shock 
energy is expected to be converted to cosmic rays. This process naturally leads to a CR spectrum with a power law of $E^{-2}$, which is close to the observed behavior. However, 
instead of a single power law, there are a number of interesting features in the observed spectrum that suggest a change in either acceleration site or mechanism. For instance, a 
steepening in the spectrum occurs at about $3 \cdot 10^{15}$ eV (see e.g.\ \citealt{Wiebel-Sooth1997}), which is referred to as the `knee'. It is believed that SNR shock fronts, 
with the aid of magnetic 
field amplification (e.g.\ \citealt{2012A&ARv..20...49V}) that boosts the CR acceleration rate \citep{2004MNRAS.353..550B, 2009MNRAS.392.1591A, 2010ApJ...709.1412Z, 2012MNRAS.427.2308D, 2014ApJ...794...46C}, can 
accelerate CRs up to this energy but likely not beyond it. Similarly, there is a flattening of the spectrum at $\approx 10^{18}$ eV (see e.g.\ \citealt{Eichmann_2018}), which is 
called the `ankle'. At 
energies beyond the ankle, the acceleration site is believed to be extragalactic, though the exact energy transition region from galactic to extragalactic CRs is not quite known. 

In between the `knee' and the `ankle', which we refer to as the `shin', the CR source is still unknown \citep{1984ARA&A..22..425H, 2011ARA&A..49..119K, 2014CRPhy..15..329B}. A 
possible class of accelerators is Galactic wind termination shocks (GTS), where large-scale outflows of material, probably driven by a combination of thermal, cosmic 
ray, and radiation pressure, shocks with the surrounding intergalactic medium (IGM). \cite{1985ApJ...290L...1J} and \cite{1987ApJ...312..170J} first proposed these shocks as 
sources of CRs well past the `ankle'; however, current estimates of the available acceleration time (i.e.\ the wind, and hence shock, duration) lower these energy estimates by orders 
of magnitude. \Change{\cite{VOELK1996} have shown that termination shocks formed by starburst galaxies and clusters of galaxies are also able to accelerate particles. However, we will focus in this work on the termination shock of the Milky Way because that gives the best comparison to observations.}

Using the wind model of \cite{2016ApJ...819...29B}, \cite{2017ApJ...835...72B} relate some fundamental parameters of thermal wind driving to CR acceleration rates. Under fairly optimistic assumptions, they estimate CR acceleration to shin energies, but probably only to $\approx 10^{17}$ eV for starburst galaxies. This maximum energy can shift up or down depending on the acceleration time, which depends on the burstiness of star formation in the galaxy, assumptions about the CR acceleration rate, and the pressure adopted for the circumgalactic or intergalactic medium. Taking the 
best-fit Milky Way galactic wind model of \cite{2010ApJ...711...13E}, they estimate a maximum CR energy just above the knee for a theoretical Milky Way termination shock.

In general, it seems promising that termination shocks could theoretically accelerate CRs to shin energies. The next crucial step, which we undertake in detail in this followup 
paper, is to determine whether these accelerated CRs, specifically those accelerated by a Milky Way termination shock, can contribute significantly to the flux in the shin region. The interplay between shock-accelerated CRs and the progenitor of 
the termination shock, the expanding galactic wind, determines whether these CRs are stored for a significant time in the Galactic halo, possibly interacting with thermal gas to 
produce gamma rays and neutrinos \citep{2014PhRvD..89j3003T}, or are blown out of the Galaxy and into the circumgalactic medium (CGM). Past studies have illuminated the importance 
of this interplay \citep{1993A&A...267..372B, 1997A&A...321..434P, 2017PhRvD..95b3001T}. Unlike the pioneering work of \citet{1987ApJ...312..170J}, which modeled the transport of ultra high energy
CRs---up to rigidities of $\approx 10^{19}$~V---in a Galactic wind, we focus in this work on CRs with energies just above the knee ($E\leq 10^{16}$~eV), but our analysis provides insight for lower energy CRs as well. 

We motivate this work by semianalytic estimates in Section \ref{semiestimates} and then discuss
the importance of various outcomes. In Section \ref{transport}, 
we describe the sophisticated transport model we employ in this paper, specifically focusing on the addition of advection and adiabatic energy change to the CRPropa 
software\footnote{www.crpropa.desy.de} \citep{crpropa30}. With these modifications, we are able to study the time-dependent propagation of cosmic rays in an expanding galactic 
wind, including estimates for arrival direction, arrival time, and CR flux at an observer sphere 10 kpc from Galactic center. Further studies of propagation into the inner regions 
of the Galaxy, which contains a much more complicated magnetic field geometry, are left to future work. In Section \ref{simulation}, we describe the two simulation setups we will explore. Unlike other 
similar studies, which primarily focus on the Fermi Bubbles and local outflows or fountains originating in the Central Molecular Zone \citep{2011PhRvL.106j1102C, 2014MNRAS.444L..39L, 
2015MNRAS.453.3827S, 2017PhRvD..95b3001T}, our complementary work assumes a spherically symmetric, \emph{global} wind with monotonically increasing velocity up to 600 $\mathrm{km}\,\mathrm{s}^{-1}$. The 
asymptotic wind velocity is motivated by the best-fit models of the Milky Way's hybrid thermal and CR-driven wind \citep{2010ApJ...711...13E} (see Fig.\ \ref{fig:WindProfile} 
for details). In Section \ref{greensfunction}, we describe the Green's function technique, which allows us to use the same simulations to describe different physical scenarios. As an example of this flexible method, we analyze an instantaneous burst of CR acceleration. This builds intuition for a more realistic, continuous emission of CRs, which is modeled in Section \ref{continuoussource}. In addition to the resulting CR proton properties, we calculate the resulting neutrino flux and compare it to the observed flux at different energies. In Section \ref{conclusions}, we state our conclusions and give an outlook on the possible impact of this work and future studies.

\section{Preliminary estimates of the return probability}
\label{semiestimates}
\begin{figure}[htbp]
\centering
\includegraphics[width=.43\textwidth]{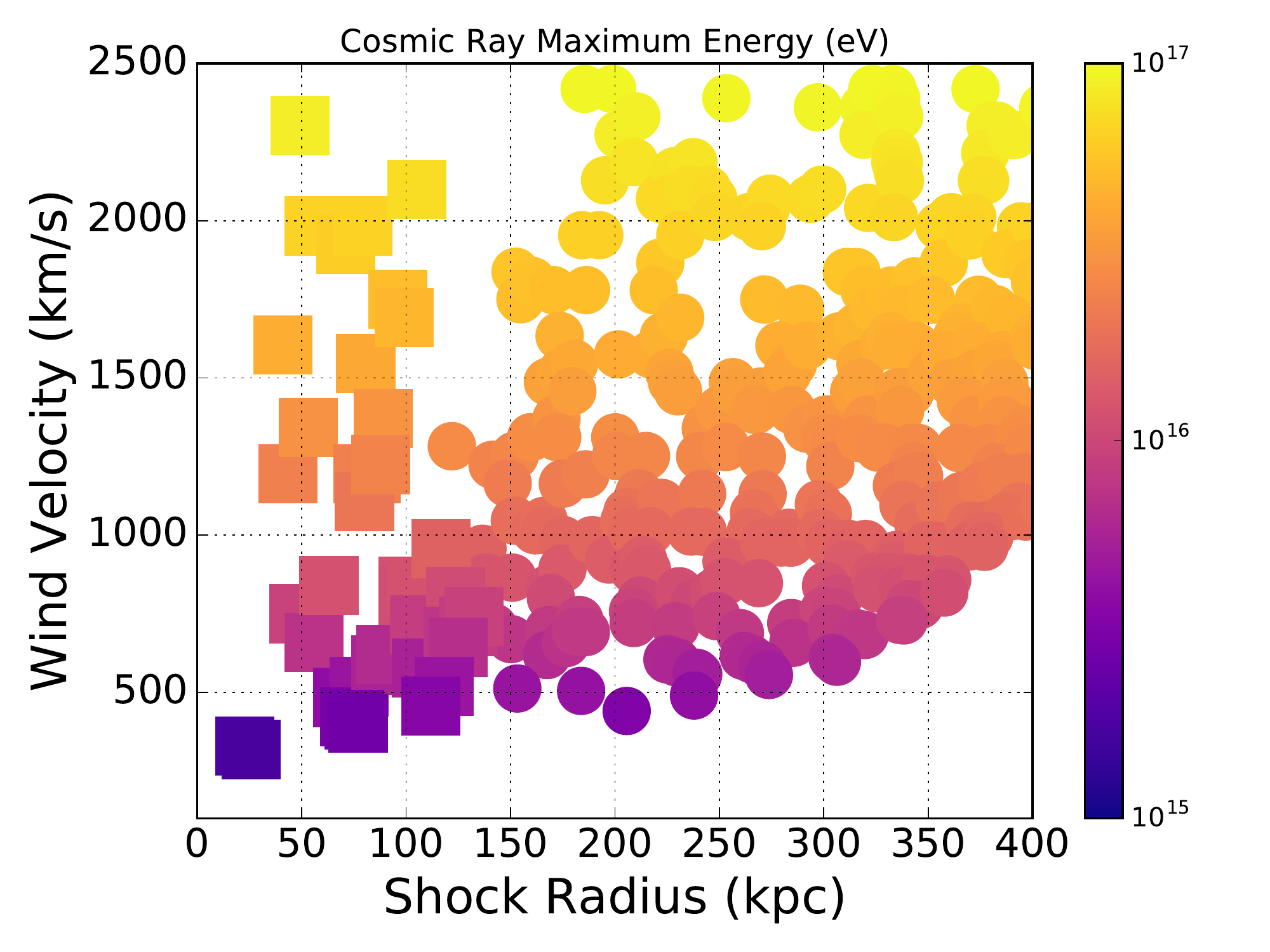}
\caption{Cosmic ray maximum energies for theoretical wind models from \cite{2017ApJ...835...72B}. Square markers indicate models that generate diffusion-dominated CRs at that 
maximum energy, i.e.\ $\mathcal{R}_{CR} < 1$. For this assumed diffusion exponent of $\delta = 0.4$, most winds lead to advection-dominated CR propagation.}
\label{fig:CR_Reynolds_Number}
\end{figure}

To estimate whether CRs can diffuse back to their host galaxy in opposition to an outflow, \cite{2017ApJ...835...72B} define a CR Reynolds number,  
\begin{equation}
\label{reynolds}
\mathcal{R}_{\rm CR} \equiv \frac{R_{\rm shock}V_{\rm shock}}{\kappa(E)} ,
\end{equation} 
which is the ratio of the diffusion time, $\tau_{\rm diff} = R_{\rm shock}^{2}/\kappa$, to the advection time, $\tau_{\rm adv} = R_{\rm shock}/V_{\rm shock}$, and $\kappa(E)$ is the 
energy-dependent diffusion coefficient of CRs. For an illustration, let's assume $\kappa(E) = D_{0} \cdot 10^{28} \rm cm^{2} s^{-1} E_{GeV}^{\delta}$ and choose $\delta = 0.4$ 
and $D_{0} = 5$ (\cite{2007ARNPS..57..285S}). We then generate a number of wind models using the fiducial model of \cite{2016ApJ...819...29B} and plot in Fig.\ 
\ref{fig:CR_Reynolds_Number} the wind velocity, shock radius, and maximum CR energy assuming the IGM pressure is $P_{\rm IGM} = 10^{-14}$ ergs $ \rm cm^{-3}$ (see Eqn.\ 20 of 
\cite{2017ApJ...835...72B}). Each point denotes a different wind model, while the square markers show a subset of scenarios for which $\mathcal{R}_{CR} < 1$, i.e.\ cases 
where CR diffusion may overcome advection with the outflow. Fig.\ \ref{fig:CR_Reynolds_Number} shows that most CRs are likely not able to diffuse back to the galaxy and are, 
instead, blown out into the IGM. It needs to be emphasized that diffusion scales positively with energy, hence if a CR with some maximum estimated energy is advection-dominated, so are 
all CRs of lower energy. Similarly, all square markers in Fig.\ \ref{fig:CR_Reynolds_Number} are diffusion-dominated at that maximum energy but not necessarily for lower energy 
CRs. 
If we instead use $\delta = 0.5$ in Fig.\ \ref{fig:CR_Reynolds_Number}, \emph{all} of the points satisfy $\mathcal{R}_{CR} < 1$ at that maximum CR energy. For the Milky Way wind 
parameters of \cite{2010ApJ...711...13E}, the minimum energy of diffusion-dominated CRs is $\approx 2.8 \cdot 10^{16}$ eV for $\delta = 0.4$ and $\approx 9.2 \cdot 10^{14}$ eV 
for $\delta = 0.5$ \citep{2017ApJ...835...72B}. 
Therefore, if the diffusion exponent is large, the inward flux of CRs may be large enough to contribute to the measured flux in the shin region. These CRs may also interact with 
denser gas in the Galaxy's inner regions and produce a substantial neutrino flux that may be detectable by IceCube. 

On the other hand, if many CRs are advection-dominated, termination shocks may represent another energy source for the IGM and CGM. Cosmic rays that would have 
otherwise adiabatically lost energy as they expanded into the IGM can be rejuvenated by a conversion of shock kinetic energy flux to CR energy flux. This energy can then be 
transferred to the surrounding gas. \cite{2017ApJ...835...72B} estimate that, if a tenth of the shock energy is given to CRs, high velocity shocks may lead to CR luminosities of 
order $10^{43}$ ergs/s, which is $\approx 10 \%$ of the total Milky Way luminosity. This extra energy generation may have interesting implications for not just the Milky Way but 
also starburst galaxies, which can expel very strong galactic winds (e.g.\ M82), and the cocoons of AGN jets, for which the same termination shock arguments apply. On the other hand, if the intergalactic pressure were lower, the termination shock would be further from the galaxy ($R_{shock}\propto P_{IGM}^{-1/2}$), and the returning fraction would be lower.

To properly calculate the fraction of CRs expelled or retained in the galaxy, it is important to accurately model CR propagation in opposition of an expanding velocity field. In this paper, we substitute the CR Reynolds number estimate from \cite{2017ApJ...835...72B} with a detailed treatment of CR propagation, which is described in the following section.

\section{Transport Model}
\label{transport}
The transport of CRs is typically described in a sample averaged sense by the so called Parker transport equation:
\begin{align}
\frac{\partial n}{\partial t} + \vec{u}\cdot\nabla n&= \nabla\cdot(\hat{\kappa}\nabla n) + \frac{1}{p^2}\frac{\partial}{\partial p}(p^2\kappa_{pp} \frac{\partial n}{\partial p}) 
\notag \\ 
&+ \frac{p}{3}\nabla\cdot\vec{u}\frac{\partial n}{\partial p} + S \quad . \label{eq:ParkerEquation}
\end{align}
This equation includes anisotropic spatial diffusion described by a diffusion tensor $\hat{\kappa}$, momentum diffusion $\kappa_{pp}$, 
advection due to e.g.\ galactic winds with wind speed $\vec{u}$ and losses caused by the adiabatic expansion due to the advection. If the scattering is produced by Alfvén waves, then the momentum diffusion coefficient is smaller than the scattering coefficient by a factor of $(v_A/c)^2$ so we neglected the momentum diffusion in this work. Sources are described by $S$. Collisional and radiative losses can be included where appropriate. If the cosmic rays are streaming relative to the gas, as in the self confinement model \citep{2017PhPl...24e5402Z}, then $\vec{u}$ includes the streaming speed, but we ignore that possibility here, as cosmic rays in the TeV to PeV range are not generally thought to be self confined. \Change{We also ignore particle drifts, like curvature or gradient drifts, due to inhomogeneities in the magnetic fields, since the estimated drift velocities are more than an order of magnitude smaller than the wind speeds and the displacement due to drifts in a typical propagation time is small. For a detailed review of the drifts in our model, see the appendix \ref{app:Drift}.}

In the following we review briefly how Eqn.\ \ref{eq:ParkerEquation} can be solved using stochastic differential equations (SDEs). For anisotropic spatial diffusion this is explained in detail in \cite{MER17}. In this paper we concentrate on the application of the implemented advection and adiabatic cooling modules, which were implemented into CRPropa for the purpose of this work. For the complete 
technical details the reader is referred to an upcoming paper on the next version of the software CRPropa \citep{crpropa32}.

\subsection{Transport equation and stochastic differential equations}
\label{SDE}
Every Fokker-Planck like equation has a corresponding set of equivalent SDEs (see e.g.\ \citep{gardiner1985}).
\begin{align}
 \frac{\partial n(\vec{x}, t; \vec{y}, t')}{\partial t} &= -\sum_i \frac{\partial}{\partial x_i}[A_i(\vec{x}, t)n(\vec{x}, t; \vec{y}, t')] \notag \\
  &+ \frac{1}{2}\sum_{i,j}\frac{\partial^2}{\partial x_i \partial 
x_j}[B_{ij}(\vec{x}, t)n(\vec{x}, t; \vec{y}, t')] \label{eq:generalFPE} \quad , 
\end{align}
where $A_i(\vec{x}, t)$ is the drift vector, $B_{ij}(\vec{x}, t)$ is the diffusion tensor and $n(\vec{x}, t; \vec{y}, t')$ is the density at place $\vec{x}$ and time $t$ depending 
on the density at place $\vec{y}$ 
and time $t'$. Here, $\vec{x}$ and $\vec{y}$ are in principle higher dimensional phase space vectors. The corresponding SDE, which can be seen as the equation of motion of the 
phase space elements $\vec{x}$, can be written as (here: three spatial and one momentum dimension):
\begin{align}
 {\rm d}r_\nu = A_\nu\,{\rm d}t+ D_{\nu\mu}\,{\rm d}\omega^\mu \label{eq:SDE} \quad ,
\end{align}
where ${\rm d}t$ is the time increment, $r_\nu$ is a 4-dimensional vector $(\vec{r}, ||\vec{p}||)$, and ${\rm d}\omega^\mu = \sqrt{{\rm d}t}\,\eta^\mu$ stands for a 4-dimensional 
Wiener process with Gaussian noise. Equation \ref{eq:SDE} can be solved for example using the Euler-Maruyama \citep{CAM96} scheme which can be compared to the conventional Euler-Forward 
algorithm (e.g.\ \citealt{BUT03}), known from ordinary differential equations. In contrast to the conventional Euler scheme the Euler-Maruyama scheme has a larger region of convergence, leading to stable solution for most of our simulation setups. The calculation of the tensor $D_{\nu \mu}$ from the physical diffusion tensor $\hat{\kappa}$ as well as the solution of a general SDE is explained in detail in \cite{MER17}.

The advantage of this \emph{ansatz}, compared to the more conventional grid based solvers like GALPROP \citep{strong_propagation_1998}, DRAGON2 \citep{EVO17} or PICARD 
\citep{kissmann2014}, is the independence of the single phase-space-elements 
(or pseudo-particle) trajectories. This allows for---among other things---a very efficient and trivial possibility of parallelizing the computation.

\subsection{Advection}
\label{advection}

The major improvement of this update of the CRPropa software is the implementation of the advection module. This new module allows the user to include the proper handling of 
advective processes---described by the vector $A_\nu$ in Eqn.\ \ref{eq:SDE}---such as galactic winds. The implementation is done via a simple addition of a deterministic part 
into the pseudo-particle propagator:
\begin{align}
\vec{x}_{n+1} &= \vec{x}_n \underbrace{+ \vec{u}{\rm \Delta}t}_{\text{new part}} + D_r\,{\rm \Delta}\vec{\omega}_r \quad , \label{eq:EulerMaruyama}
\end{align}
where $\vec{u}$ is the advection vector, ${\rm \Delta}t$ is the time increment, $D_r$ is the spatial part of the diffusion tensor $D_{\nu \mu}$, and ${\rm \Delta}\vec{\omega}_r$ is 
a three dimensional Wiener process in the basis of the local trihedron\footnote{The local trihedron defines an orthonormal basis consisting of the tangential, normal and 
binormal vector of the magnetic field.}, defined by the curvature of the coherent magnetic field line (see \citealt{MER17}) and the appendix \ref{appendix} for details).

\subsection{Adiabatic energy change}
\label{cooling}

The implementation of advection makes a consistent treatment of energy changes due to an adiabatic expansion or compression of the CRs mandatory. When a gas expands adiabatically, 
so without external heat exchange, it loses
energy because of the decreasing pressure. The opposite effect, namely an energy gain, is observed when a gas is compressed. 
Mathematically this energy change can be written as a change in momentum (see also Eqn.\ \ref{eq:ParkerEquation}):
\begin{align}
\frac{\mathrm{d}p}{\mathrm{d}t} = -\frac{p}{3}\nabla\cdot\vec{u}(\vec{r}) \quad .
\end{align}
Here, it becomes clear that the sign of the divergence of the advection field $\nabla\cdot\vec{u}(\vec{r})$ determines if the particles gain or lose energy.

This adiabatic cooling is handled with the new energy loss module \texttt{AdiabaticCooling}. It is implemented, like all other loss processes in CRPropa, using an Euler-Forward \emph{ansatz}:
\begin{align}
p_{n+1} &= p_n\cdot\left(1 - \frac{\nabla\cdot\vec{u}(\vec{r}_n)}{3}\Delta t\right) \quad .
\end{align}
The reader is referred to the appendix \ref{appendix} for the validation of these two modules.

\section{Simulation setup}
\label{simulation}

In this section the simulation set up is explained that was used to answer the question of whether CRs are able to propagate back into the Galaxy once they have been accelerated at the termination shock. 

Here, a more sophisticated approach than in \citep{2017ApJ...835...72B}, where only the Reynolds number was used as a measure of the return probability, is developed. To describe 
the CR transport properly, not only advection (galactic wind streaming out of the Galaxy) and diffusion in the Galactic magnetic field but also adiabatic cooling due to the 
Galactic wind have to be taken into account. If a significant fraction of the accelerated particles can make it back into the Galaxy the time scale becomes interesting. It is not 
clear beforehand if the transport of particles takes place on a reasonable time scale compared for example with the lifetime of the termination shock or even the Galaxy itself. This is what is tested here in a quantitative way.

\paragraph{Spherical model } The simplest model, which includes all processes mentioned above, is radially symmetric. Although this quasi-one-dimensional model is probably 
not correct it can be seen as an upper bound of this transport problem. Every other, more complicated model (e.g.\ including a more complex magnetic background field) is very 
likely leading to a decreased CR flux at the Galactic boundary as long as the model covers $4\pi$ sr of the sky. In a Galactic wind model that does not cover the full sphere, 
things might be different. In such a scenario CRs can escape the wind, when perpendicular diffusion is allowed, and diffuse back into the Galaxy unimpeded.

For the first simulation a spatially constant diffusion coefficient $D=5\cdot 10^{28} (\rho/GV)^\alpha~\mathrm{cm}^2 \mathrm{s}^{-1}$ was used. Here, the diffusion index $\alpha=(0.3, 0.4, 0.5, 0.6)$ was varied because it is not clear what the power spectrum of the magnetic turbulence looks like in the Galactic halo; this is the range of exponents usually quoted \citep{2007ARNPS..57..285S}. 

\begin{figure}[htbp]
\centering
\includegraphics[width=\linewidth]{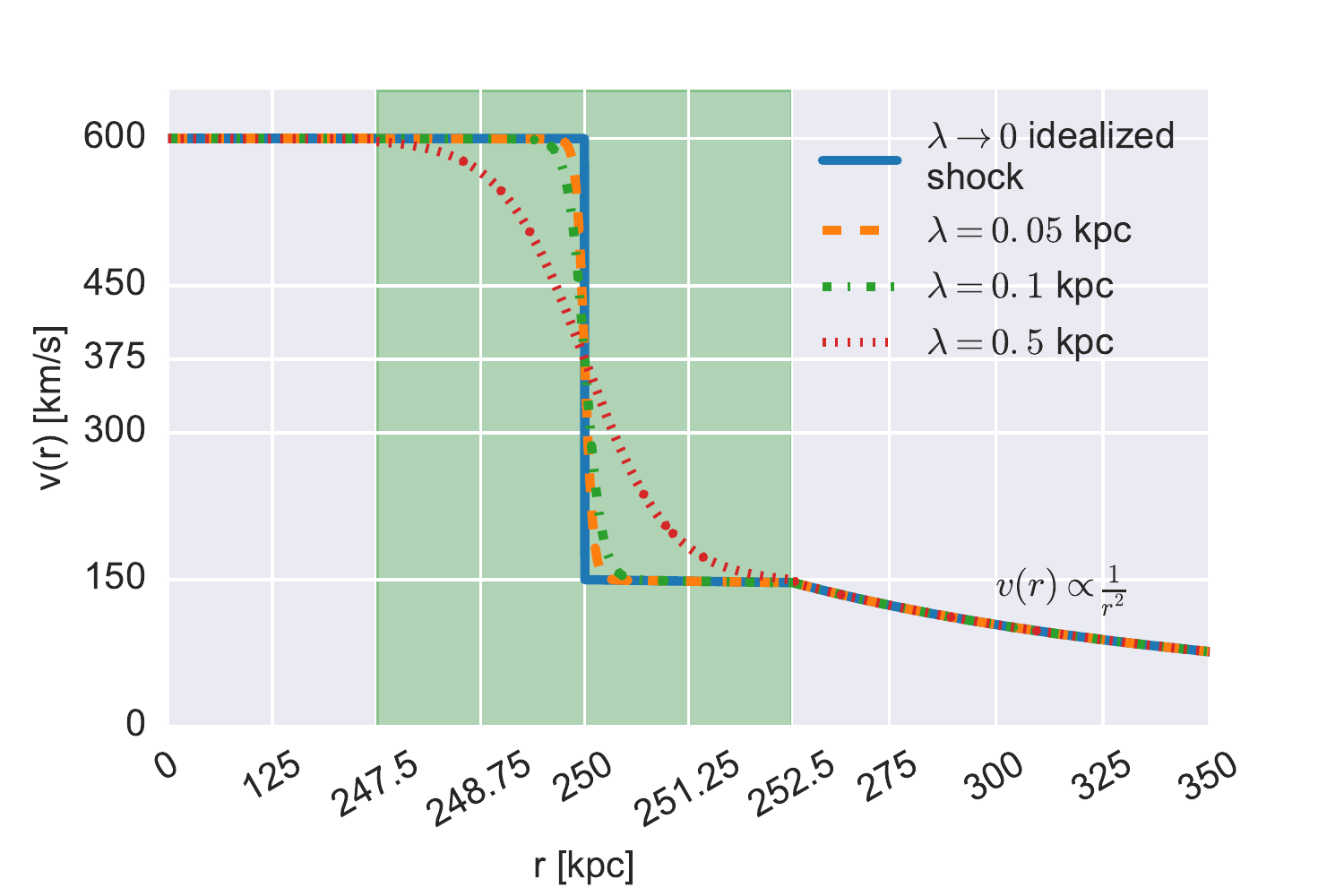}
\caption{The radial component of the galactic wind model used in this work. The ideal shock is shown in blue. Different realizations of Eqn.\ \ref{eq:WindShock} are shown for 
comparison. In this work $\lambda=0.05$~kpc is used. Note that different scales are used for pre-shock, shock (green shaded), and post-shock region.}
\label{fig:WindProfile}
\end{figure}

Although \cite{2017ApJ...835...72B} computed the velocity of a steady, spherically symmetric, radial outflow as a function of $r$, here we adopt a simple analytic fit with the main 
features of the flow:
\begin{align}
u(r) = u_0\left[1+\left(\left(\frac{r_0}{2r}\right)^2-1\right)\frac{1}{1+\mathrm{e}^{-\frac{r-r_0}{\lambda}}}\right] \quad .\label{eq:WindShock}
\end{align}
Here, $u_0$ is the constant\footnote{For simplicity the increase in wind velocity near the center of the galaxy is not modeled in this work, e.g. \citep{2016ApJ...819...29B}.} wind speed for small radii, $r_0$ is the position of the termination shock and $\lambda$ is the shock thickness. Figure \ref{fig:WindProfile} shows the wind profile used for this work. After the shock, the wind velocity decreases with $u(r>r_0)\propto 1/r^2$ which corresponds to a vanishing divergence. In the shock region itself the divergence is negative, which leads to possible re-acceleration of CRs, whereas the divergence in the downstream region ($r<r_0$) is positive ($\nabla\cdot\vec{u}(r)=2u_0/r$) which corresponds to a continuous energy loss.

\paragraph{Archimedean spiral model } It is very likely that the spiral structure of the Galactic magnetic field also has some influence on the magnetic field in the halo and therefore also on cosmic ray propagation. As a first attempt to account for this, we decided to include an Archimedean-spiral as a background field (see e.g.\ \citealt{1987ApJ...312..170J}). Even for this simple magnetic field model:
\begin{align}
\frac{\vec{B}}{B_0} = \underbrace{\left[1-2S\left(\theta - \frac{\pi}{2}\right)\right]}_{\text{change direction at z=0}}\left(\frac{r_{\mathrm{ref}}^2}{r^2}\vec{e}_r - 
\frac{\Omega r_{\mathrm{ref}}^2 \sin(\theta)}{r v_w} \vec{e}_\phi \right) \quad , \label{eq:ArchmedeanSpiral}
\end{align}
where, $\Omega r_\mathrm{ref} \sin(\theta)=200$~$\mathrm{km}\,\mathrm{s}^{-1}$ is the rotational velocity at the reference level $r_{\mathrm{ref}}=10$~kpc and $v_w=600$~$\mathrm{km}\,\mathrm{s}^{-1}$ is the constant wind velocity, 
we expect a significant dependence of the CR flux on the galactic latitude. Since only the magnetic field direction is relevant for this analysis the magnetic field strength at the 
reference level is set to $B_0=1$~T. (The field strength would, of course, be relevant if we explicitly modeled particle gyro-orbits).

To assure that the magnetic field lines ($\vec{B}/B$) and the wind directions ($\vec{u}/u$) are parallel, which is expected from the frozen flux theorem in steady-state MHD, we 
include an additional azimuthal component in our wind model:
\begin{align}
\vec{u}_\phi(r) = \vec{u}(r)\cdot\vec{e}_\phi = u_{\phi, 0}\frac{r_{\mathrm{ref}}}{r} \quad . \label{eq:WindAzimuthal}
\end{align}
Here, the azimuthal wind speed $u_\phi(r_\mathrm{ref})=200$~$\mathrm{km}\,\mathrm{s}^{-1}$ is fixed to the same value as the rotation speed of the magnetic field. 

The CR source is modeled as an infinitesimally thin shell at $r_0=250$~kpc. We do not take the actual acceleration of the CRs into account but most likely the CRs are 
accelerated by diffuse shock acceleration (DSA) leading to an energy spectrum at the source of the form $\mathrm{d}n/\mathrm{d}E\propto E^\gamma$. In this work the index of the 
energy spectrum at the source is fixed to $\gamma=-2$, known from the conventional DSA. 

The simulation volume is confined by two free escape boundaries at $r_{\mathrm{min}}=10$~kpc and $r_{\mathrm{max}}=350$~kpc. Cosmic rays which cross these boundaries are seen as lost into the Galaxy or into the IGM/CGM respectively.

\paragraph{Simulation mode }Depending on the symmetry (one- or three-dimensional) and the spectral index of the diffusion coefficient $\delta$ 
between $N=10^7-2.5\cdot 10^8$ pseudo 
particles are injected into the simulation. As mentioned above, the calculations for the pseudo-particle trajectories are independent of each other. For each particle that enters 
the 
observer sphere at the Galactic boundary $r_\mathrm{obs}=r_\mathrm{min}=10$~kpc the CR properties like the propagation time $T$, the start- and end-position ($\vec{x}_0, \vec{x}$) and the initial and final energy ($E_0, E$) are recorded. For selected simulations in addition, the column density a CR would accumulate under the assumption that the target density scales as $n_\mathrm{target}\propto 1/r^2$ is estimated. These parameters can be used to calculate relevant physical observables (energy spectrum, total luminosity, arrival direction, and neutrino flux) as explained in Sec.\ \ref{greensfunction} and Sec.\ \ref{continuoussource}.

\section{Green's function}
\label{greensfunction}
One of the biggest advantages of the SDE approach is that it is independent of a specific cosmic ray source function. As mentioned above, this allows the same simulation data 
to be used to construct solutions for different physical source scenarios, which helps save computation time. This \emph{ansatz} is explained in more detail using as an example the time 
evolution of the source. The reader is also referred to \cite{MER17} for more information on the construction of stationary solutions.

\begin{figure}[htbp]
\centering
\includegraphics[width=\linewidth]{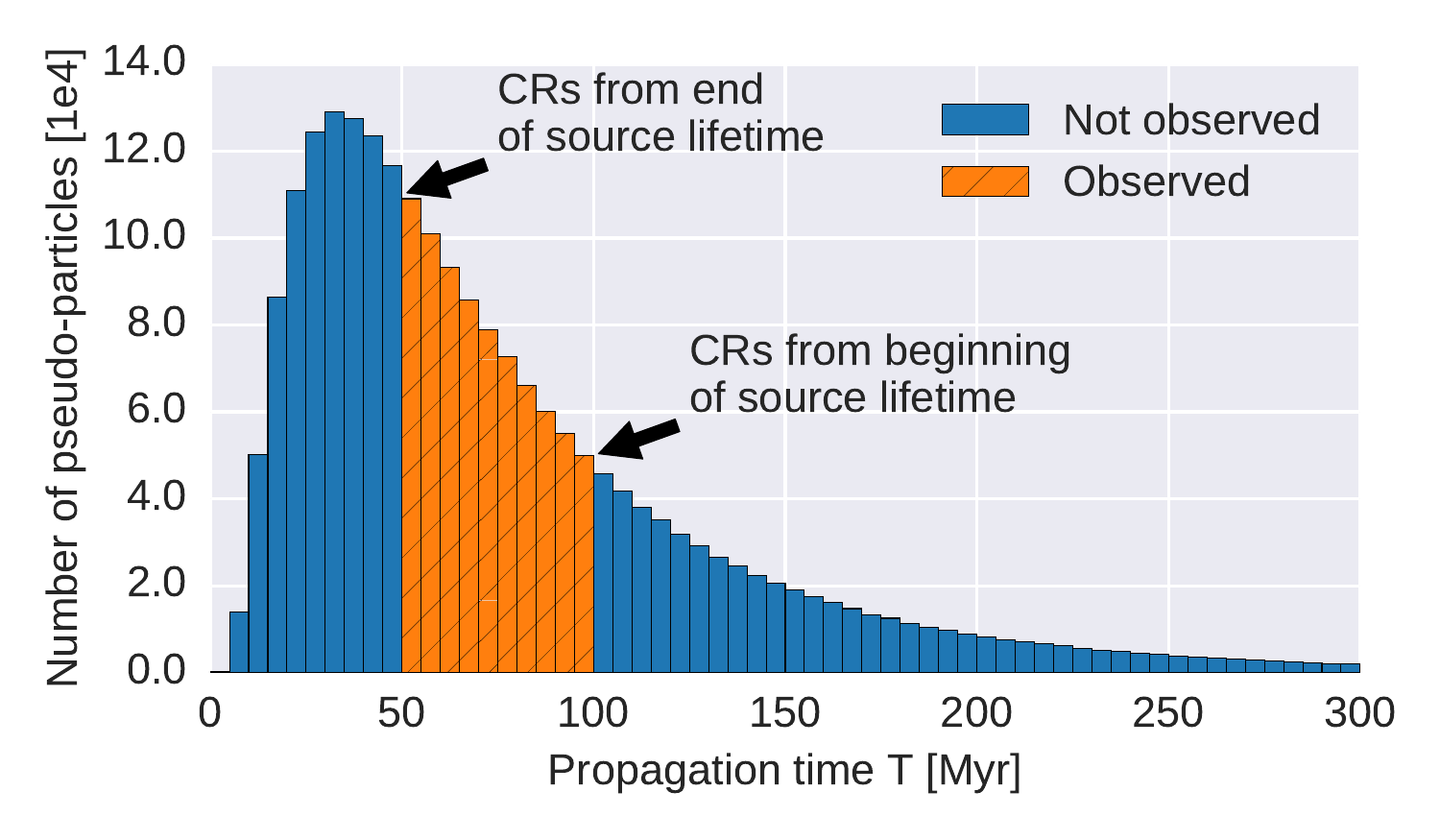}
\caption{The histogram shows number of CRs per propagation time using a diffusion index $\delta=0.5$. This corresponds to the number of observed CRs at time $T$ for a burst like 
injection of particles (Eqn.\ \ref{eq:Source_deltaT}). For a source  with finite duration (Eqn.\ \ref{eq:Source_finite}), active between $t=0$ and $t=50$~Myr, those particles 
that could be observed at $t=100$~Myr are 
shown in (dashed) orange bars. [Tab.\ \ref{tab:Simulations}, Sim.\ 1]}
\label{fig:Green}
\end{figure}

\paragraph{Bursting source } Pseudo-particles in this simulation (see Tab.\ \ref{tab:Simulations}, Sim.\ 1) are injected simultaneously.
This corresponds to a source distribution:
\begin{align}
S_\mathrm{burst}(\vec{r}, p, t) = S_0(\vec{r}, p)\delta(t-t_0) \quad , \label{eq:Source_deltaT}
\end{align}
with a sharp $\delta$-injection at the point in time $t=t_0$. Here, $S_0$ is the number of injected particle. Since each pseudo-particle propagates on an 
independent trajectory they will reach the observer after very different 
propagation times. Figure \ref{fig:Green} (ignore the different colors for now) shows the distribution of the propagation time as a histogram in the case of a Kraichnan 
turbulence 
spectrum with a diffusion spectral 
index 
of $\delta=0.5$ \citep{1965PhFl....8.1385K}. The simulation parameter of this and all other simulation can be found in Table\ \ref{tab:Simulations} in the Appendix. 
From this data we can conclude that a stationary observer would detect a fast increase of the CR flux ($n(r_\mathrm{obs}, t)$) until the maximum is reached at about 
$t_\mathrm{max}\approx 35$~Myr. After that the CR flux is slowly decreasing.

For a closer look at the observed flux of a bursting source, the mean flux for different time bins is calculated, which allows for a qualitative 
assessment of the time evolution. Figure 
\ref{fig:Green} suggests that time bins with equal width lead to very poor statistics for late points in time. To avoid this problem we decided to choose the bin edges such that 
each bin contains the same number of pseudo-particles. Each bin is then characterized by a mean propagation time $\langle T \rangle$ and its width $\Delta T$.

\begin{figure}[htbp]
\centering
\includegraphics[width=\linewidth]{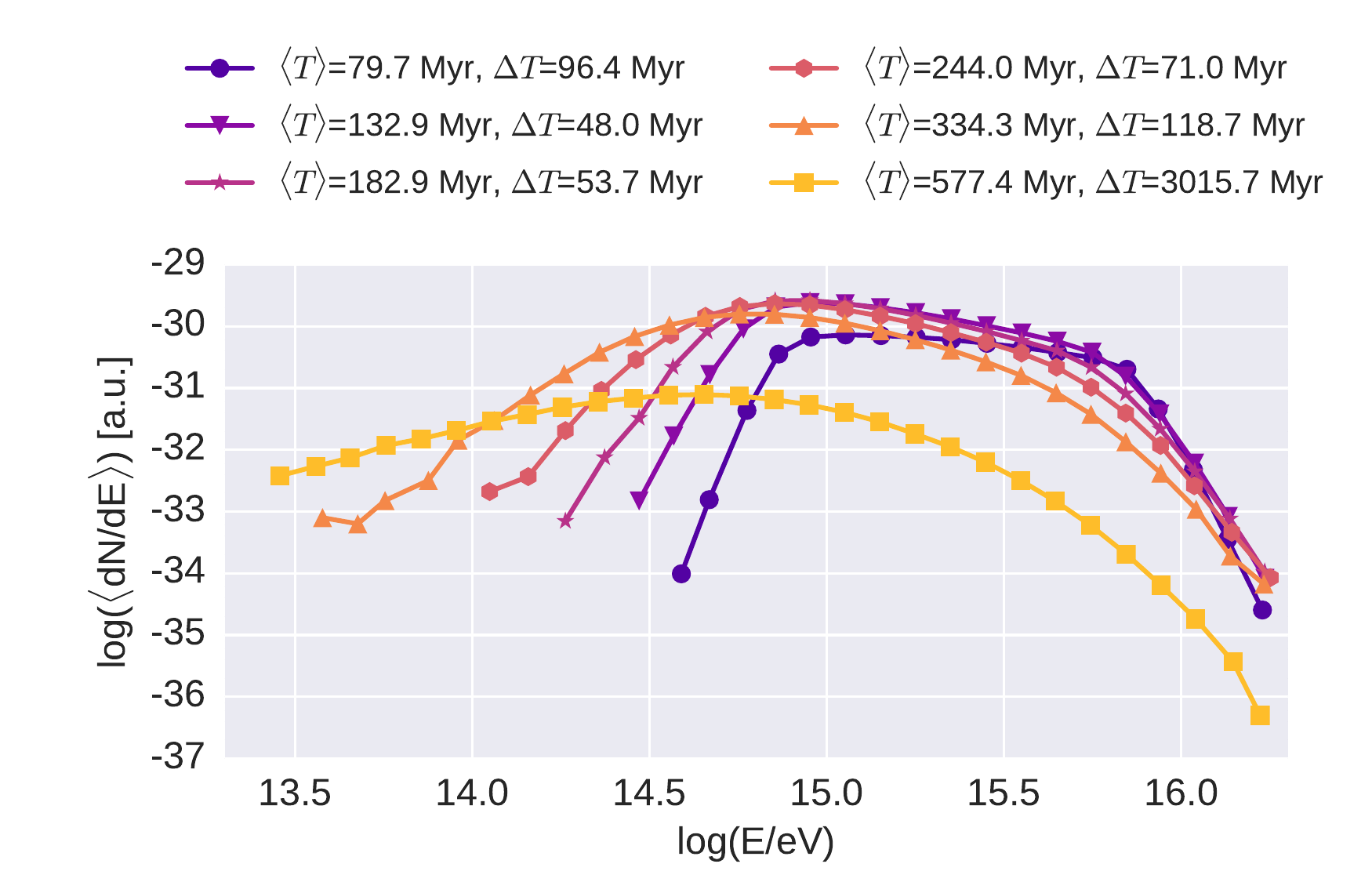}
\includegraphics[width=\linewidth]{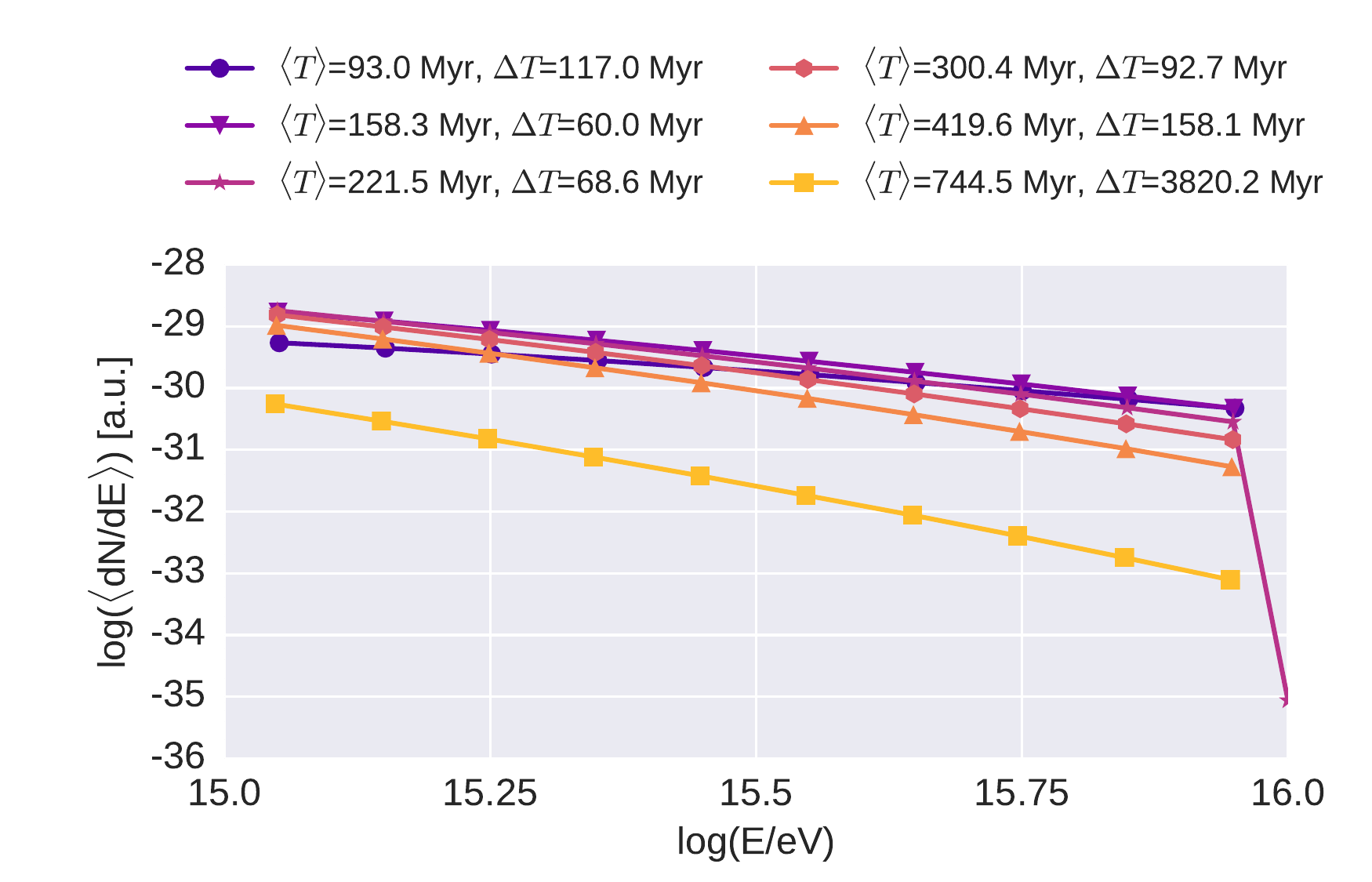}
\caption{Time evolution of a burst like source event with particles injected in the range $10^{15}$-$10^{16}$~eV. Here the diffusion index is $\delta=0.4$. The six energy spectra 
(dark-blue---early, bright-yellow---late) show the change in 
the slope and the cooling of the CRs with time. Upper panel is the time development including a galactic wind [Tab.\ \ref{tab:Simulations}, Sim.\ 2] whereas the lower panel shows the results without winds [Tab.\ \ref{tab:Simulations}, Sim.\ 3].}
\label{fig:TimeEvolution04}
\end{figure}

Figure \ref{fig:TimeEvolution04} shows the time evolution of the observed flux $n(t)$ at $r_\mathrm{obs}$ for a burst like source injection for six time bins. It should be 
noted that the bin width is increasing with time. It is 
clearly visible that, as expected, the CRs are cooled more strongly with increasing propagation time. Furthermore, the energy range does not change if no wind is included in 
the simulation. However, a slight softening of the CR flux $n$ with increasing time is visible. This can be explained by the energy dependence of the diffusion time 
scales---higher energetic particles propagate faster.

\paragraph{Finite source }Nevertheless, such a burst like injection of particles is not very likely for the termination shock. Indeed it is more likely that the shock is active 
for at least several million years and injects CRs 
rather continuously. Let us assume that the source is active for a period in time $\Delta t=50$~Myr starting at $t_0=0$ and injects CRs at a constant rate.  Which CR flux 
($n(r_\mathrm{obs}, t_\mathrm{obs}$)) can be observed at $t_\mathrm{obs}=100$~Myr? The solution can be constructed by the sum of all pseudo-particles with a propagation time 
between $50\leq t_\mathrm{prop}/\mathrm{Myr} \leq 100$ (Orange dashed bars in Fig.\ \ref{fig:Green}). All pseudo-particles that propagated faster than $t_\mathrm{prop}\leq 50$~Myr 
have passed the observer already and all particles that needed more than $t_\mathrm{prop}\geq 100$~Myr could not have reached the observer yet, since the source was not active that 
long ago. In other words, for a source distribution:
\begin{align}
S_\mathrm{finite}(\vec{r}, p, t) = \tilde{S}_0(\vec{r}, p)\Theta(t-t_0)\Theta(t_1-t) \quad , \label{eq:Source_finite}
\end{align}
where $\Theta(t)$ is the Heaviside-step-function, all pseudo-particles with a propagation time $t-t_1\leq t_\mathrm{prop} \leq t-t_0$ have to be taken into account. Here, 
$\tilde{S}_0$ is a particle rate so that the units of $S_\mathrm{finite}$ and $S_\mathrm{burst}$ have the same unit. This method is 
applied in Sec.\ \ref{continuoussource} to calculate the flux for different source scenarios.

\subsection{Energy change}
\label{sec:EnergyChange}

The chosen GTS model described leads naturally to a re-acceleration of the CRs. The large change (factor 4) in the wind velocity around the shock at $r_0$ leads to 
a strong compression of the medium. This compression leads to an energy gain for particles staying in this region. After the CRs leave the shock region they will either lose
energy (propagating into the direction of the Galaxy) or maintain their energy (propagating outwards).
Beforehand it is not possible to determine the energy change for a specific 
pseudo-particle. In addition, such a single particle treatment is not physically meaningful in the context of SDE. However, the distribution of energy gains or losses for 
a particle ensemble might give some insights into the re-acceleration  of CRs.

\begin{figure}[htbp]
\centering
\includegraphics[width=\linewidth]{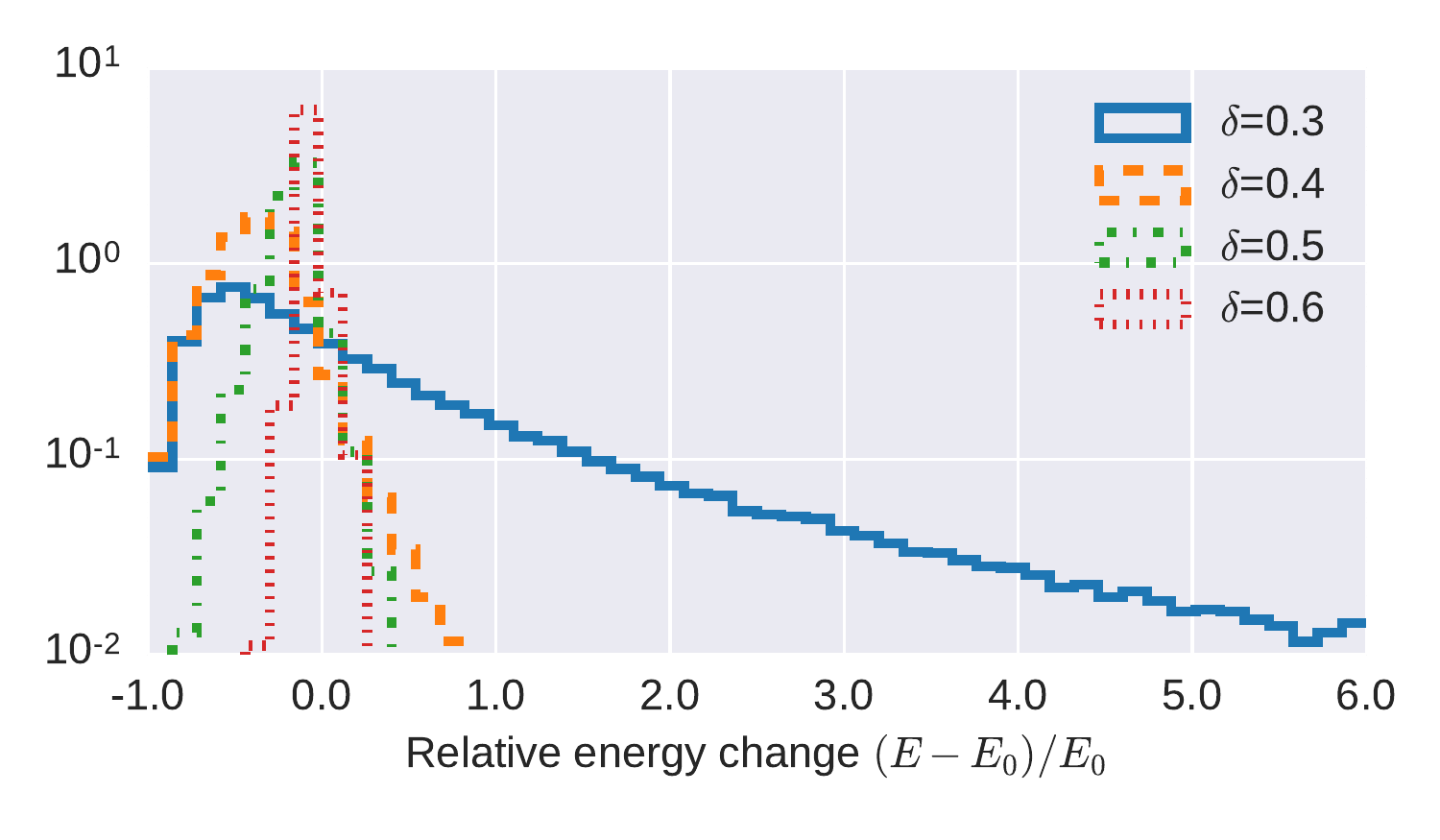}
\includegraphics[width=\linewidth]{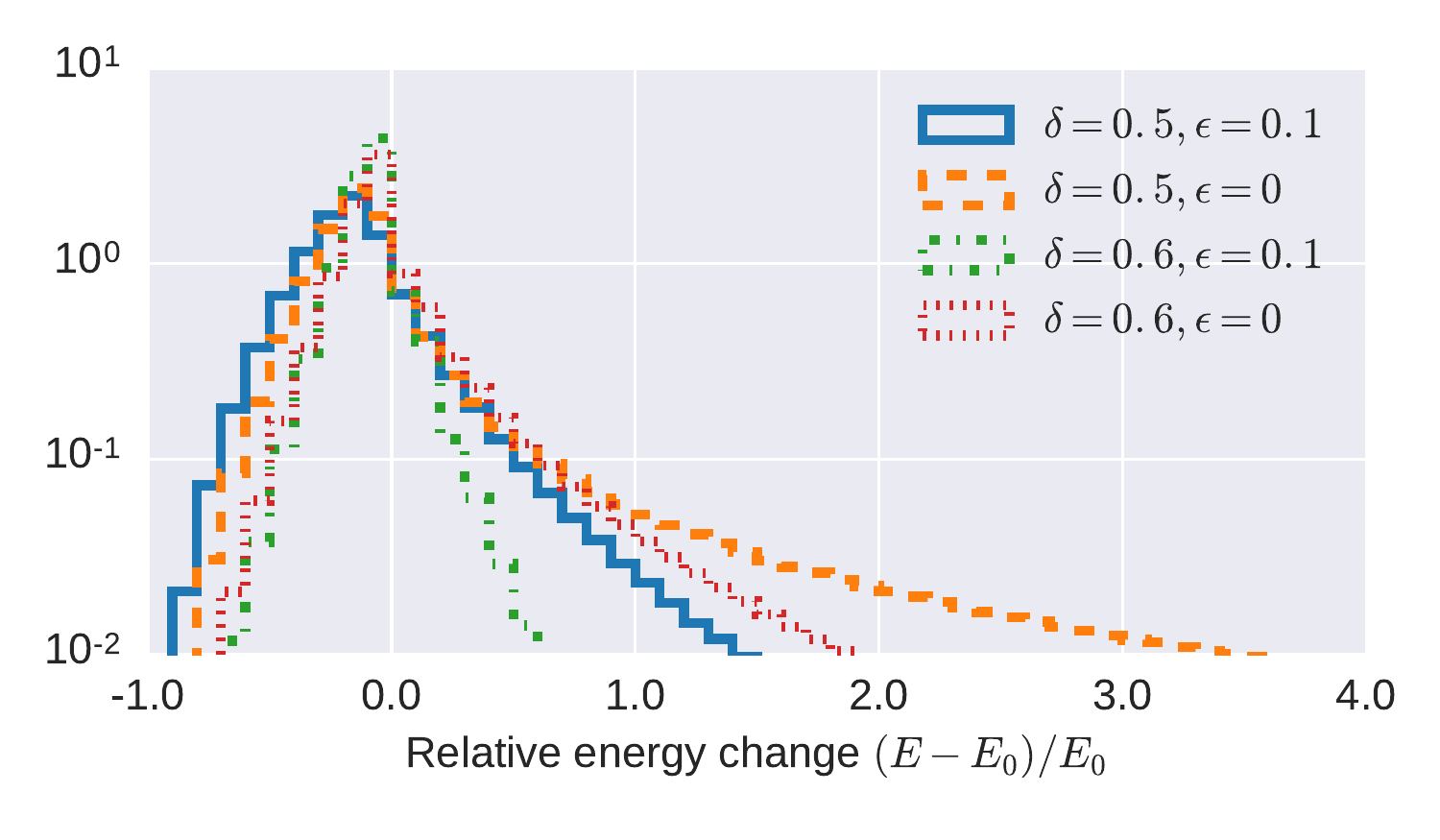}
\caption{Relative energy change of the CR for the one dimensional scenario (upper) [Tab.\ \ref{tab:Simulations}, Sim.\ 1, 2, 4, and 5] and the three dimensional symmetry (lower) [Tab.\ \ref{tab:Simulations}, Sim.\ 6-9]. Here, only those CRs that reach 
the Galactic observer are considered. It is clearly visible that most of the CRs loose energy. But despite the large volume with an expanding wind some CR gain a significant 
amount 
(up to a factor 2) of energy. The effect is strongest for small diffusion indices $\delta$ and parallel diffusion ($\epsilon=0$).}
\label{fig:EnergyChange}
\end{figure}

Figure \ref{fig:EnergyChange} shows the relative energy change of CRs that reached the Galaxy for different diffusion models---Tab.\ \ref{tab:Simulations}, Sim.\ 1, 2, 4, and 5: upper panel; Tab.\ \ref{tab:Simulations}, Sim.\ 6-9: lower panel. The wind velocities are chosen as specified above. 
Most of the CRs lose
energy because of the outward directed wind. Nevertheless, some CRs are found that nearly double their initial energy by this 
re-acceleration process (see e.g.\ \citealt{2016A&A...595A..33T}). How large this maximum energy gain is and how many CRs on average gain energy during the total transport depends on the time they spend in the different 
regions of the wind. The more time is spent in the shock region, the higher is the net energy gain. This explains the huge difference of the energy-change-distributions for 
different diffusion indices $\delta$. The smaller the mean free path in the shock region is, the larger is their energy gain. Since high energy CRs diffuse faster they gain less energy on average.

\section{Continuous source emission}
\label{continuoussource}
In this section we choose a continuous source emission as one example to demonstrate the flexibility of the SDE approach. 
In total we examine four different simulation scenarios.
Two sets of models, one with a simple one dimensional (or radially symmetric) magnetic background field and the other with a more sophisticated three dimensional 
(Archimedean spiral) background field are studied, each one with and without a Galactic wind. One might argue that a vanishing Galatic wind is not 
realistic because without a wind no termination shock forms. This is correct but two situations may make such considerations interesting. First, the termination shock and the wind 
structure might not cover the full $4\pi$ sphere but only parts. Here, the real observation would be a superposition between the cases with and without wind. Secondly, the CR 
source may not be active any more because the wind has shut down. Here, already accelerated CR are still able to diffuse back into the Galaxy but are no longer advected outward. 
Due to the lack of statistics, only steep diffusion models ($\delta\geq 0.5$) can be tested for the three-dimensional scenario. In addition, two different diffusion tensors with 
pure parallel ($\epsilon:=\kappa_\perp/\kappa_\parallel=0$) and strong perpendicular diffusion ($\epsilon=0.1$) are tested.

As explained in Sec.\ \ref{greensfunction} it is straightforward to identify those pseudo-particles that have to be taken into account depending on the source duration and the 
observation point in time. Here, the calculation of the correct normalization is discussed. The total cosmic ray luminosity of the termination shock is assumed to be:
\begin{align}
 L_\mathrm{CR} &= \int_{E_\mathrm{min}}^{E_\mathrm{max}} \frac{\mathrm{d} N}{\mathrm{d} E} E \; \mathrm{d}E 
 = \int_{E_\mathrm{min}}^{E_\mathrm{max}} N_0 E^{-2} E \; \mathrm{d}E \\
 &= N_0 \log\left(\frac{E_\mathrm{max}}{E_\mathrm{min}}\right) \stackrel{!}{=} 10^{40}~\frac{\mathrm{erg}}{\mathrm{s}} \quad , \label{eq:Luminosity}
\end{align}
where a spectral index of the energy power law $\gamma=-2$ is assumed. Furthermore, the minimum cosmic ray energy is $E_\mathrm{min}=10^{9}$~eV and the maximum energy is 
$E_\mathrm{max}=10^{16}$~eV, which is based on the upper limit of the CR energy given in \cite{2017ApJ...835...72B}.
Since not the whole energy range but only those CRs with energies above $E_1=10^{15}$~eV are simulated the fraction of the luminosity in the simulated energy range is 
$f_L = L_\mathrm{CR}/L_\mathrm{sim}=1/7$. Now the weighting $w$ of the simulated pseudo-particles can be defined: $w := L_\mathrm{sim}/\sum_i E_i$. Here, the total energy of all 
simulated pseudo-particles $E_\mathrm{tot}=\sum_i E_i$ is used to calculate the weight $w$ which has the unit of $[w]=\mathrm{time}^{-1}$. The flux at the observer sphere per 
energy bin yields:
\begin{align}
\frac{\mathrm{d}N}{\mathrm{d}E}_\mathrm{sim} (E+\Delta E, t) = \sum_{\epsilon, \tau} \frac{w}{\Delta E\cdot 4\pi \cdot4\pi r_\mathrm{obs}^2} \quad , \label{eq:ObservedFlux}
\end{align}
where all pseudo-particles with observation time $\tau\in [t-\Delta t, t]$ and energy $\epsilon \in [E, E+\Delta E]$ are summed and $r_\mathrm{obs}$ is the radius of the observer 
sphere. These values have units $(\mathrm{TeV}\,\mathrm{m}^2\,\mathrm{s}\,\mathrm{sr})^{-1}$. The data should not be compared with observational data directly, since the flux is 
calculated for an observer at the edge of the Galaxy and not at Earth. We assume that the total energy budget of the CRs of the GTS is not altered too much during the additional propagation through the Galaxy. However the the shape of the energy spectrum as well as the arrival are likely to change significantly.

A source duration of $\Delta t=100$~Myr was used for the results of this work. This is probably a reasonable duration for a starburst-driven outflow and resulting termination shock \citep{2010ApJ...724...49M, 2017arXiv170504692M}, but the shock stability may be affected by shorter duration ``flickers" of the outflow. It should be noted that a different source duration---or even time dependent source evolution---does not require a new simulation but only a different 
weighting of the existing data, but this is beyond of the scope of this paper.

\begin{figure*}[htbp]
\centering
\begin{minipage}{.49\textwidth}
\includegraphics[width=\linewidth]{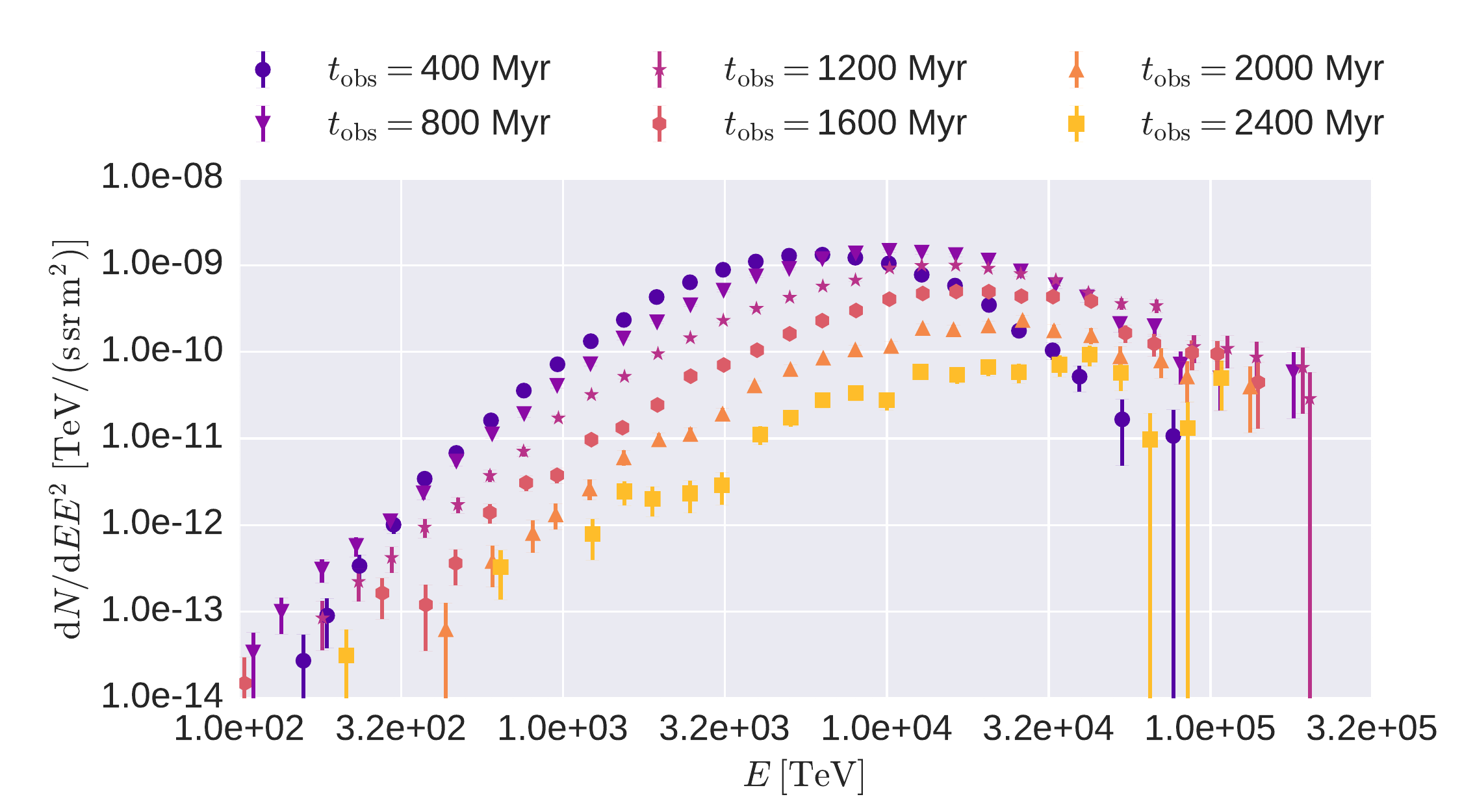}
\includegraphics[width=\linewidth]{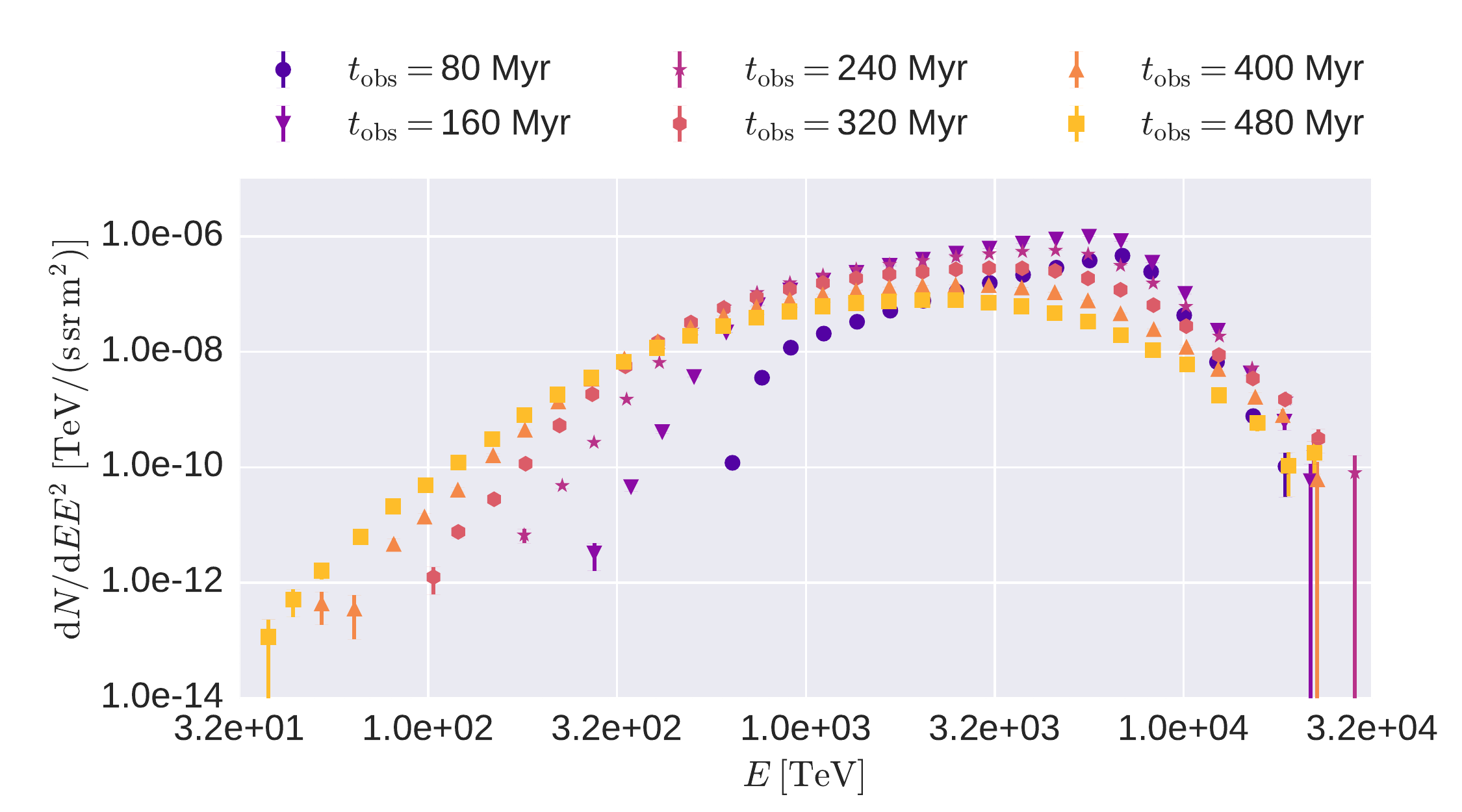}
\includegraphics[width=\linewidth]{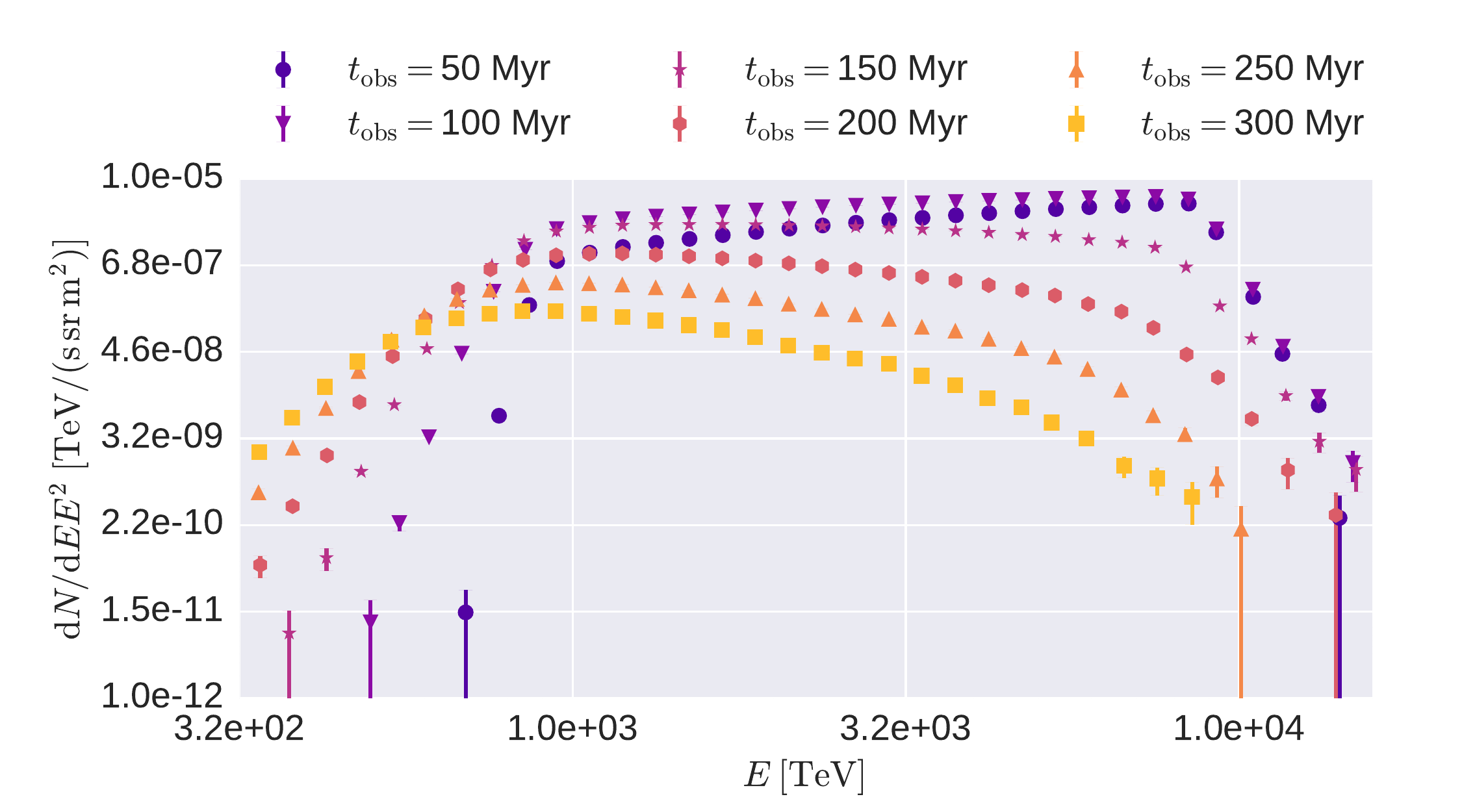}
\includegraphics[width=\linewidth]{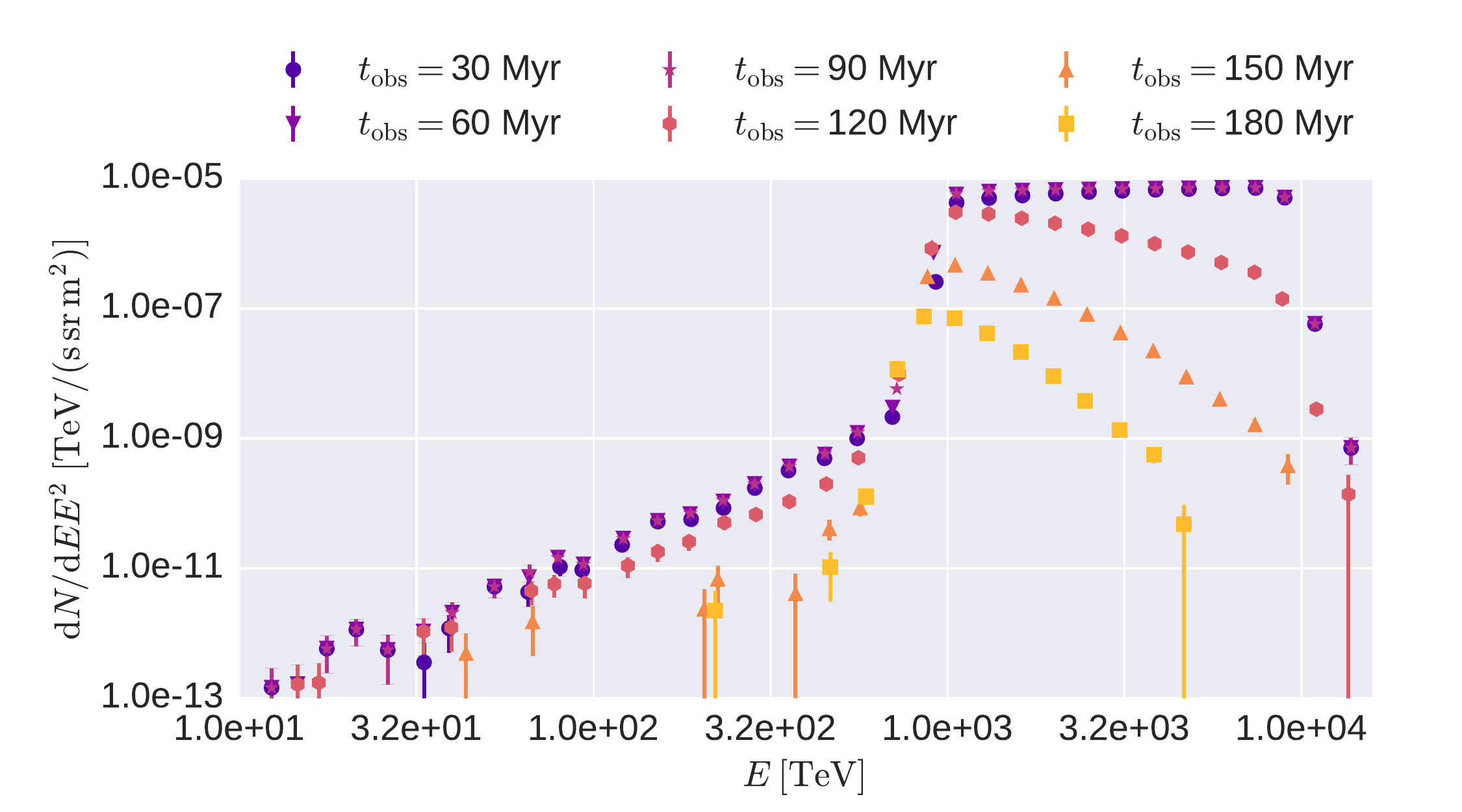}
\end{minipage}%
\begin{minipage}{.49\textwidth}
\includegraphics[width=\linewidth]{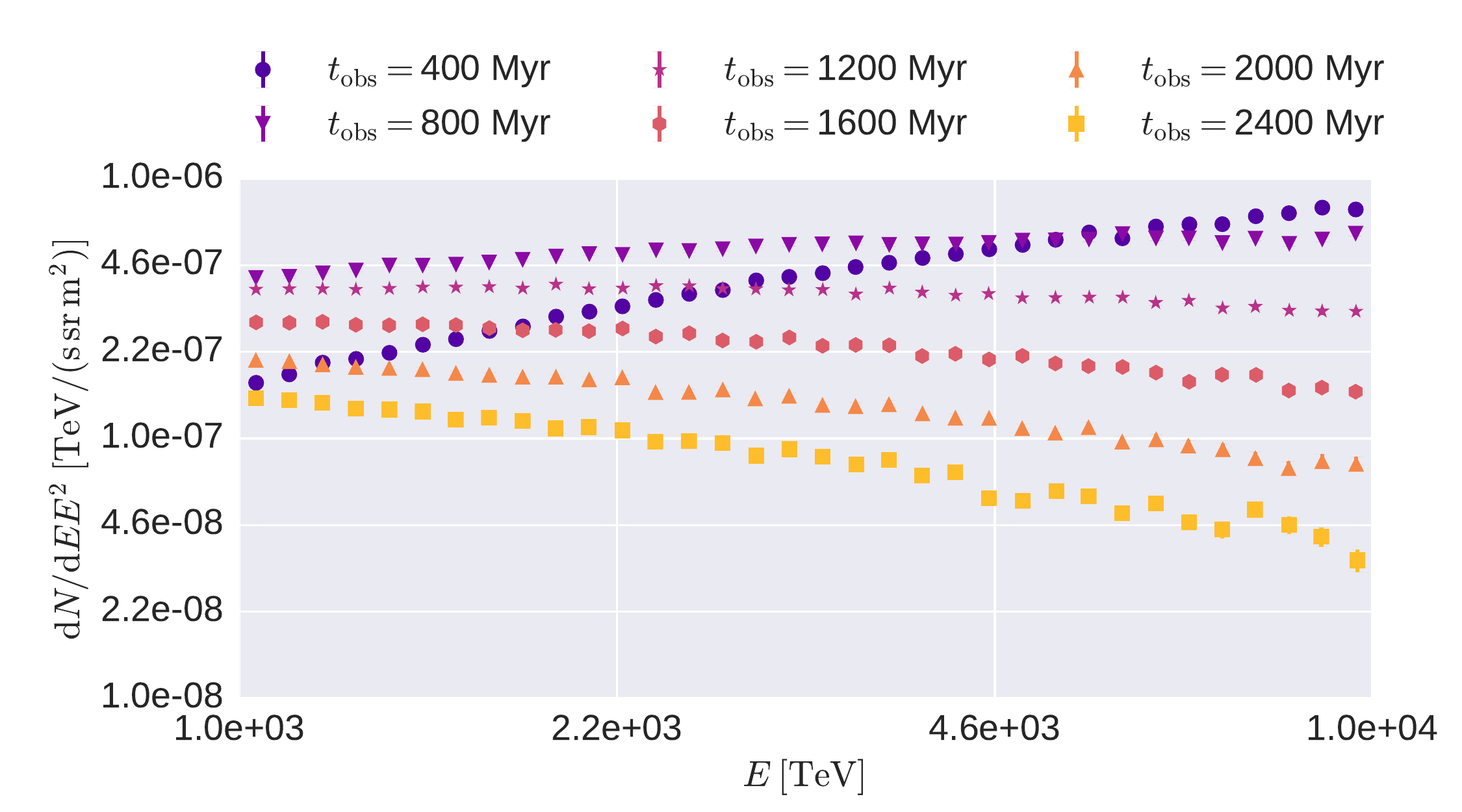}
\includegraphics[width=\linewidth]{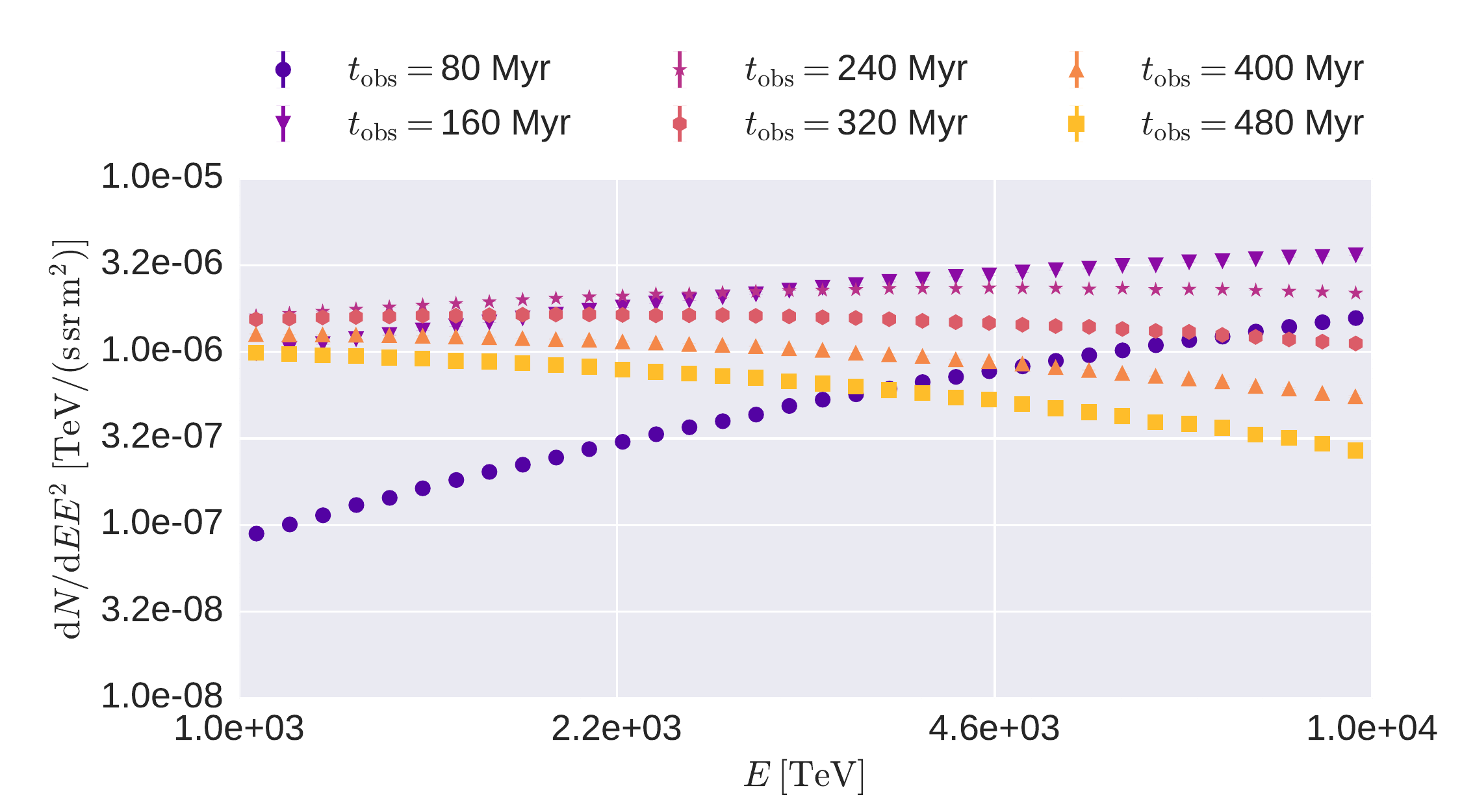}
\includegraphics[width=\linewidth]{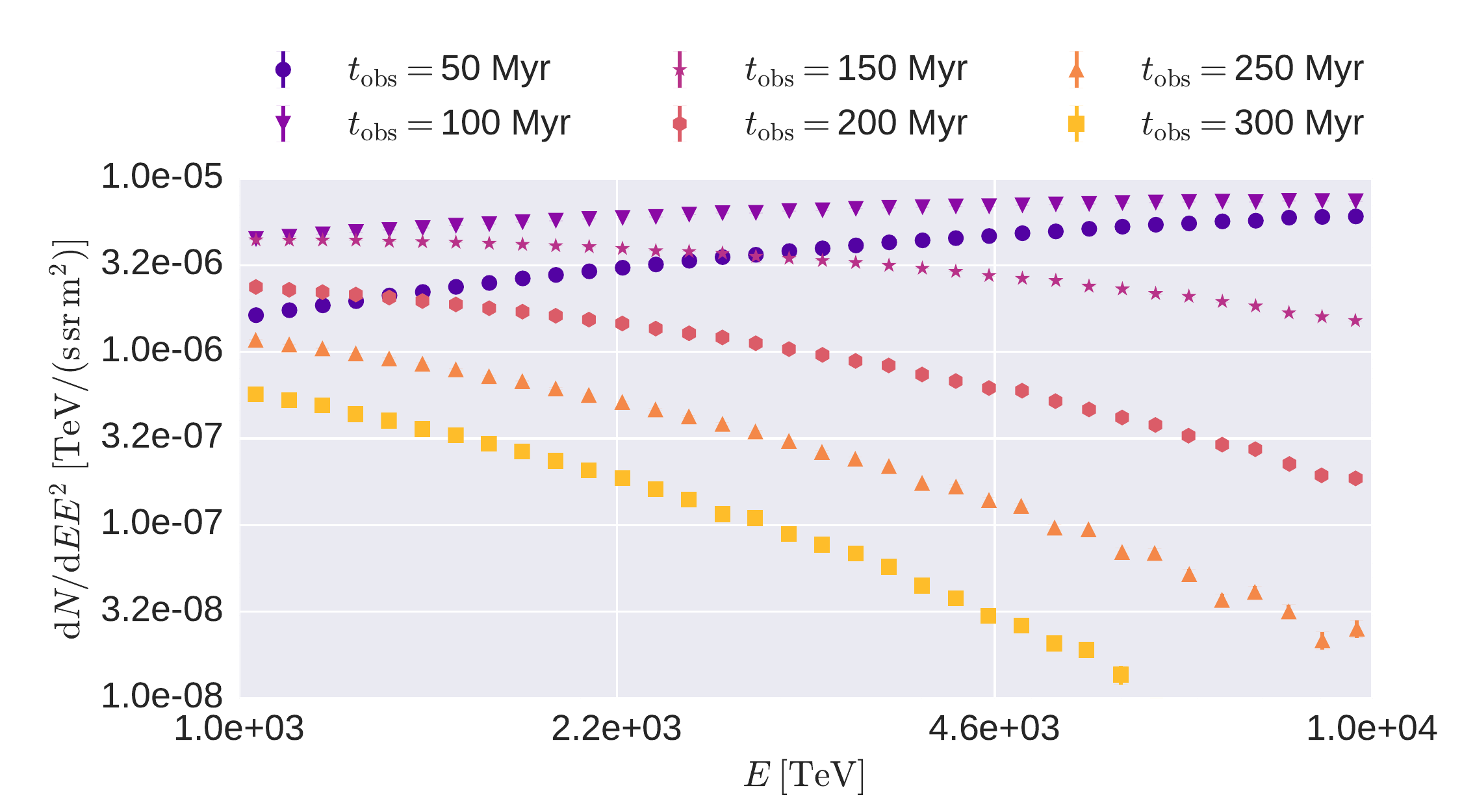}
\includegraphics[width=\linewidth]{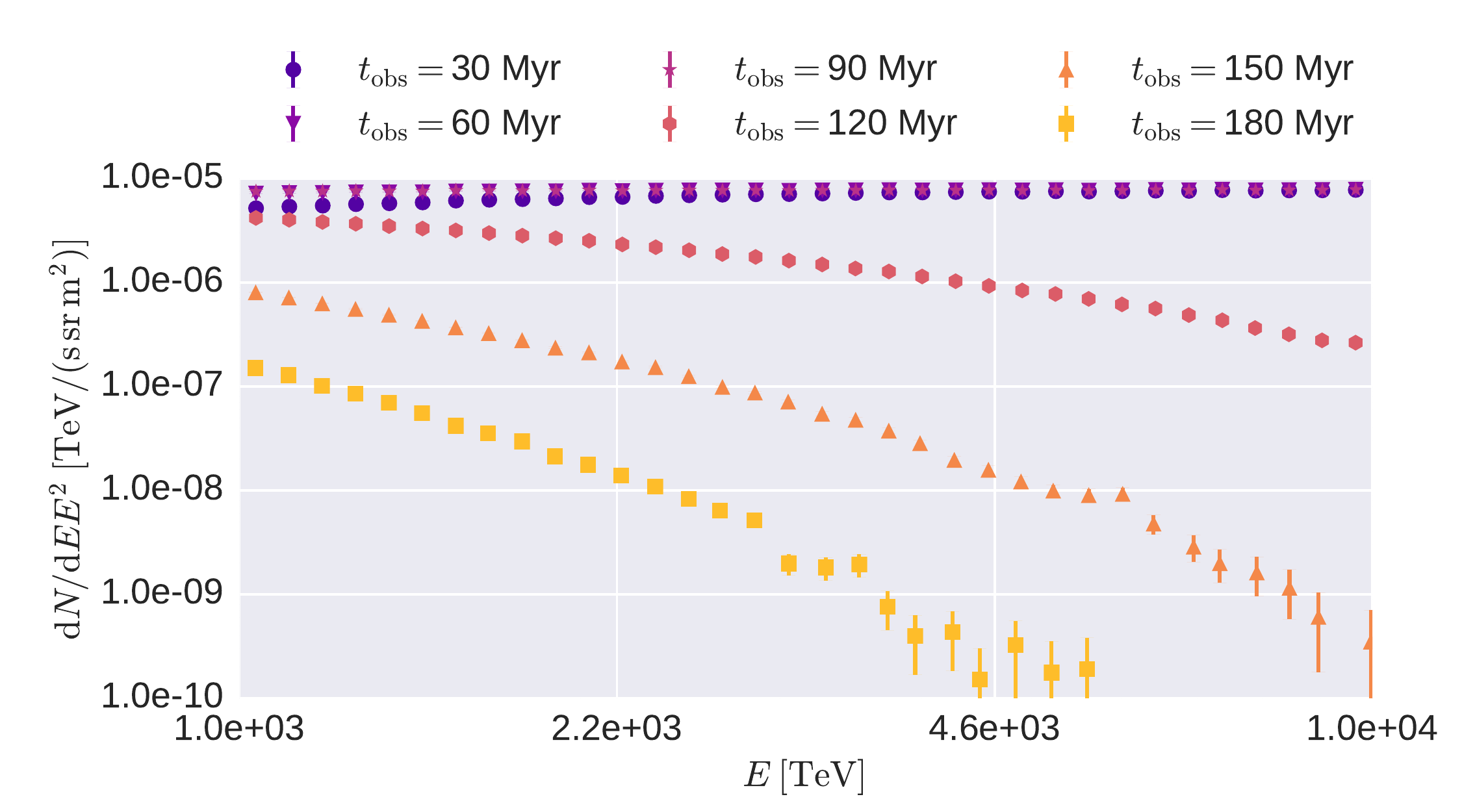}
\end{minipage}
\caption{Cosmic ray flux evolution for the one-dimensional scenario. From top to bottom the spectral index is increasing $\delta=(0.3, 0.4, 0.5, 0.6)$. Left column shows the 
results with wind [Tab.\ \ref{tab:Simulations}, Sim.\ 1, 2, 4, and 5] and the right column the results without a galactic wind [Tab.\ \ref{tab:Simulations}, Sim.\ 3, and 10-12], respectively. Note the different time and energy scaling.}
\label{fig:Spectrum_1Dim}
\end{figure*}

Figures \ref{fig:Spectrum_1Dim} shows the time evolution of the observed flux at the Galactic boundary $r_\mathrm{obs}=10$~kpc. Here, four simulations with different diffusion 
indices $\delta$ including (left column) and neglecting (right column) a Galactic wind are shown. If CRs are described by diffusion only, the most obvious observation is the change 
in slope of the energy spectrum with time. The spectrum is softening with time, which is expected since higher energetic particles propagate faster. This feature is universal 
and does not depend on the diffusion index $\delta$. Nevertheless, the time scale and the range of the difference in the slope does depend strongly on $\delta$. For Kolmogorov 
diffusion, the spectral index changes from $\gamma(t_\mathrm{obs}=400\,\mathrm{Myr})\approx -1.3$ to $\gamma(t_\mathrm{obs}=2.4\,\mathrm{Gyr})\approx -2.6$, but for a very steep 
diffusion index $\delta=0.6$ the spectral slopes changes from $\gamma\approx-1.8$ to $\gamma\approx-5.8$ within only 150 million years.

The energy spectra become more complex when, in addition, advection by the Galactic wind is considered. First it can be noted that, as discussed before, CRs may gain or lose energy due to the adiabatic energy change, which leads to an energy spectrum reaching beyond the injection limits: $E\leq10^{15}$~eV and $E\geq10^{16}$~eV. Furthermore, a simple description of the spectral shape using a single spectral index $\gamma$ is not adequate any more. It is hard to characterize universal features that are present in all simulation models at all times. But if we pick for example the energy spectrum for $\delta=0.4$ at $t_\mathrm{obs}=160$~Myr we can identify three different ranges of the energy spectrum. The low energy part ranging from $10^{2.5}\leq E/\mathrm{TeV} \leq 10^3$ is loss dominated due to adiabatic cooling. The medium part is diffusion dominated and shows a similar time evolution as the simulation without advection. The last, high energy part, starting at $E\approx 6\cdot 10^3$~TeV is much steeper than the injection energy spectral index $\gamma_\mathrm{init}=-2$ and is related to re-acceleration in the shock region. The boundaries and slopes of these regions are changing with time and diffusion index $\delta$.

\begin{figure*}[htbp]
\centering
\begin{minipage}{.49\textwidth}
 \includegraphics[width=\linewidth]{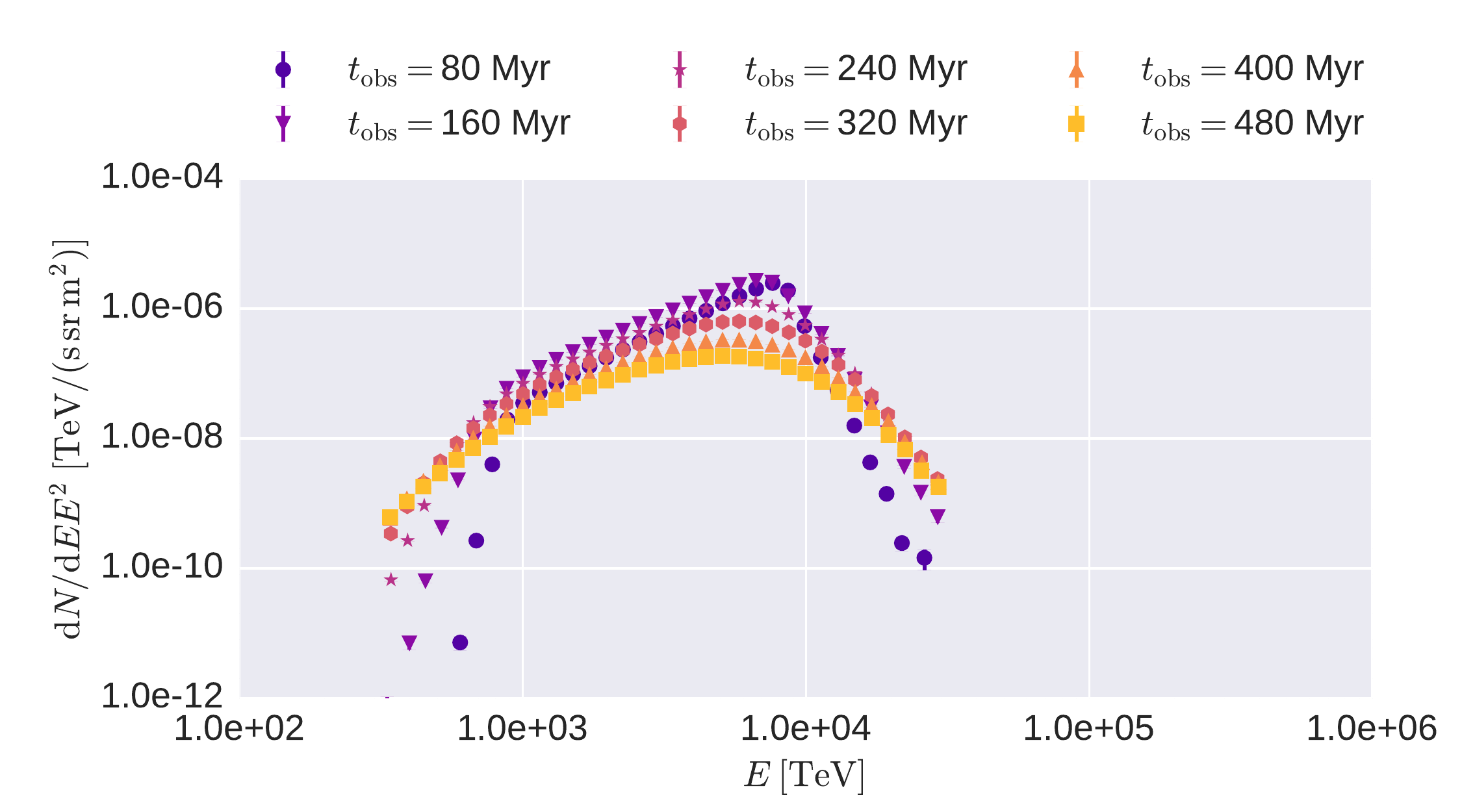}
 \includegraphics[width=\linewidth]{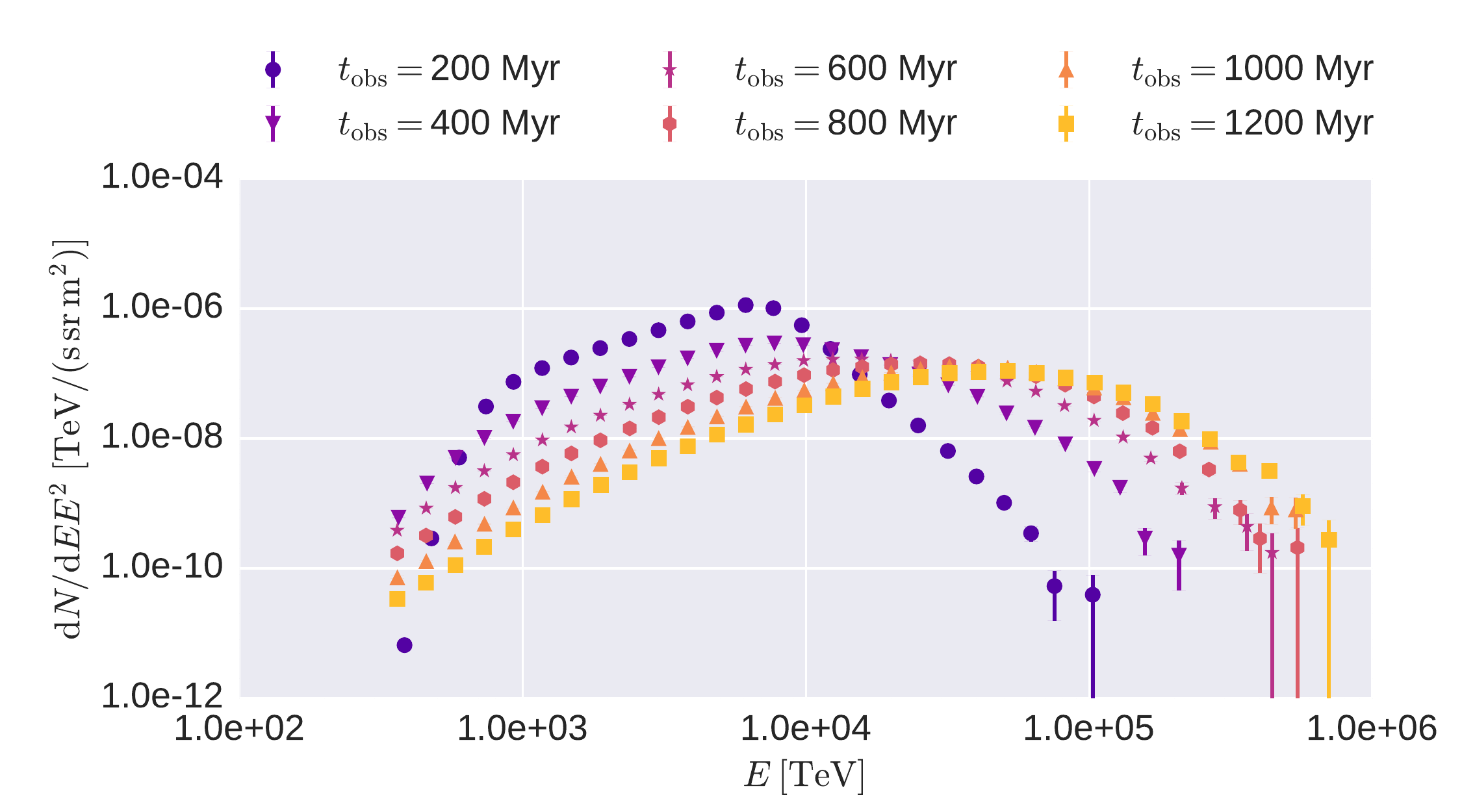}
\end{minipage}
\begin{minipage}{.49\textwidth}
 \includegraphics[width=\linewidth]{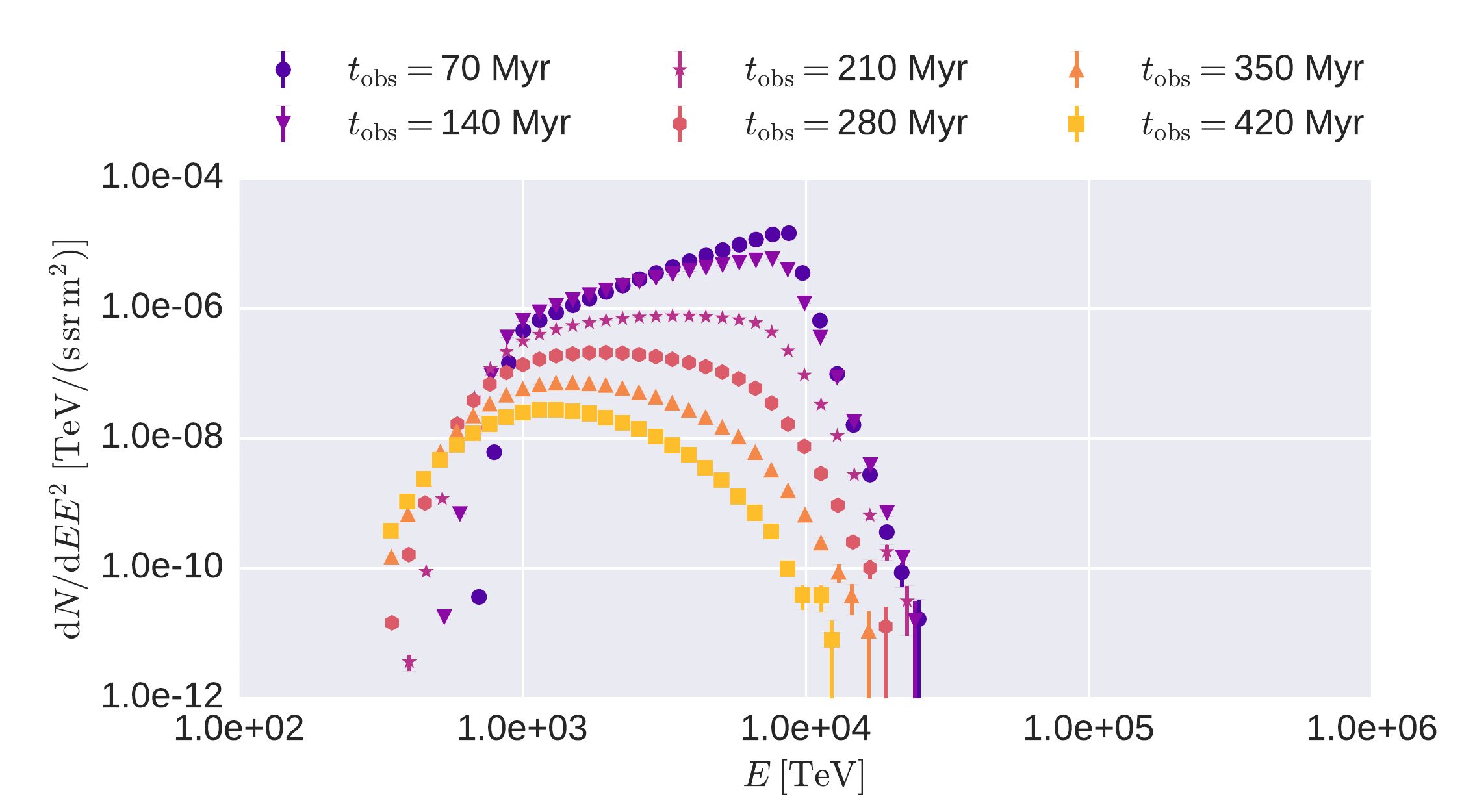}
 \includegraphics[width=\linewidth]{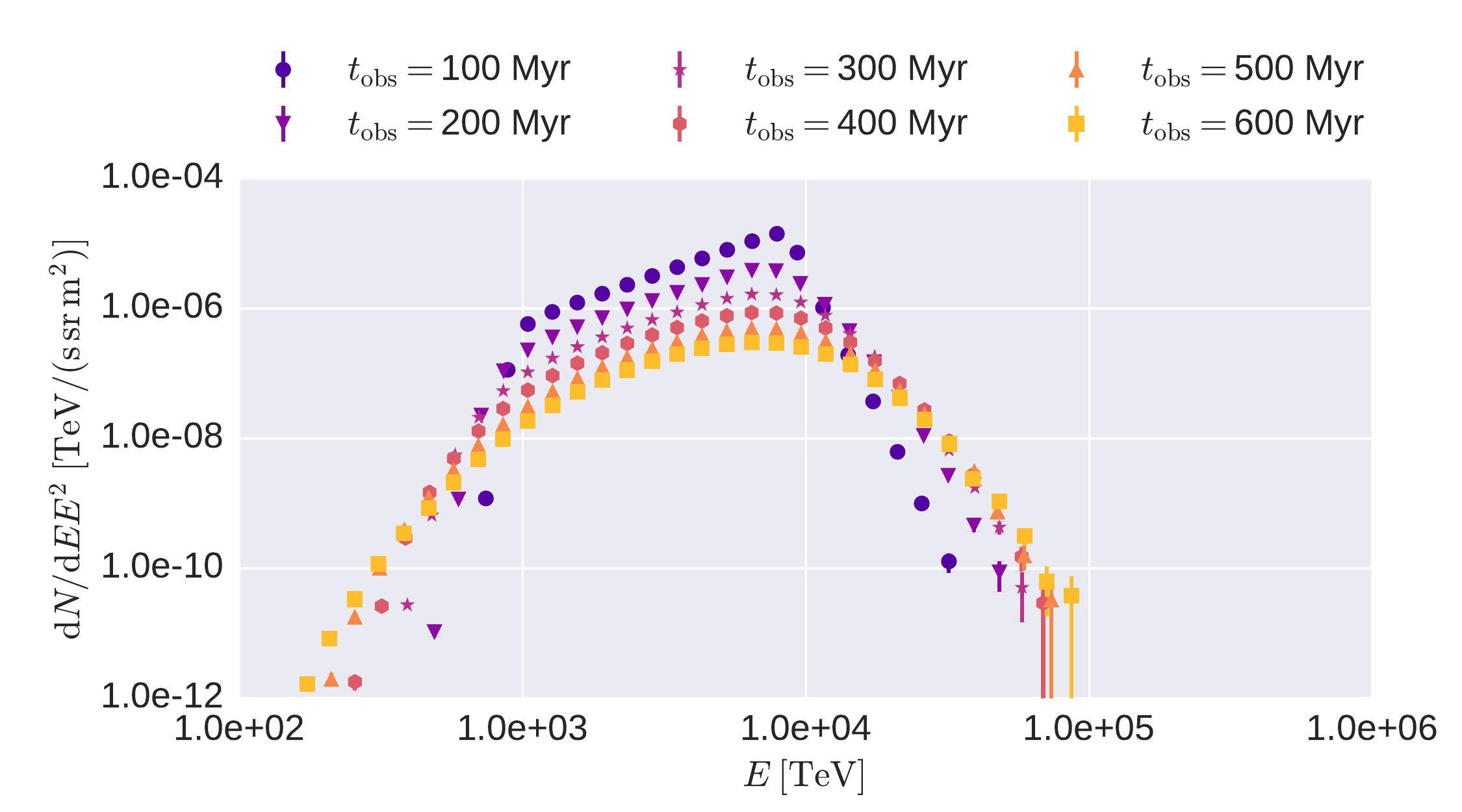}
\end{minipage}
\caption{Time evolution (color coded) of the CR energy spectrum for the three dimensional model. From top left to bottom right the following simulations are shown: Tab.\ \ref{tab:Simulations}, Sim.\ 6, 8, 7, and 9. So top row shows perpendicular and bottom row shows pure parallel diffusion. Left column is for $\delta=0.5$ and right column corresponds to $\delta=0.6$, respectively. Note the different time scales for all four simulations.}
\label{fig:Spectrum_3Dim_equalBinning}
\end{figure*}

For the three dimensional diffusion model the same analysis of the energy spectrum was done as in the one dimensional case. Here, we concentrate on the simulations including the 
full transport process (diffusion, advection, and adiabatic cooling). Figure \ref{fig:Spectrum_3Dim_equalBinning} shows that the general shape of the spectra can be compared with the one dimensional model. As before, three different, more or less 
pronounced, sections can be 
identified in the energy spectrum. Generally, one may notice that a purely parallel diffusion process ($\epsilon=0$) leads to a narrower 
energy spectrum. In particular, the loss- and gain-regions are increased for the parallel compared with the perpendicular ($\epsilon=0.1$) diffusion model. So the spectrum becomes 
broader and the maximum flux is decreased compared with the perpendicular model. The diffusion dominated (medium energy) region of the energy spectrum is consistent with energy 
range of the injected CRs. The spectrum becomes smoother with time and as before this process depends very much on the diffusion time scale: Meaning the process is much faster for 
a steeper diffusion spectrum.

\begin{figure}[htbp]
 \centering
 \includegraphics[width=\linewidth]{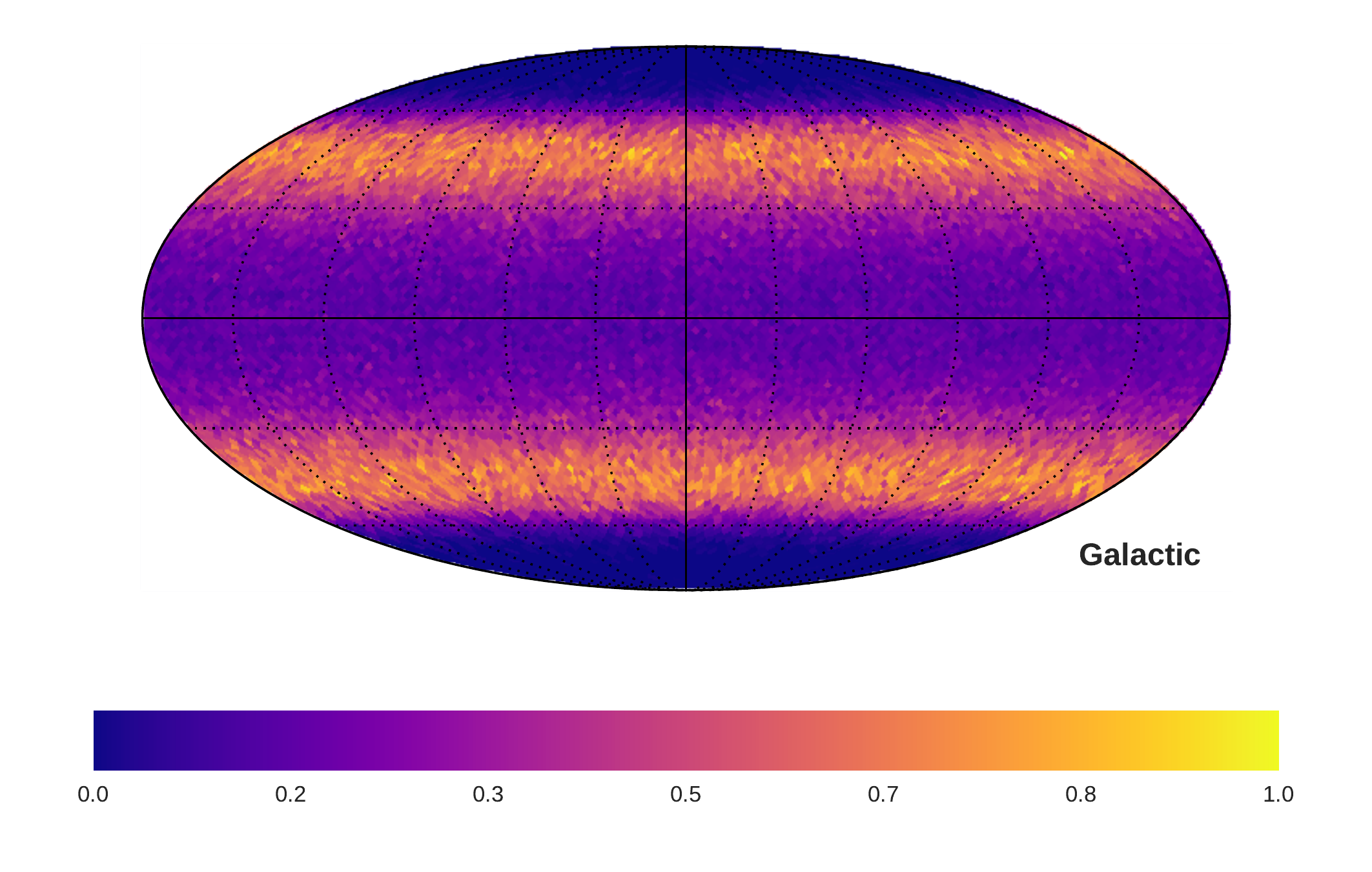}
 \caption{Arrival direction of CRs as seen at the $r_\mathrm{obs}=10$~kpc observer sphere at the edge of the galaxy. The total proton luminosity per solid angle is shown in galactic 
coordinates, where lighter colors refer to a higher flux. Here, a pure parallel diffusion ($\epsilon=0.$) with a steep diffusion index ($\delta=0.6$) is shown for an observation 
point at time $t_\mathrm{obs}=300$~Myr. [Tab.\ \ref{tab:Simulations}, Sim.\ 9]}
 \label{fig:Arrival_2d}
\end{figure}

For the full three dimensional analysis not only the energy spectrum but also the arrival direction is of interest.\footnote{Here, `arrival direction' always refers to the observed 
CR position at $r_\mathrm{obs}=10$~kpc. The results must not be compared with experimental data observed at Earth because the galactic propagation is not simulated in this work.} 
Figure \ref{fig:Arrival_2d} shows the density of CRs arrival direction in galactic projection using the HEALPix\footnote{http://healpix.sourceforge.net/}-equal-area pixelization. Here, the cosmic ray distribution for a steep, pure parallel ($\delta=0.6, \epsilon=0.$) diffusion model observed 200 million years after the accelerator shut down 
($t_\mathrm{obs}=300$~Myr) is shown. The arrival distribution has a prominent double-ring feature where the maximum CR flux is detected in two narrow bands with galactic latitudes 
between $30^\circ\lesssim \Delta l \lesssim 60^\circ$. Furthermore, almost no flux reaches the Galaxy at the poles while the equator region is reached by a small smeared out flux.

\begin{figure*}[htbp]
\centering
\begin{minipage}{.49\textwidth}
 \includegraphics[width=\linewidth]{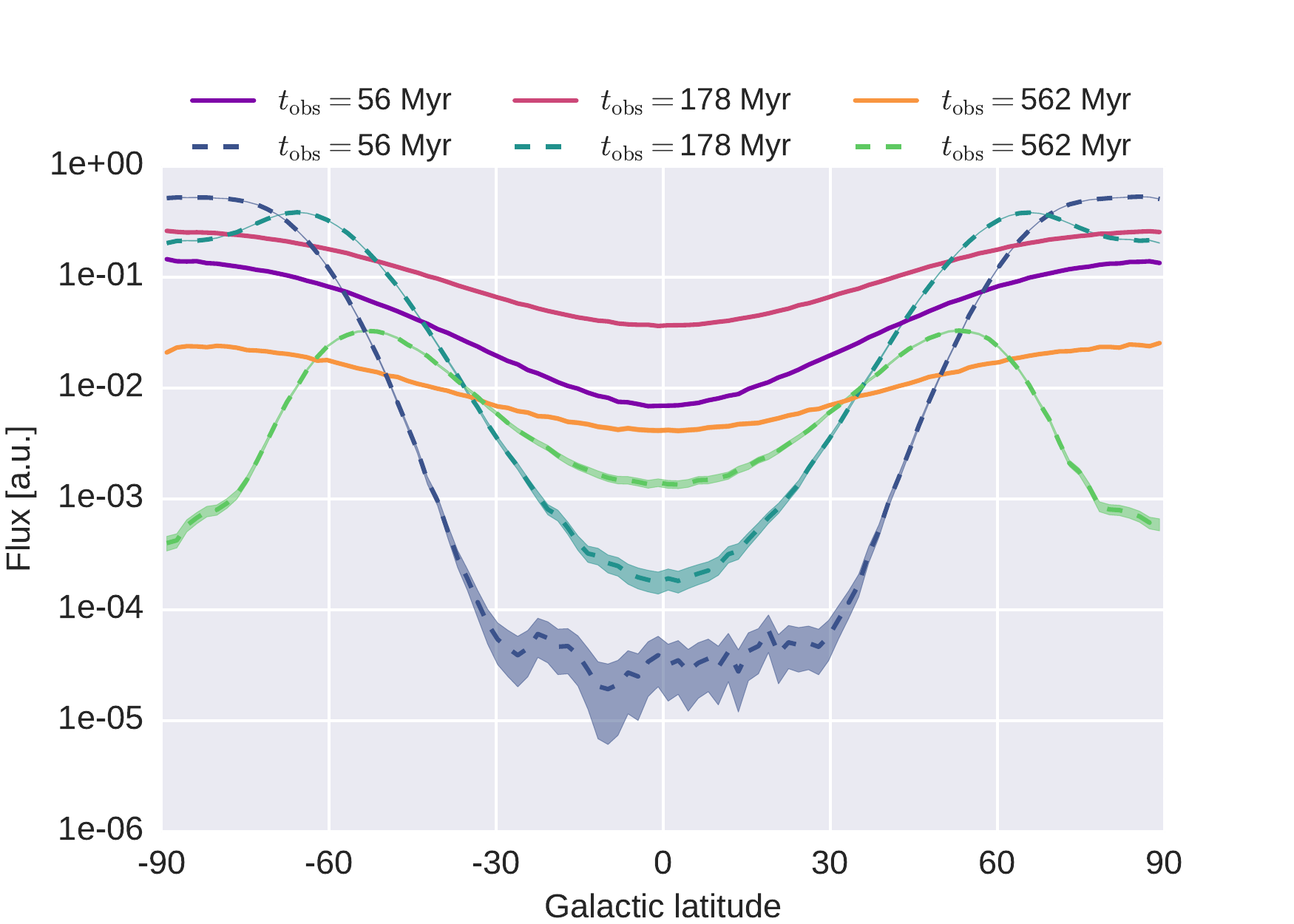}
 \includegraphics[width=\linewidth]{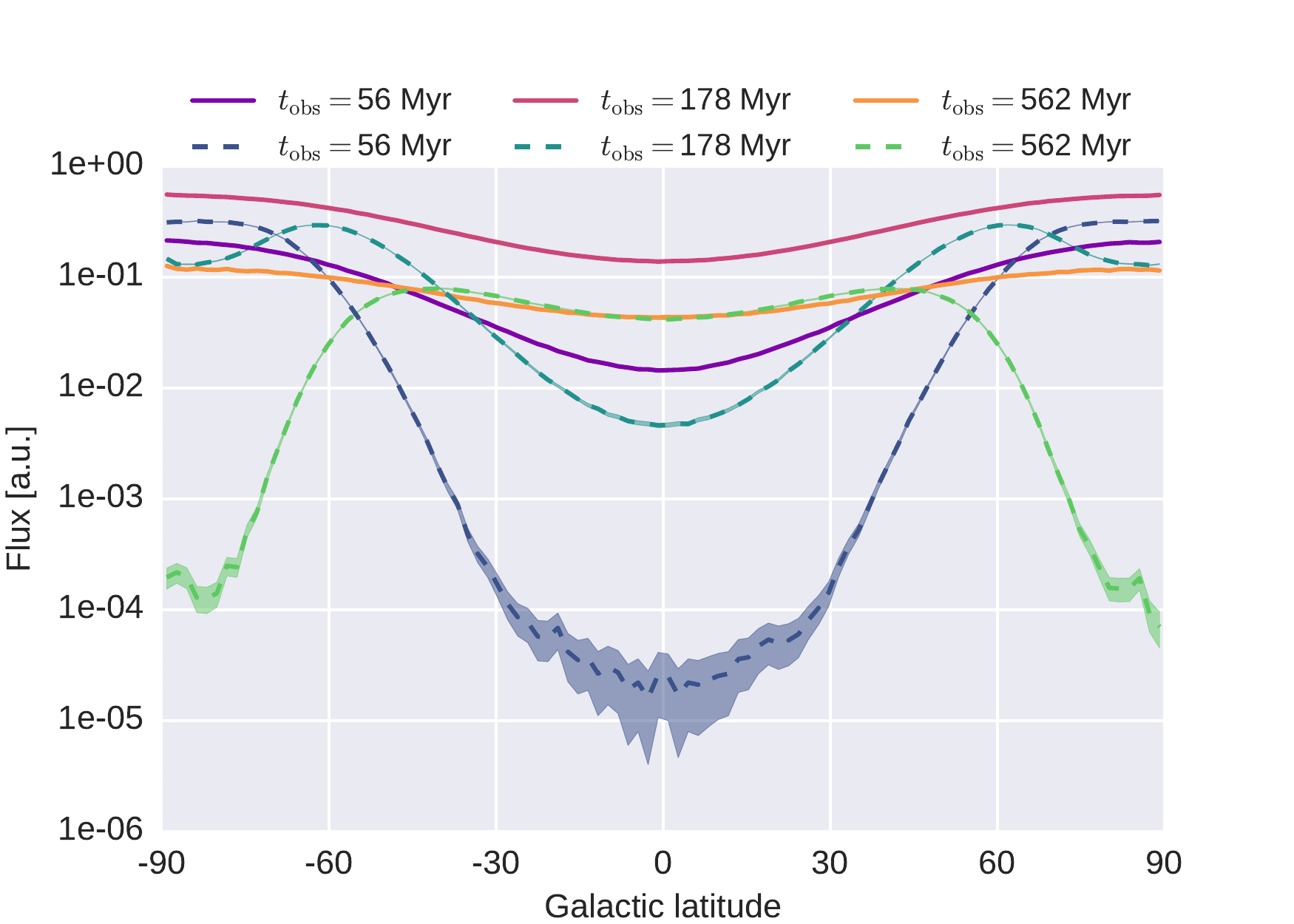}
\end{minipage}
\begin{minipage}{.49\textwidth}
\includegraphics[width=\linewidth]{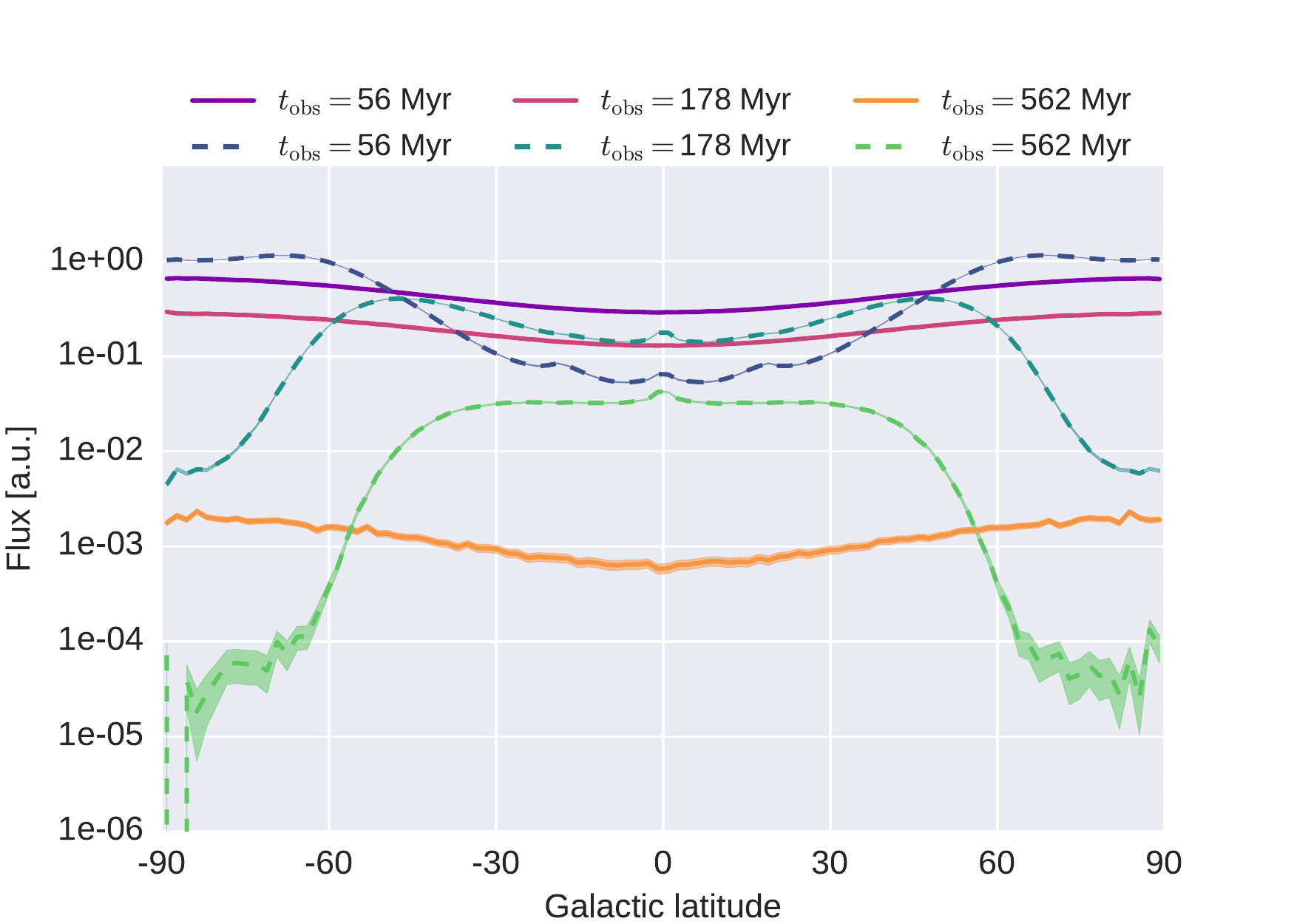}
\includegraphics[width=\linewidth]{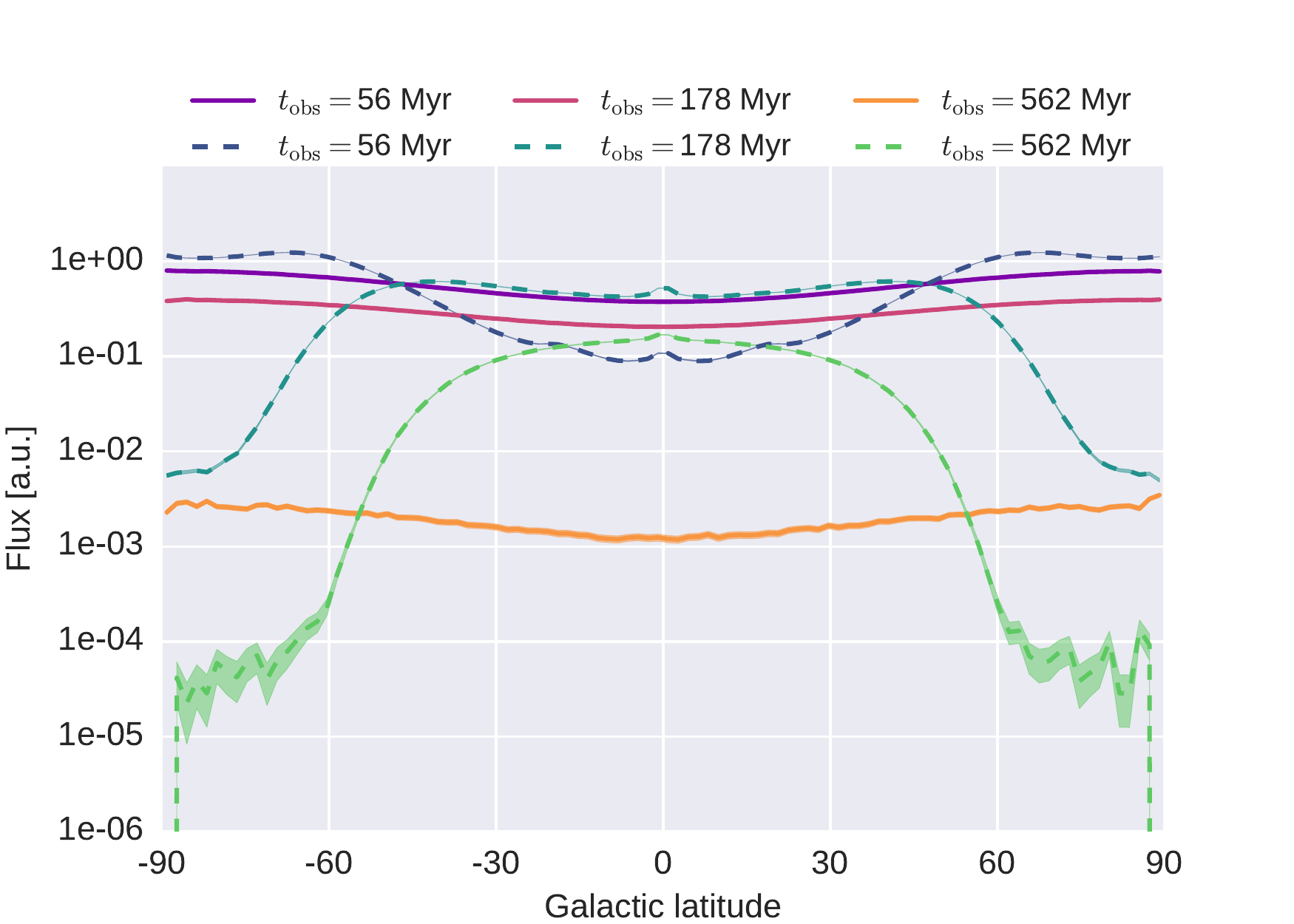}
\end{minipage}
\caption{Time evolution (color coded) of the arrival direction of CRs as seen at the $r_\mathrm{obs}=10$~kpc observer sphere at the edge of the galaxy. The results are averaged in 
azimuthal direction making use of the rotation symmetry of the simulation. Top row show results including the galactic wind [Tab.\ \ref{tab:Simulations}, Sim.\ 6-9] bottom row neglects any influence of a wind [Tab.\ \ref{tab:Simulations}, Sim.\ 13-16]. The left 
panels show a diffusion index $\delta=0.5$ and the right column represents the steep diffusion spectrum ($\delta=0.6$). Parallel (dashed lines) and perpendicular diffusion (solid 
lines) models are shown in a single plot)}
\label{fig:Arrival_combined}
\end{figure*}

The obvious symmetry of this problem makes it possible to average over the azimuthal coordinate of the arrival direction. The remaining variable, the 
galactic latitude $l$, can be displayed in a compact form, using weighted histograms, where we account for the decreasing areas for increasing galactic latitudes. Figure 
\ref{fig:Arrival_combined} shows the results for all simulated three dimensional scenarios. In each panel the data for perpendicular diffusion (reddish, solid lines) are 
compared with the pure parallel model (greenish, dashed lines) applying the same but arbitrary normalization. All models share one universal feature independent of the wind and 
the specific diffusion index: The latitude-distribution is significantly smoother when perpendicular diffusion is included. This is not surprising since a non vanishing $\epsilon$ 
allows for jumps between individual field lines. \Change{It should be noted that we use a spatially constant diffusion coefficient with $D_0=5\cdot 10^{28}$~cm$^2/s$, assuming that the diffusion coefficient at the termination shock is the same as in the Milky Way. A change in this number simply applies a change in normalization. For smaller coefficients, the flux at Earth will be reduced further, for larger coefficients, the flux will be enhanced.}

The double-ring structure that is visible in Fig.\ \ref{fig:Arrival_2d} can be found as two bumps in the case of parallel diffusion and is even more pronounced for Kraichnan 
diffusion ($\delta=0.5$). The maxima move with time from the poles to the equator region, while the total flux is decreasing. This is not surprising due to two facts: First, in 
the case of parallel diffusion the mean propagation time of the CRs depends on the diffusion coefficient $\kappa_i$ and the length of the magnetic field line. Secondly, the 
length of the magnetic field line---using the Archimedean spiral as the background field---depends on the latitude. In fact the polar field line of the Archimedean spiral is five 
times shorter than the equatorial one. Therefore, the maximum of the CR flux is expected to shift from the poles to smaller latitudes over time.

Comparing the two different diffusion coefficients, a faster time evolution of all processes is found for $\delta=0.6$, once again. In addition, the bumps are broader for the 
steep diffusion coefficient which eventually results in a plateau structure for later points in time.

On the other hand, we do not find any significant differences in the arrival direction comparing the simulation with and without a galactic wind (top and bottom row). The fluxes 
are a little bit higher, as one would expect, if the wind is neglected, but the overall morphology is unchanged. This is because the magnetic field and fluid flow are parallel, by construction.

Since all elements of the diffusion tensor depend on energy ($\kappa_\perp \propto \kappa_\parallel \propto E^\delta$) the arrival direction was analyzed for an energy 
dependence. For a fixed observation time ($t_\mathrm{obs}=250$~Myr) the energy dependent latitude distribution is shown in Fig.\ \ref{fig:Arrival_enery}. Here a binning was used 
that ensures equal number of CRs in each energy bin, creating non-equidistant energy bins. The energy dependence is small compared with the time evolution of the 
arrival direction. Once again, perpendicular diffusion (top row) leads to a smoother distribution. Here, only a small concentration of higher energetic CRs in the equatorial 
region is visible. This effect becomes more pronounced for parallel diffusion where a quite similar double bump structure is found. A plateau is formed only for $\delta=0.6$.

\begin{figure*}[htbp]
\centering
\begin{minipage}{.49\textwidth}
 \includegraphics[width=\linewidth]{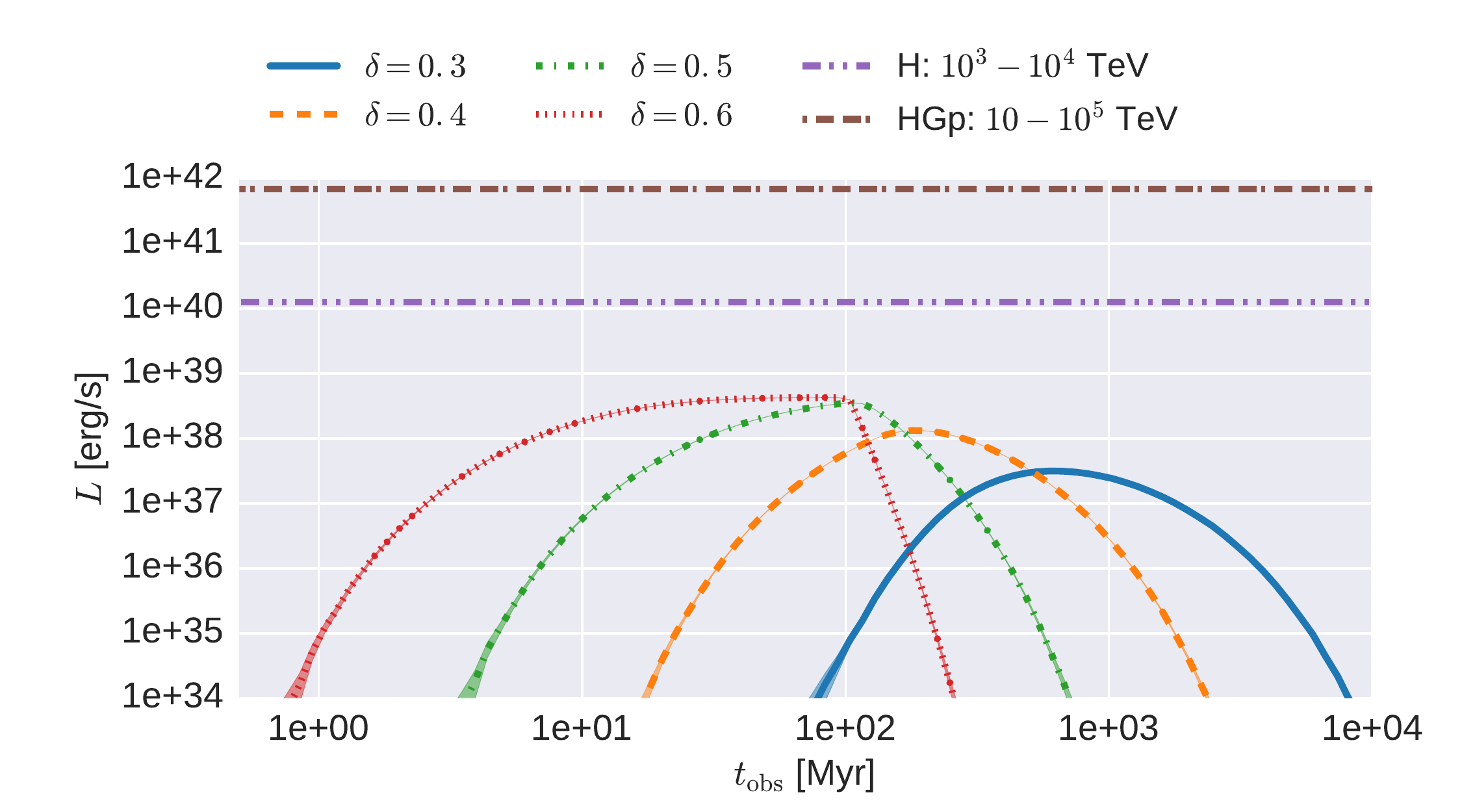}
 \includegraphics[width=\linewidth]{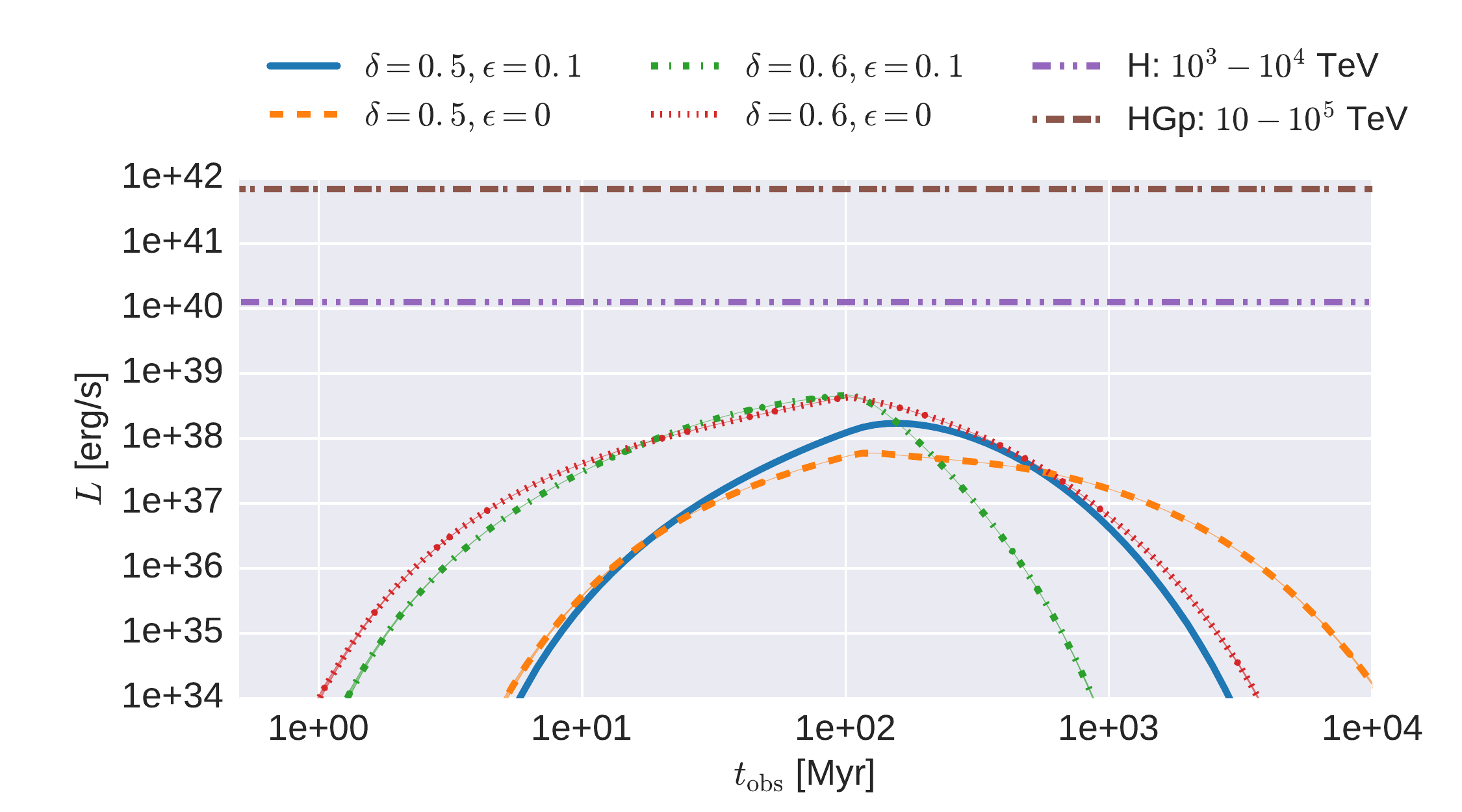}
\end{minipage}
\begin{minipage}{.49\textwidth}
\includegraphics[width=\linewidth]{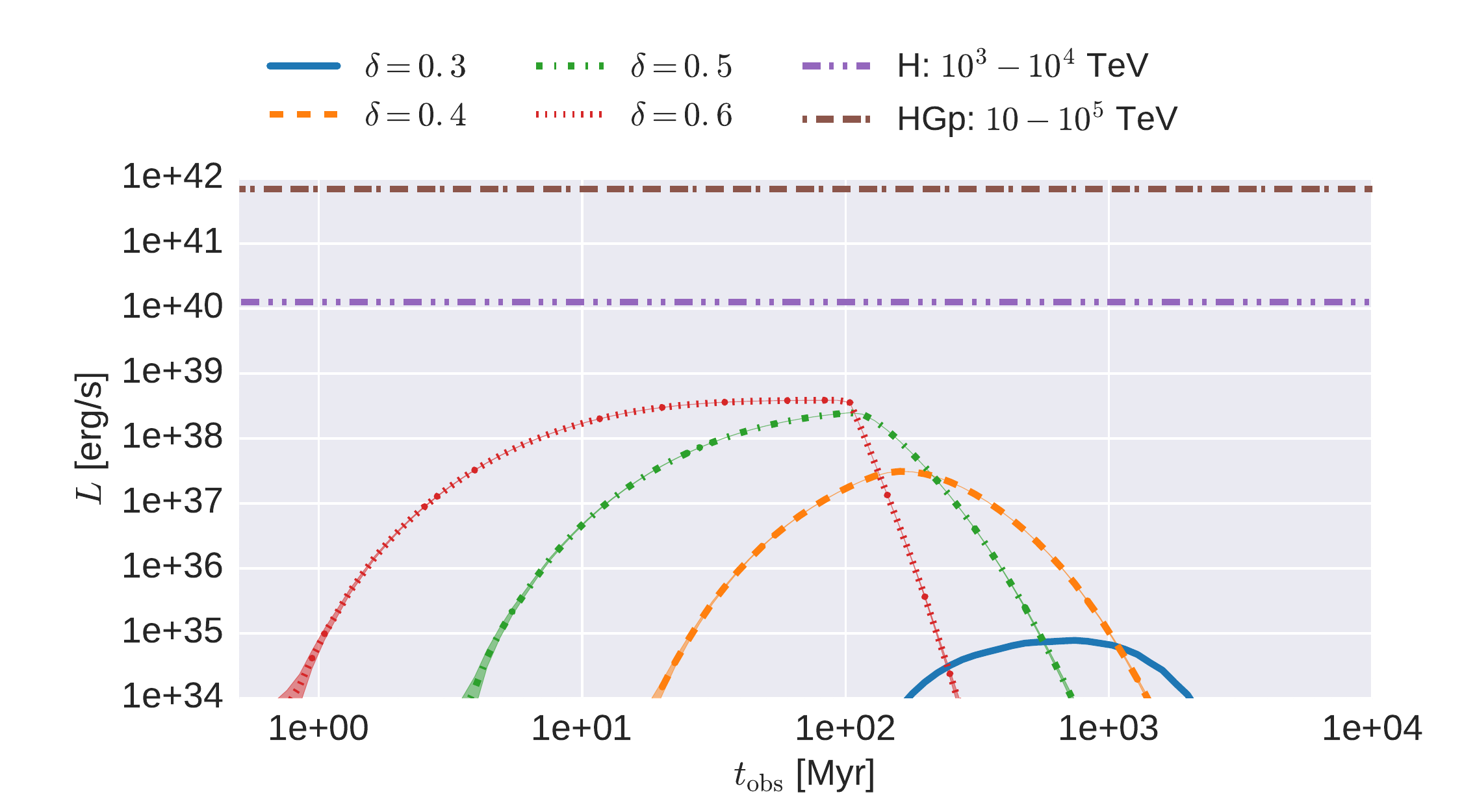}
\includegraphics[width=\linewidth]{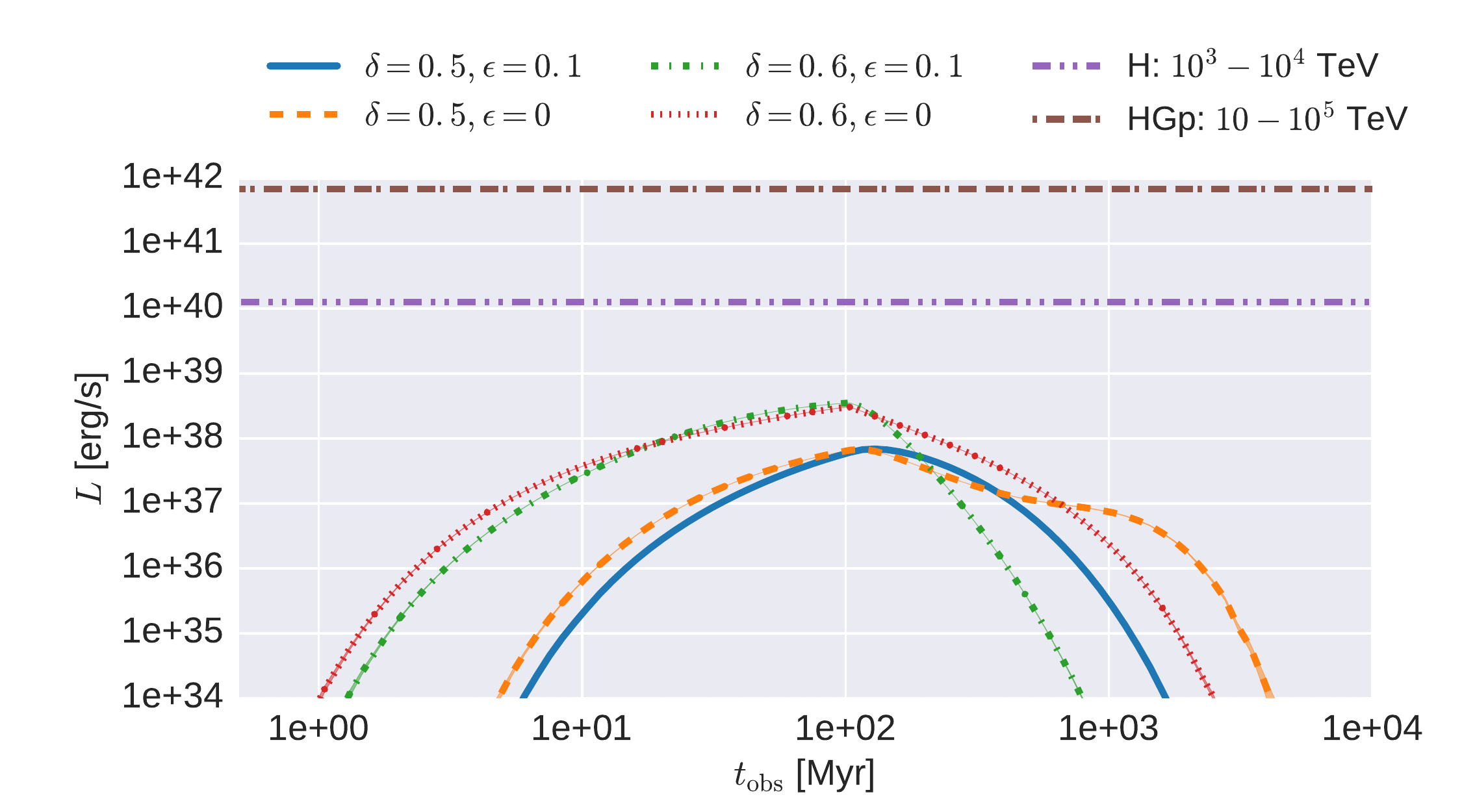}
\end{minipage}
\caption{Time evolution of the total CR luminosity from the GTS. It is seen that a Galactic wind (top right---[Tab.\ \ref{tab:Simulations}, Sim.\ 1, 2, 4, and 5]) slows down the CR propagation compared to a scenario without a wind (top left---[Tab.\ \ref{tab:Simulations}, Sim.\ 3, and 10-12]). 
However, even more striking is the suppression of the total proton luminosity for small diffusion indices. The plots in the lower two panels show that a three dimensional simulation does 
change the time dependence significantly (bottom left---[Tab.\ \ref{tab:Simulations}, Sim.\ 13-16]; bottom right---[Tab.\ \ref{tab:Simulations}, Sim.\ 6-9]). The shaded bands give the $3$-$\sigma$-uncertainty range of the luminosity. The horizontal lines represent analytical expectations, 
following  H--\cite{HOE03} and HGp--\cite{GAI12}, where the energies refer to integration boundaries. }
\label{fig:TimeEvolutionLuminosity}
\end{figure*}

Next we consider the total time evolution of the  CR-luminosity at the edge of the galaxy. We compare the one-dimensional models as well as the three-dimensional models with and without wind. The one-dimensional model without wind (upper left panel in Fig.\ \ref{fig:TimeEvolutionLuminosity}) clearly shows the different time scales of the transport process, e.g.\ the luminosity for $\delta=0.6$ has nearly vanished before the luminosity at $\delta=0.3$ has reached the 10 percent level. This difference in the time evolution can be found as a general feature in all simulations. When we then compare the models with and without wind of the one-dimensional simulations (upper row in Fig.\ \ref{fig:TimeEvolutionLuminosity}) a clear suppression of the flux for small diffusion indices ($\delta<0.5$) including wind is found, whereas the luminosity is nearly undisturbed for a diffusion index ($\delta\geq 0.5$). The lower panel in Fig.\ \ref{fig:TimeEvolutionLuminosity} clearly indicates a broader time evolution for pure parallel diffusion. This is true independent of the diffusion index and of the wind. 

The total proton luminosity can be compared with different analytical models of the CR proton flux (see below for the list of published models). The luminosity $L^i_\mathrm{ana}$ is then calculated from the given energy spectrum $\mathrm{d}N/\mathrm{d}E_i$ using: $L^i_\mathrm{ana}=(4\pi r_\mathrm{obs})^2\int_{E_\mathrm{min}}^{E_\mathrm{max}} (\mathrm{d}N/\mathrm{d}E_i) E\,\mathrm{d}E$. The expected analytic flux depends of course strongly on the chosen integration boundaries ($E_\mathrm{min}, E_\mathrm{max}$). One may think of two reasonable choices: First, the boundaries refer to the injected spectrum $E_\mathrm{init}=(10^3-10^4)$~TeV. Second, the boundaries refer to minimum and maximum observed energy $E_\mathrm{final}\approx(10^2-10^{4.5})$~TeV. The minimum luminosity $L^\mathrm{H}_\mathrm{ana}$ for the initial energy range $E_\mathrm{init}$ and the maximum luminosity $L^\mathrm{HGp}_\mathrm{ana}$ for the final energy range $E_\mathrm{final}$ using models from \cite{HOE03} and \cite{GAI12}, respectively. These values are displayed for comparison in Fig.\ \ref{fig:TimeEvolutionLuminosity}. Furthermore, we checked the models by \cite{GAI02}, \cite{ZAT06}, \cite{PamelaFlux2011}, and \cite{FEY12}, but they all lie within the two former ones. This means CRs from the GTS cannot be sole source of the of CRs in the shin region. However, they may contribute on the level of a few percent.

\subsection{Neutrino flux}
\label{neutrinos}
Another interesting aspect for CRs accelerated at the GTS is the production of neutrinos during the propagation. Under the assumption that the density of target material for 
proton-proton-collision 
$n_\mathrm{target}$ does not vanish completely at the galactic boundary a resulting neutrino flux is expected. The calculation of the exact neutrino flux is complicated and beyond 
the scope of this paper. Nevertheless, we are able to give rough estimates on the flux assuming that the optical depth $\tau = N\sigma_\mathrm{pp}$ for the inelastic 
collision is small $\tau<1$.

This allows for an approximation of the neutrino flux following \cite{BECTJU14}, \cite{KEL06}, and references therein. The flavor dependent neutrino flux can be calculated via:
\begin{align}
 \left. \frac{\mathrm{d}n}{\mathrm{d}E} \right|_{\nu_i} &\approx 1.6 N \int_{E}^\infty j_p(E_p) \sigma_\mathrm{pp}(E_p) F_i(E/E_p, E_p) E_p^{-1} \,\mathrm{d}E_p
\end{align}
Here, $j_p$ is the proton flux $\sigma_\mathrm{pp}$ is the total inelastic proton proton cross section and $F_i$ is the production rate of neutrinos with energy $E$ for a given 
primary  $E_p$. The details of this equation are explained in \cite{KEL06}, where analytical expressions for $\sigma_{pp}$ are also given. The total neutrino flux is then the sum of all three neutrinos produced in the decay of each pion.

\begin{figure}[htbp]
 \centering
 \includegraphics[width=\linewidth]{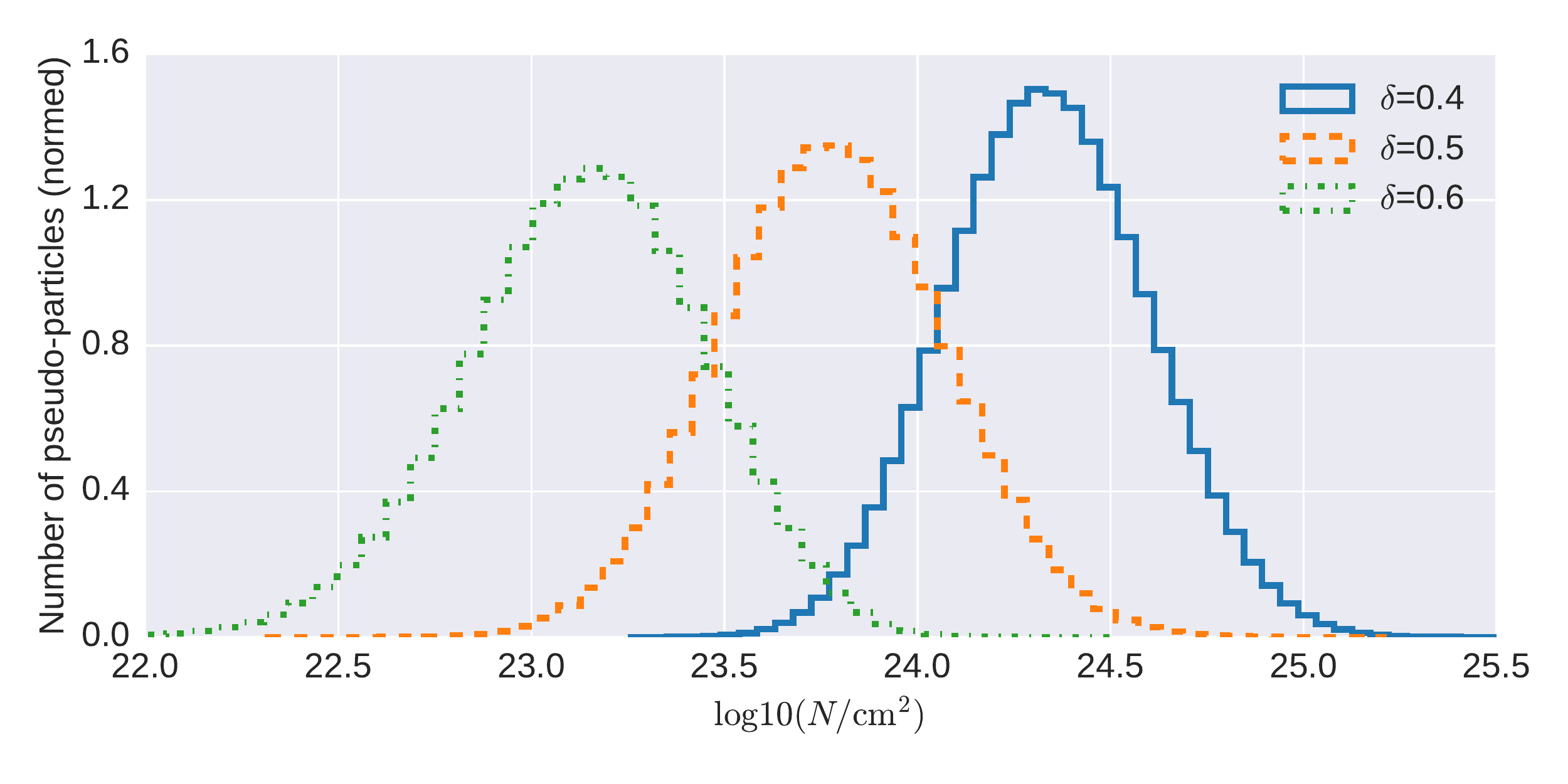}
 \caption{Distribution of column densities of all particles that arrive at the observer. The mean column density is increasing with decreasing diffusion index $\delta$. [Tab.\ \ref{tab:Simulations}, Sim.\ 17-19]}
 \label{fig:CD}
\end{figure}

To estimate the column density we use a simple model for the target 
particles with $n_\mathrm{target}\propto 1/r^2$, which is a good approximation once $u$ is near its asymptotic value. Then the column density $N$ is defined as:
\begin{align}
N = \int n_\mathrm{target}(s) \;\mathrm{d}s \approx n_0\sum_{i}\left(\frac{10\mathrm{kpc}}{r_i}\right)^2\cdot c \cdot \Delta t_i \quad \label{eq:ColumnDensity}
\end{align}
where $n_0=n_\mathrm{target}(10\mathrm{kpc})$ is the normalization and the sum goes over all integration steps of the particle transport with step width $\Delta t$. Figure 
\ref{fig:CD} shows the column density $N$ for the one-dimensional diffusion model including wind with three different diffusion indices $\delta=(0.4, 0.5, 0.6)$. The 
logarithm of accumulated column density $\log(N)$ is nearly normal distributed with increasing mean for decreasing diffusion index. This is not surprising as the column density 
scales with propagation time which is increasing for lower diffusion indices, as pointed out before.

The existing data does not allow calculation of the mean neutrino production distance. To do so, time resolved position data of the primary CRs during the whole propagation process would be needed, which is not practical due to computational and memory restrictions. Instead, we can estimate an upper limit assuming that all neutrinos are produced when the primary cosmic ray is observed at $r_\mathrm{obs}=10$~kpc. This is probably a fairly good approximation since the target density is not constant but rapidly decreasing with distance. So most of the column density is accumulated 
in the close vicinity of the observer sphere.

This approximation allows us to calculate the neutrino flux at a given observation time $t_\mathrm{obs}$ from the corresponding proton flux $\left. \mathrm{d}n/\mathrm{d}E\right|_\mathrm{p}$ as defined in Eqn.\ \ref{eq:ObservedFlux}. The cross section $\sigma_\mathrm{pp}(\langle E\rangle)$ is approximated by the time dependent mean energy of the proton distribution following the semi-analytical equation given in \cite{KEL06}. The same averaging is applied on the column density $N$ using Eqn.\ \ref{eq:ColumnDensity}.

\begin{figure}[htbp]
 \centering
 \includegraphics[width=\linewidth]{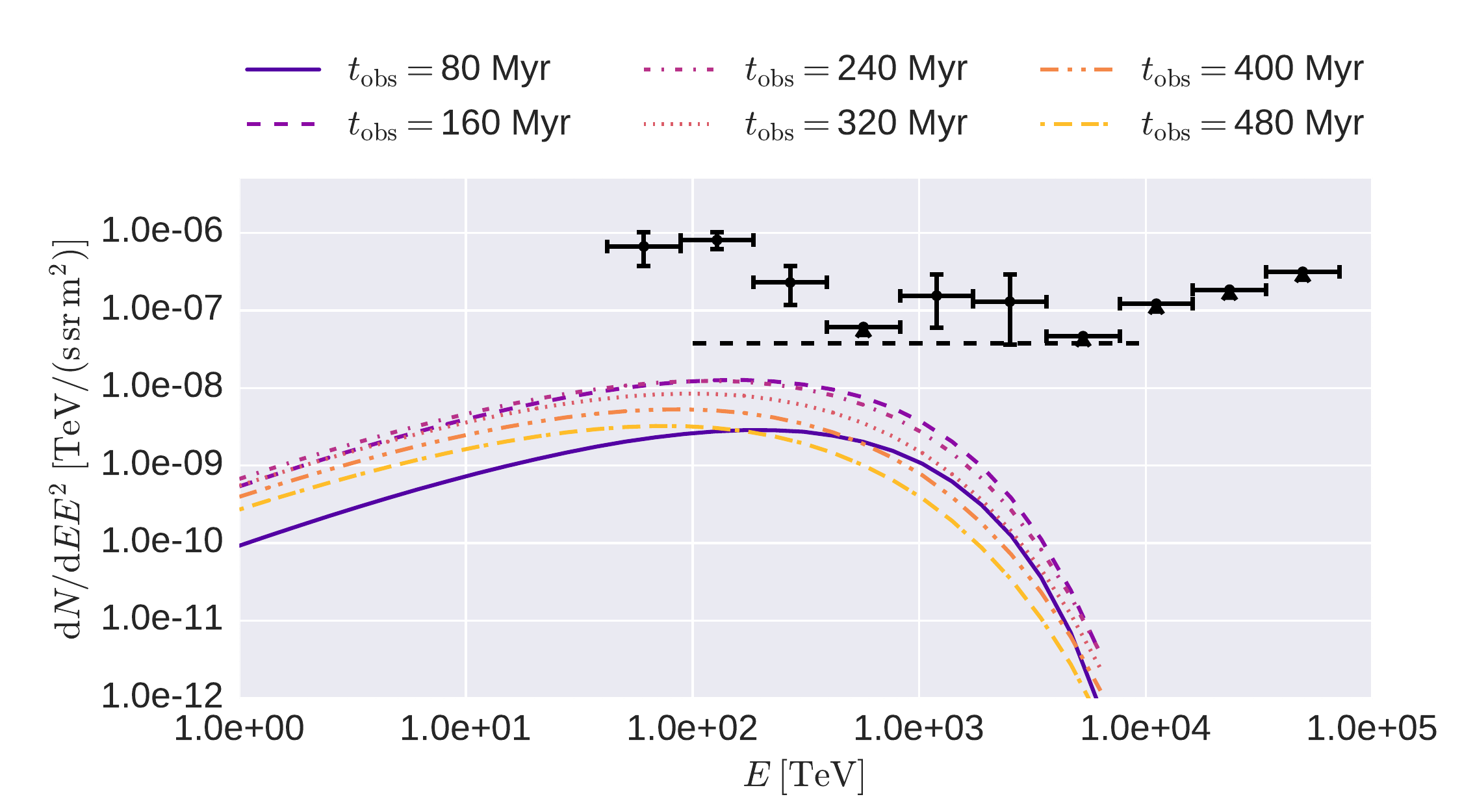}
 \caption{Time evolution of the neutrino flux produced in proton-proton-interaction of CRs accelerated at the GTS. A diffusion index of $\delta=0.4$ is used 
in a spherical symmetric model including advection and adiabatic cooling. For comparison the measured neutrino flux by IceCube (black points) \citep{HESE-6-years} (We multiplied the provided single-flavor data points by a factor of three, assuming a 1-1-1 flavor ratio) and the 1-year 
diffuse flux limit from km3Net (black dashed line) are shown \citep{KM3NeT}. [Tab.\ \ref{tab:Simulations}, Sim.\ 17]}
 \label{fig:Neutrinos_04}
\end{figure}

Figure \ref{fig:Neutrinos_04} shows the time evolution of the neutrino flux for the spherically symmetric model. Here, advection and the corresponding adiabatic cooling are included as well as parallel diffusion ($\delta=0.4, \epsilon=0$).\footnote{The corresponding proton flux can be found in Fig.\ \ref{fig:Spectrum_1Dim}.} For comparison the total neutrino flux measured by IceCube \citep{HESE-6-years} (black data points) and the expected one-year diffuse neutrino limit by KM3Net \citep{KM3NeT} (black dashed line) are shown. First, one may note that the neutrino flux is much smoother than the corresponding primary proton flux. This is simply due to the fact that the primary flux is convolved with the smooth production rate $F_i$. As expected the maximum neutrino energy is less than one order of magnitude below the maximum primary energy. We note that a simple monochromatic approach for the calculation of the neutrino flux as in \cite{BECTJU14} would have led to a spectral shape of the neutrinos reproducing the primary one. Nevertheless, this approach lacks in the description of the highest energetic neutrinos, because it assumes an unbroken power-law for the primary cosmic rays, which is not seen in this work. Secondly, the temporal variation is 
much less pronounced compared with the protons.
Figure \ref{fig:Neutrinos_0506} shows the neutrino flux for two additional diffusion indices ($\delta=0.5, 0.6$). Here once again, one may note that the neutrino flux is very 
stable with respect to different diffusion models, which is in contrast to the primary flux. This unexpected stability might be explained by the fact that the normalization of 
the neutrino 
flux depends on two counteractive parameters: the proton flux, which is on average higher for steep diffusion spectra, and the accumulated column density, which is increased 
for lower diffusion indices. So the product of these two components leads to a quite stable neutrino flux. This means that the expected neutrino flux is more or less independent of the specific diffusion model and also of the exact observation time. However, it depends still on the luminosity of the GTS and its duration. A shorter GTS would not accelerate as many primaries as a longer shock, leading to a reduced neutrino flux.

\section{Conclusions}
\label{conclusions}
The main goal of this work is to evaluate the contribution of cosmic rays accelerated at the GTS to the observed flux in the so called shin-region between the `knee' and the 
`ankle'. In doing so, we use the CRPropa framework to simulate the transport of accelerated cosmic rays including diffusion, advection and the corresponding adiabatic energy 
changes. Two different morphologies for the background magnetic field and flow---a pure spherically symmetric one and an Archimedean spiral---are tested, as well as different models of the diffusion tensor, 
including different diffusion indices $\delta$ and different ratios between parallel and perpendicular diffusion $\epsilon$. 

Regarding the total luminosity of cosmic rays from the termination shock, as discussed in Section  \ref{continuoussource}, we can draw the following main conclusions: (1) Cosmic 
rays accelerated at the termination shock are able to propagate back into the galaxy also when advection, driven by the galactic wind, is taken properly into account. For some of 
the tested models, these CRs may even contribute significantly to the observed flux on a percent level. Nevertheless, none of the evaluated models is able to explain the total cosmic ray flux in the shin 
region using the assumed parameters like the total luminosity and position of the GTS ($L_\mathrm{tot}=10^{40}$~$\mathrm{erg}\,\mathrm{s}^{-1}$, $r_0=250$~kpc). (2) The amount of 
cosmic rays that reach the observer depends strongly on the chosen transport model. In general a larger diffusion index $\delta$ leads to an increased flux and a faster time 
evolution.

The analysis of the energy spectra has shown that these results depend strongly on the observation point in time. This is very reasonable, since the propagation time of the cosmic 
rays depends on their energy. Even if we neglect the fact that the correct diffusion model is unknown it is very hard to predict a concrete spectral behavior of the cosmic rays, 
since the current state---Is it active? When has it shut down?---of the GTS is unclear. Nevertheless, if the diffusion model can be restricted, either by theoretical estimates or 
independent observations, the unique temporal evolution of the spectra might also help us gain new insights on the GTS. 

The most favorable scenario might be one in which there is a wind that subtends a fairly large angle such that cosmic rays are accelerated by its shock but can diffuse back through the windless part of space.  

The differences in the arrival direction comparing pure parallel ($\epsilon=0$) and strong perpendicular ($\epsilon=0.1$) diffusion might be used to distinguish between these 
models. Here, at least one problem arises: It is unclear how the arrival direction pattern is changed by the additional propagation in the Galaxy. But one can imagine that a 
pronounced double-ring structure, as 
expected for pure parallel transport, will have some impact on the arrival direction compared with the very smooth arrival direction 
distribution in the case of perpendicular transport. Of course also the true background field is not known, but we expect that the general behavior---a dependency between 
$\epsilon$ and the isotropy level of arrival direction---will be present independent of the chosen background model. \Change{Furthermore, we note that the produced anisotropy level at the Galactic boundary might also be used to constrain the proposed model when it is compared to observed arrival directions. Although the actual arrival direction likely changes due to the propagation in the Galaxy, the multipole-moments are supposedly more stable. This means that a pure parallel transport is very unlikely since the introduced anisotropy is too large to be smoothed out during the Galactic propagation.}

When we compare the simple estimates from Section \ref{semiestimates} with the more sophisticated simulations of this work, we can conclude the following: It is hard to draw a sharp boundary between the diffusion and the advection dominated regime. A diffusion index of $\delta=0.4$ requires a minimum CR energy of $\approx 2.8\cdot 10^{16}$~eV to be dominated by diffusion (see Eqn.\ \ref{reynolds}). Nevertheless, our simulation show that a significant amount of the accelerated CRs diffuse back into the Galaxy. On the other hand also for a configuration that should be diffusion dominated ($\delta=0.6$) not more than about 30 percent of all CRs reach the Galaxy. The CR Reynolds number gives a good estimate whether particles can diffuse back or not; however, especially in the regime of comparable time scales for diffusion and advection, a detailed simulation is needed.

Our very basic estimates of the neutrino fluxes showed that our model is not restricted by the observed neutrino flux and is unlikely to be the dominant constituent, either. The produced neutrinos might even contribute to some amount 
for the observed IceCube flux. The maximum contribution is below ten percent but for most energy bins even below the percent level. However, it has to be noted that the additional propagation of the cosmic rays through the galaxy will lead to an additional neutrino flux, so that the total contribution could even be more significant. This flux 
can easily be about an order of magnitude higher than the flux that we predict, since the target density is much higher inside the galaxy, and might also be imprinted with Galactic structure, which so far has not been confirmed in the data.

From the technical point of view, we have shown that the publicly available propagation code CRPropa is able to tackle the problem of the origin of cosmic rays in the shin region. 
It has been shown that in addition to the spatial diffusion that was already introduced in \cite{MER17} the software is now also able to take advection and the corresponding 
adiabtic cooling into account. The full technical details will be explained in an upcoming paper by the CRPropa development group \citep{crpropa32}. Furthermore, we also demonstrated the 
flexibility of the Green's method approach to construct different source evolutions using the same simulation data by applying different weights on the data.

\section{Outlook}
\label{Outlook}

This paper leads to at least two interesting ideas for future projects. First, extending the propagation of the cosmic rays from the galactic boundary at $r_\mathrm{obs}=10$~kpc through the 
Galaxy would allow for a direct comparison of simulated cosmic ray spectra with observational results. This comparison was not possible in this work since we expect 
the spectra to alter due to the propagation in the Galaxy. Such a study might also include other elements apart from protons to also address questions on the composition of the 
cosmic rays. However, such a study is beyond the scope of this work since a completely different, time intensive simulation has to be performed.

In addition, as was only briefly discussed in Section \ref{conclusions}, the cosmic ray flux into the IGM might also be of great interest \Change{(see also \cite{VOELK1996})}. Models by 
\cite{2017ApJ...835...72B} suggest that starbursting galaxies are able to accelerate cosmic rays to higher energies in their termination shock. If assumptions on the starburst duration and magnetic field strength are relaxed (see e.g. \citealt{2018arXiv180106483R, 2018arXiv180107170A}), even higher energy CRs may be produced. Due to the increased 
advection rate one can assume that a very large fraction of these cosmic rays do not diffuse back into their host galaxy but are indeed lost into the IGM. The intriguing question 
is 
if these cosmic rays can contribute to the observed flux in our galaxy. Cosmic rays from these starburst termination shocks might be able to account for the missing part of 
the flux in the shin region that is not explained by the Milky Way termination shock.

Two other things should be examined in future work: (1) Multiple bursts might lead to shocks which are closer to the Galactic center \citep{2012A&A...540A..77D}. This means that CRs would have a greater chance of diffusing back instead of being lost into the IGM, leading to an efficient increase of the observed flux. (2) The influence of a Galactic wind that is changing with time and/or not covering the full 4$\pi$-sphere is also very likely to change the results of this analysis.

Furthermore, it might also be interesting to study the re-acceleration process (see Sec.\ \ref{sec:EnergyChange})in the vicinity of the shock in more detail. For example it might be interesting to decrease the diffusion coefficient around the shock which would lead to a higher re-acceleration rate and could be explained by stronger turbulent magnetic fields in this region.

\acknowledgements
\section*{Acknowledgements}
The authors thank F.\ Halzen and A.\ Kheirandish for inspiring discussions which led to the section on neutrinos \Change{and G.\ Sigl for comments and feedback on the anisotropy of the arrival direction. Furthermore, the authors would like to thank the referee for their comments which helped to clarify certain parts of this work.} In addition, EGZ thanks C.\ Pfrommer for interesting discussions, and LM thanks B.\ Eichmann for reading early parts of this work. The authors also thank C.\ Kopper for information on the IceCube data.
LM and JBT acknowledge support from the RAPP 
Center (Ruhr Astroparticle and Plasmaphysics Center sponsored by the MERCUR project St-2014-0040) and from the Research Department of Plasmas with Complex Interactions (Bochum).
CB is supported by the National Science Foundation Graduate Research Fellowship Program under Grant No.\ DGE-1256259. Any opinions, findings, 
and conclusions or recommendations expressed in this material are those of the author(s) and do not necessarily reflect the views of the National Science Foundation. Support was 
also provided by the Graduate School and the Office of the Vice Chancellor for Research and Graduate Education at the University of Wisconsin-Madison with funding from the 
Wisconsin Alumni Research Foundation. CB and EGZ also acknowledge support from the University of Wisconsin-Madison and NSF Grant No. AST-1616037. LM thanks the Astronomy Department of the University of Wisconsin where part of this work was carried out, and EGZ thanks the Department of Astronomy and Astrophysics at the University of Chicago where it was completed.

\vspace{5mm}
\software{\Change{\texttt{matplotlib} \citep{Hunter_2007}, 
\texttt{pandas} \citep{McKinney_2012}, \texttt{numpy}, \texttt{scipy} \citep{van_der_Walt_2011}, \texttt{healpy} \citep{HealPy} and \texttt{ipython} \citep{Perez_2007}}}

\appendix
\label{appendix}

\section{Drifts}
\label{app:Drift}

\section{Validation}
\begin{figure}[htbp]
\centering
\includegraphics[width=.65\linewidth]{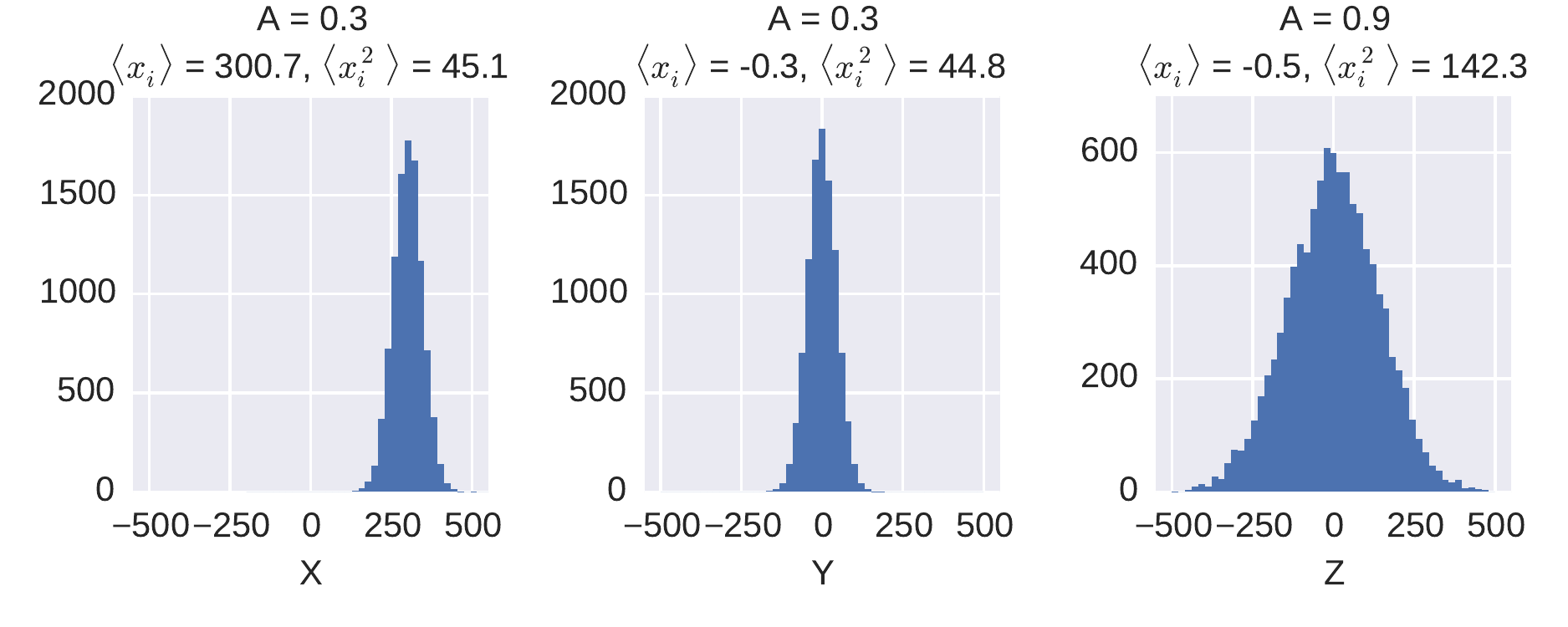}
\caption{End positions of 10,000 pseudo-particles at $t=1000$~s. The anisotropic diffusion (wider distribution in z-direction) and advection (shifted mean in x-direction) are 
clearly visible. No significant deviation from analytic expectation is found.}
\label{fig:Test_Advection}
\end{figure}

\paragraph{Testing: Advection }The validation of this new module is done by comparison with an analytic solution for a simple example. We use a homogeneous magnetic background 
field parallel to the z-axis and simple diffusion tensor with $\hat{\kappa}=\mathrm{diag}(\kappa_\perp, \kappa_\perp, \kappa_\parallel)$, where 
$\kappa_\parallel=10\kappa_\perp=10~\mathrm{m}^2\,\mathrm{s}^{-1}$. Furthermore, we implement a wind in the positive x-direction with velocity $\vec{u}=0.3~\mathrm{m}\,\mathrm{s}^{-1}\;\vec{e}_x$. The 
pseudo-particles are injected at $\vec{r}_0=0$, which corresponds to a source $S=\delta(t) \delta(\vec{r}_0) S_0$. We expect for a given point in time $t$ that the position of the 
pseudo-particles projected on the three axes follow gaussian distributions with mean $\langle E_i \rangle = u_i t$ and variance $\sigma = \sqrt{2\kappa_i t}$. Figure 
\ref{fig:Test_Advection} shows the end position of the pseudo-particles for $t=1000$~s, where no deviation of the analytic expectation can be found. 
Furthermore, we perform the Anderson-Darling-test (test value $A$ is given in Fig.\ \ref{fig:Test_Advection}) on the sample of pseudo-particle position, where also no significant 
deviation from a normal distribution is present.

\paragraph{Testing: Adiabatic Cooling } The simplest way to test adiabatic cooling is to simulate a relativistic gas which expands radially with a given wind velocity 
$\vec{u}(\vec{r}) = 
u_0\;\vec{e}_r$. For such an expanding gas it is known that the particle density is proportional to the inverse square of the radius $n(r)\propto r^{-2}$. Due to adiabatic cooling 
the energy density $w$ decreases even faster with increasing radius as $w\propto n^{4/3} \propto r^{-8/3}$.

\begin{figure}[htbp]
\centering
\includegraphics[width=.65\linewidth]{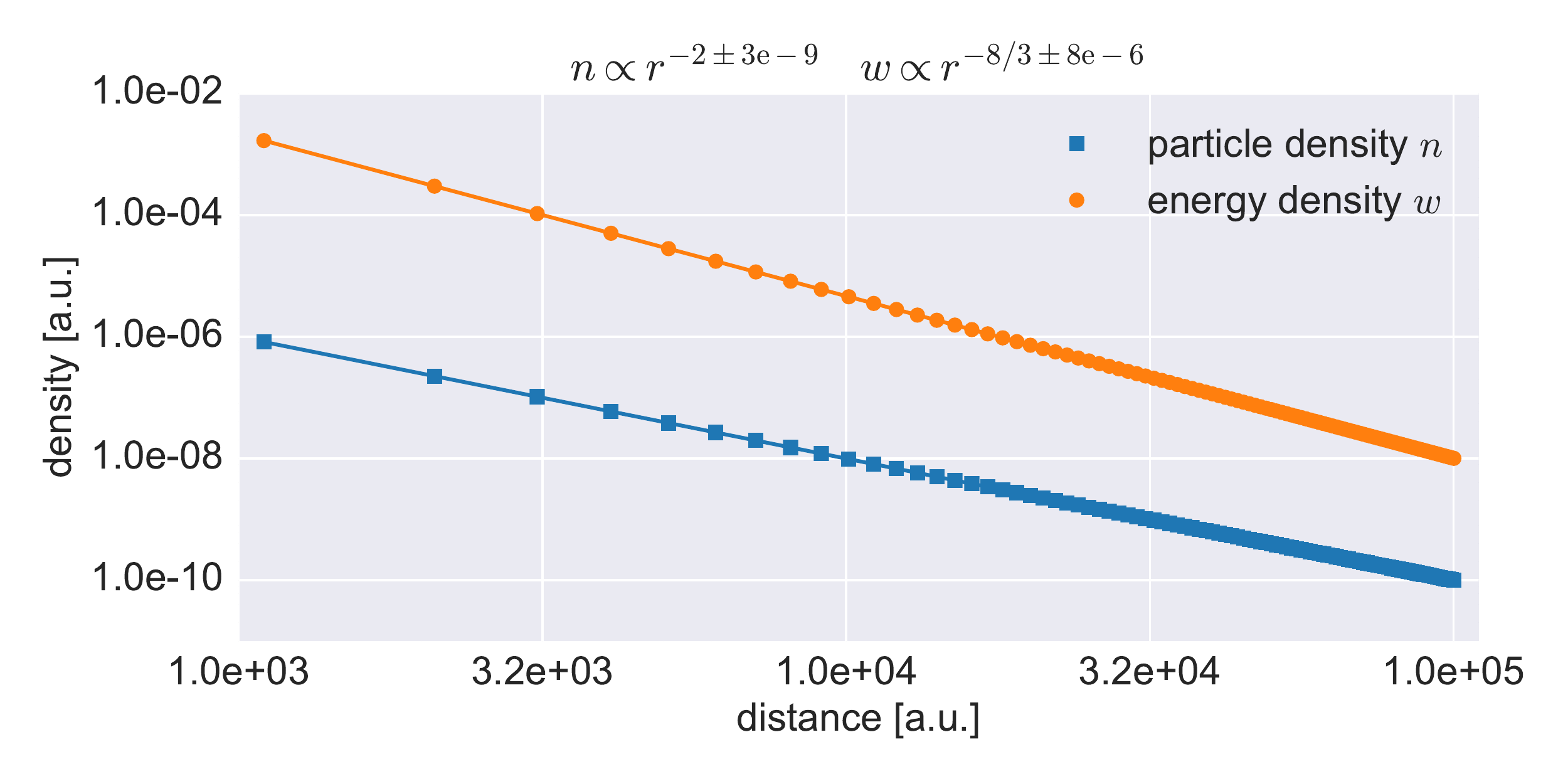}
\caption{The fitted power law indices for the particle (blue squares) and the energy (orange circles) density are in very good agreement with the theoretical expectation.}
\label{fig:Test_AdiabaticCooling}
\end{figure}

To validate the adiabatic cooling we injected particles in a shell and tracked their motion due to advection in a radial symmetric wind. From that we calculated the particle and 
energy density. The results are shown in Fig.\ \ref{fig:Test_AdiabaticCooling} where no deviation from the analytically expected results can be found.

\section{Additional Material}

\begin{deluxetable*}{cRcrrlr}[htbp]
\centering
\tablecaption{Simulation parameters. Simulations are listed in order of appearance in this paper.\label{tab:Simulations}}
\tablecolumns{7}
\tablenum{1}
\tablewidth{0pt}
\tablehead{
\colhead{\#} &
\colhead{N\tablenotemark{a}} &
\colhead{Symmetry\tablenotemark{b}} & \colhead{Diffusionindex} & \colhead{Diffusionratio} & \colhead{Wind\tablenotemark{c}} & \colhead{Figures} \\
\colhead{} & \colhead{} & \colhead{} & \colhead{($\delta$)} & 
\colhead{($\epsilon$)} & \colhead{} & \colhead{}
}
\startdata
1 & 1\times 10^7 & S & 0.5 & 0 & Yes & \ref{fig:Green}, \ref{fig:EnergyChange}, \ref{fig:Spectrum_1Dim}, \ref{fig:TimeEvolutionLuminosity}\\
2 & 4\times 10^7 & S & 0.4 & 0 & Yes & \ref{fig:TimeEvolution04}, \ref{fig:EnergyChange}, \ref{fig:Spectrum_1Dim}, \ref{fig:TimeEvolutionLuminosity} \\
3 & 4\times 10^7 & S & 0.4 & 0 & No & \ref{fig:TimeEvolution04}, \ref{fig:Spectrum_1Dim}, \ref{fig:TimeEvolutionLuminosity} \\
4 & 6.25\times 10^8 & S & 0.3 & 0 & Yes & \ref{fig:EnergyChange}, \ref{fig:Spectrum_1Dim}, \ref{fig:TimeEvolutionLuminosity} \\
5 & 1\times 10^7 & S & 0.6 & 0 & Yes & \ref{fig:EnergyChange}, \ref{fig:Spectrum_1Dim}, \ref{fig:TimeEvolutionLuminosity} \\
6 & 2.5\times 10^8 & A & 0.5 & 0.1 & Yes & \ref{fig:EnergyChange}, \ref{fig:Spectrum_3Dim_equalBinning}, \ref{fig:Arrival_combined}, \ref{fig:TimeEvolutionLuminosity}, \ref{fig:Arrival_enery}\\
7 & 2.5\times 10^8 & A & 0.5 & 0 & Yes & \ref{fig:EnergyChange}, \ref{fig:Spectrum_3Dim_equalBinning}, \ref{fig:Arrival_combined}, \ref{fig:TimeEvolutionLuminosity}, \ref{fig:Arrival_enery}\\ 
8 & 2.5\times 10^8 & A & 0.6 & 0.1 & Yes & \ref{fig:EnergyChange}, \ref{fig:Spectrum_3Dim_equalBinning}, \ref{fig:Arrival_combined}, \ref{fig:TimeEvolutionLuminosity}, \ref{fig:Arrival_enery} \\
9 & 2.5\times 10^8 & A & 0.6 & 0 & Yes & \ref{fig:EnergyChange}, \ref{fig:Spectrum_3Dim_equalBinning}, \ref{fig:Arrival_2d}, \ref{fig:Arrival_combined}, \ref{fig:TimeEvolutionLuminosity}, \ref{fig:Arrival_enery} \\
10 & 1\times 10^7 & S & 0.3 & 0 & No & \ref{fig:Spectrum_1Dim}, \ref{fig:TimeEvolutionLuminosity} \\
11 &1\times 10^7 & S & 0.5 & 0 & No & \ref{fig:Spectrum_1Dim}, \ref{fig:TimeEvolutionLuminosity} \\
12 & 1\times 10^7 & S & 0.6 & 0 & No & \ref{fig:Spectrum_1Dim}, \ref{fig:TimeEvolutionLuminosity} \\
13 & 1\times 10^7 & A & 0.5 & 0.1 & No & \ref{fig:Arrival_combined}, \ref{fig:TimeEvolutionLuminosity} \\
14 & 2.5\times 10^8 & A & 0.5 & 0 & No & \ref{fig:Arrival_combined}, \ref{fig:TimeEvolutionLuminosity} \\
15 & 2.5\times 10^8 & A & 0.6 & 0.1 & No & \ref{fig:Arrival_combined}, \ref{fig:TimeEvolutionLuminosity} \\
16 & 2.5\times 10^7 & A & 0.6 & 0 & No & \ref{fig:Arrival_combined}, \ref{fig:TimeEvolutionLuminosity} \\
17\tablenotemark{d} & 4\times 10^7 & S & 0.4 & 0 & Yes & \ref{fig:CD}, \ref{fig:Neutrinos_04} \\
18\tablenotemark{d} & 1\times 10^7 & S & 0.5 & 0 & Yes & \ref{fig:CD}, \ref{fig:Neutrinos_0506} \\
19\tablenotemark{d} & 1\times 10^7 & S & 0.6 & 0 & Yes & \ref{fig:CD}, \ref{fig:Neutrinos_0506}\\
\enddata
\tablenotetext{a}{Number of simulated pseudo-particles.}
\tablenotetext{b}{S---Spherically symmetric, A---Archimedean spiral}
\tablenotetext{c}{Including the corresponding adiabatic energy change.}
\tablenotetext{d}{Here, the column density of the primaries is also recorded.}

\end{deluxetable*}

\begin{figure*}[htbp]
\centering
\begin{minipage}{.49\textwidth}
 \includegraphics[width=\linewidth]{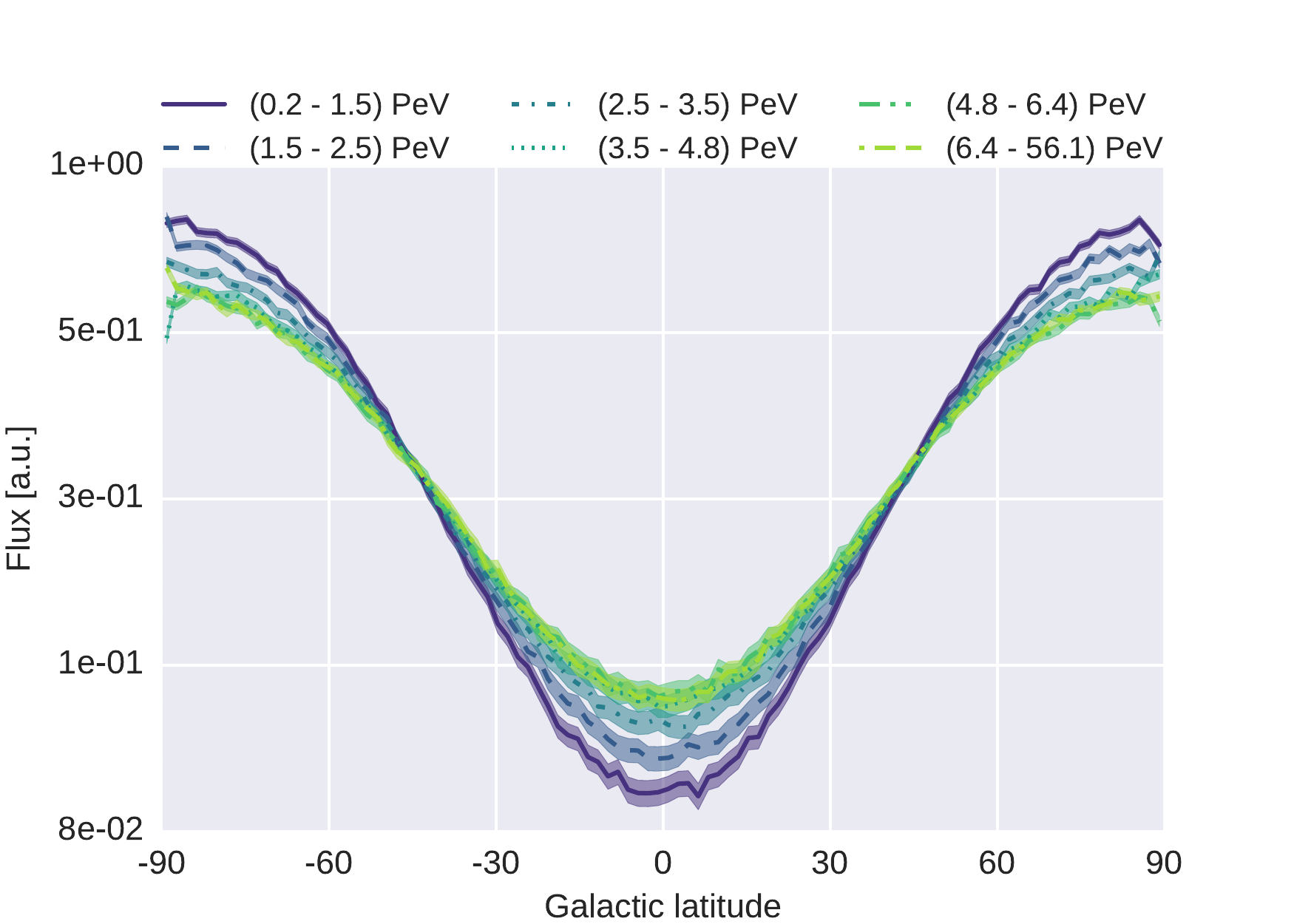}
 \includegraphics[width=\linewidth]{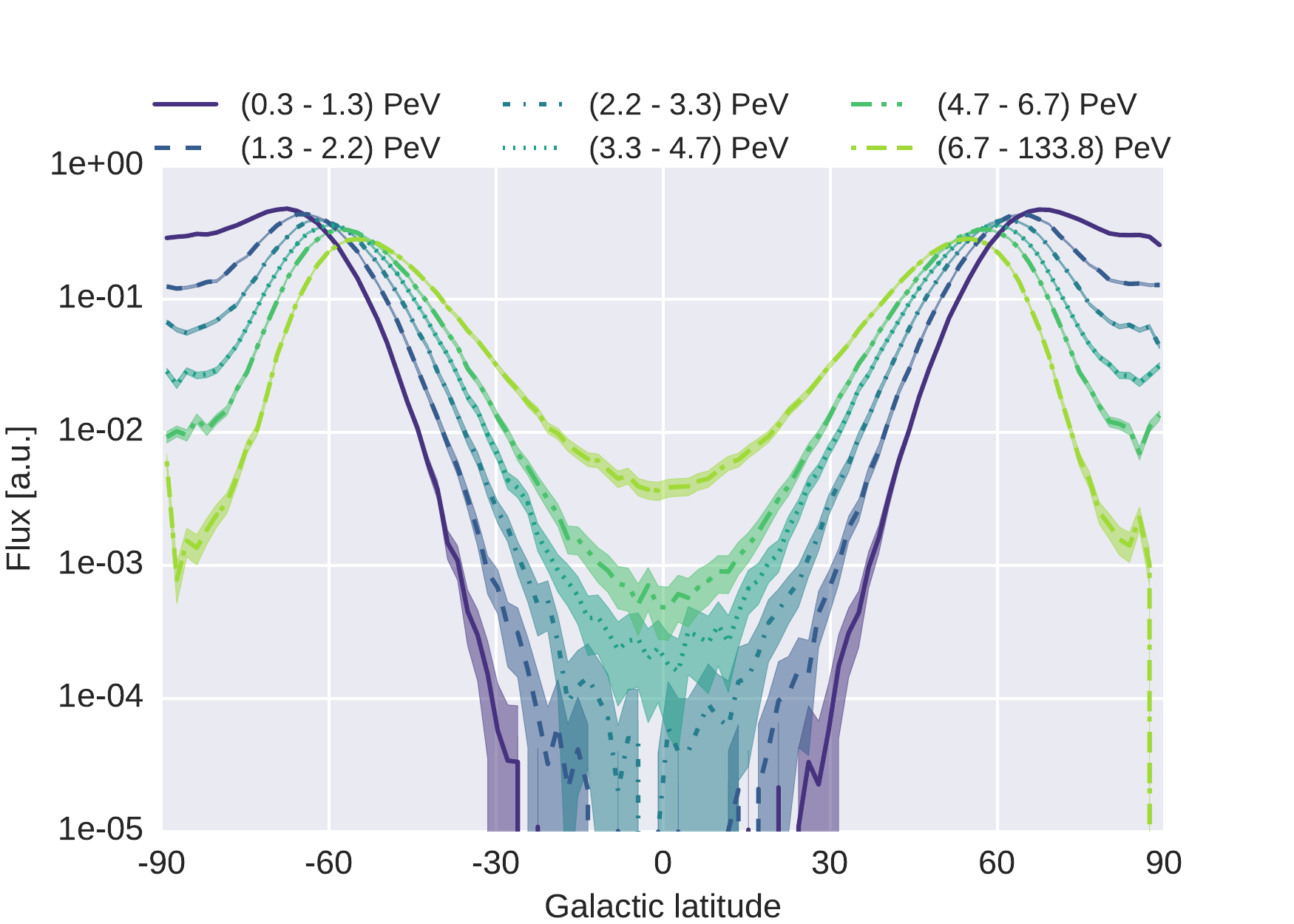}
\end{minipage}
\begin{minipage}{.49\textwidth}
\includegraphics[width=\linewidth]{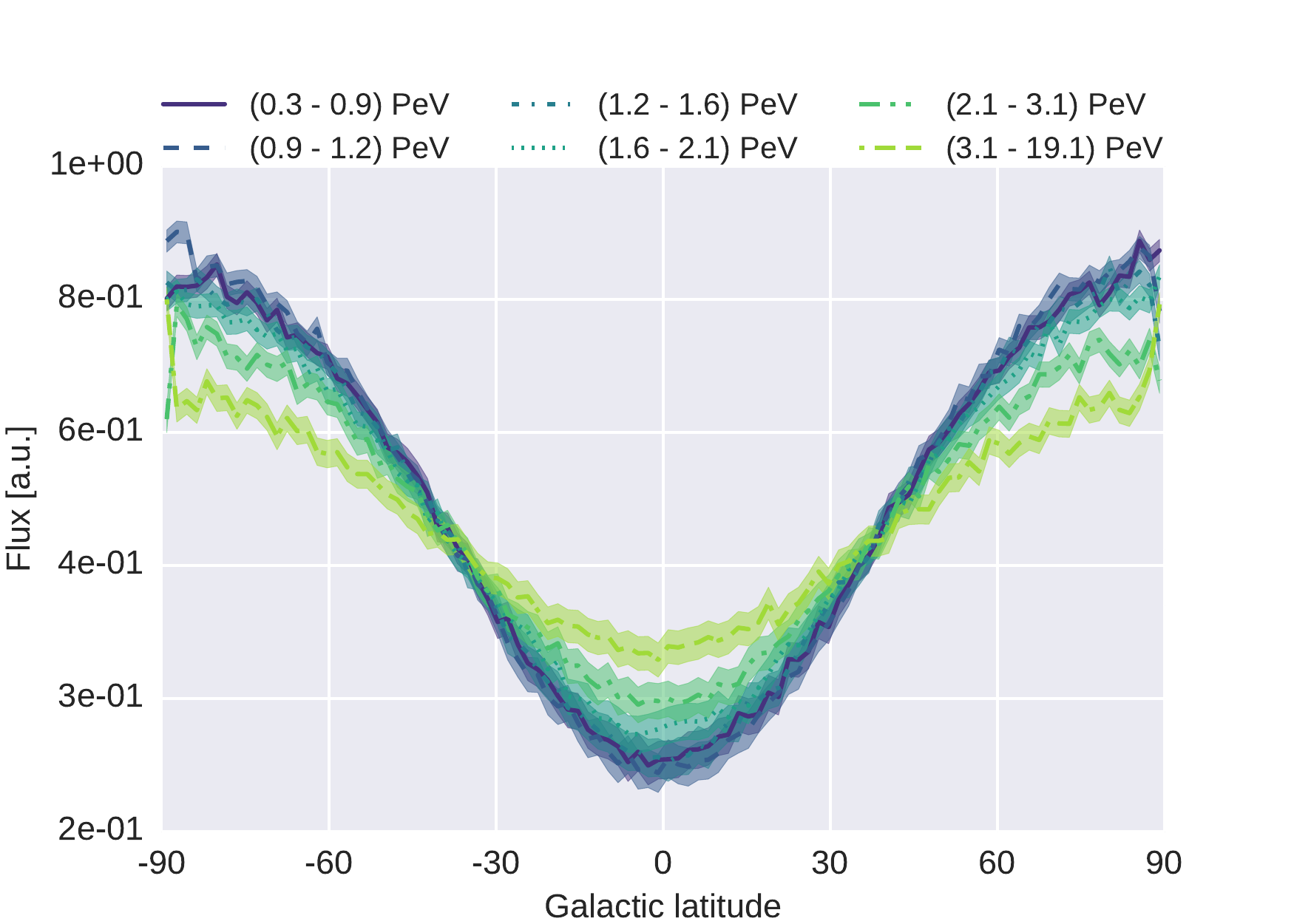}
\includegraphics[width=\linewidth]{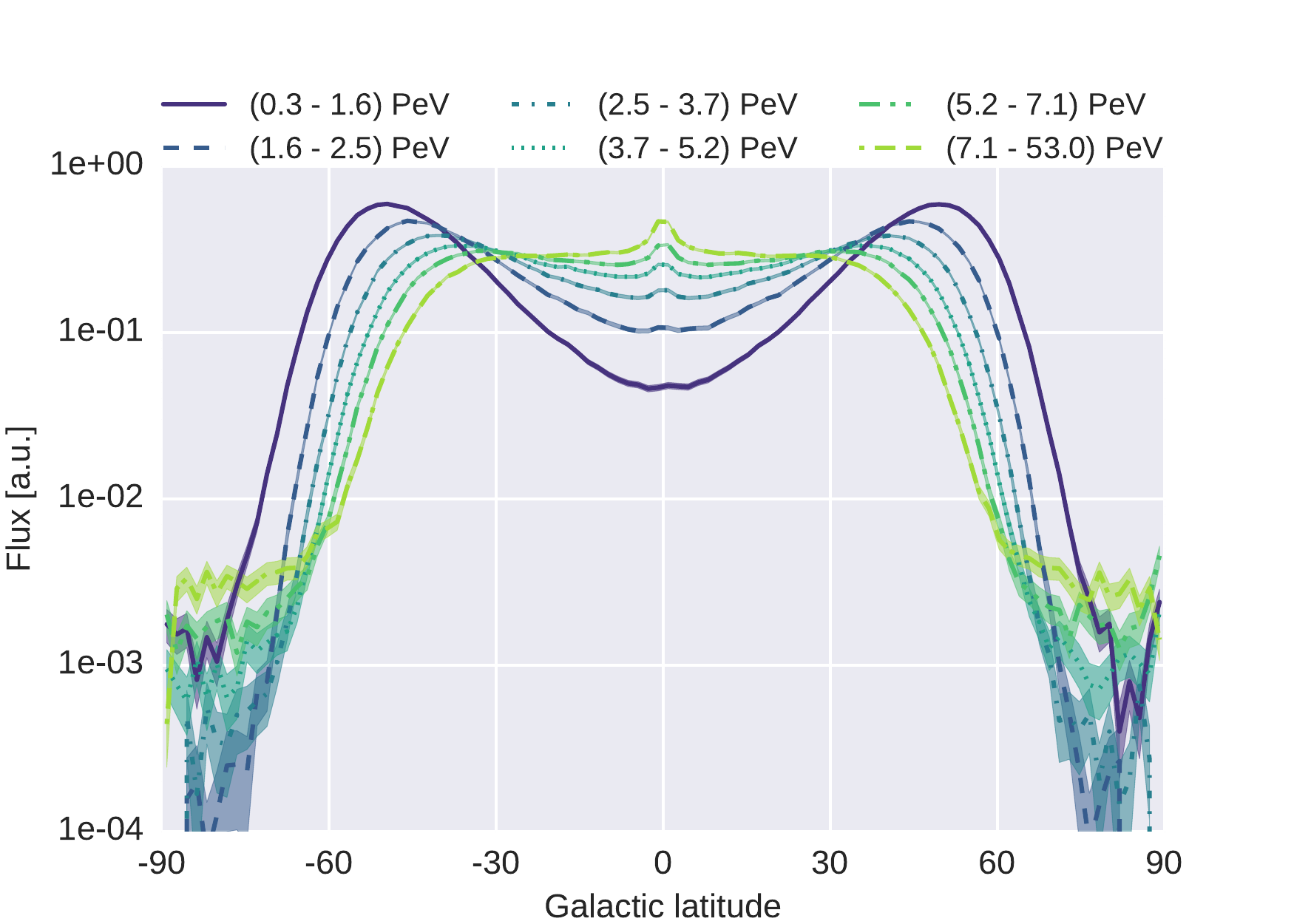}
\end{minipage}
\caption{Arrival direction of CRs binned in galactic latitude bins. Colors represent different energy bins (dark-low energy, light-high energy). Top row represents perpendicular 
diffusion ($\epsilon=0.1$) bottom row is for pure parallel diffusion ($\epsilon=0$). Left column is for Kraichnan diffusion ($\delta=0.5$) and on the right are the results for 
$\delta=0.6$. Observation point in time is for all for panels $t_\mathrm{obs}=250$~Myr. [ From top-left to bottom-right Tab.\ \ref{tab:Simulations}, Sim.\ 6, 8, 7, and 9.]}
\label{fig:Arrival_enery}
\end{figure*}

\begin{figure*}[htbp]
\centering
\begin{minipage}{.49\textwidth}
 \includegraphics[width=\linewidth]{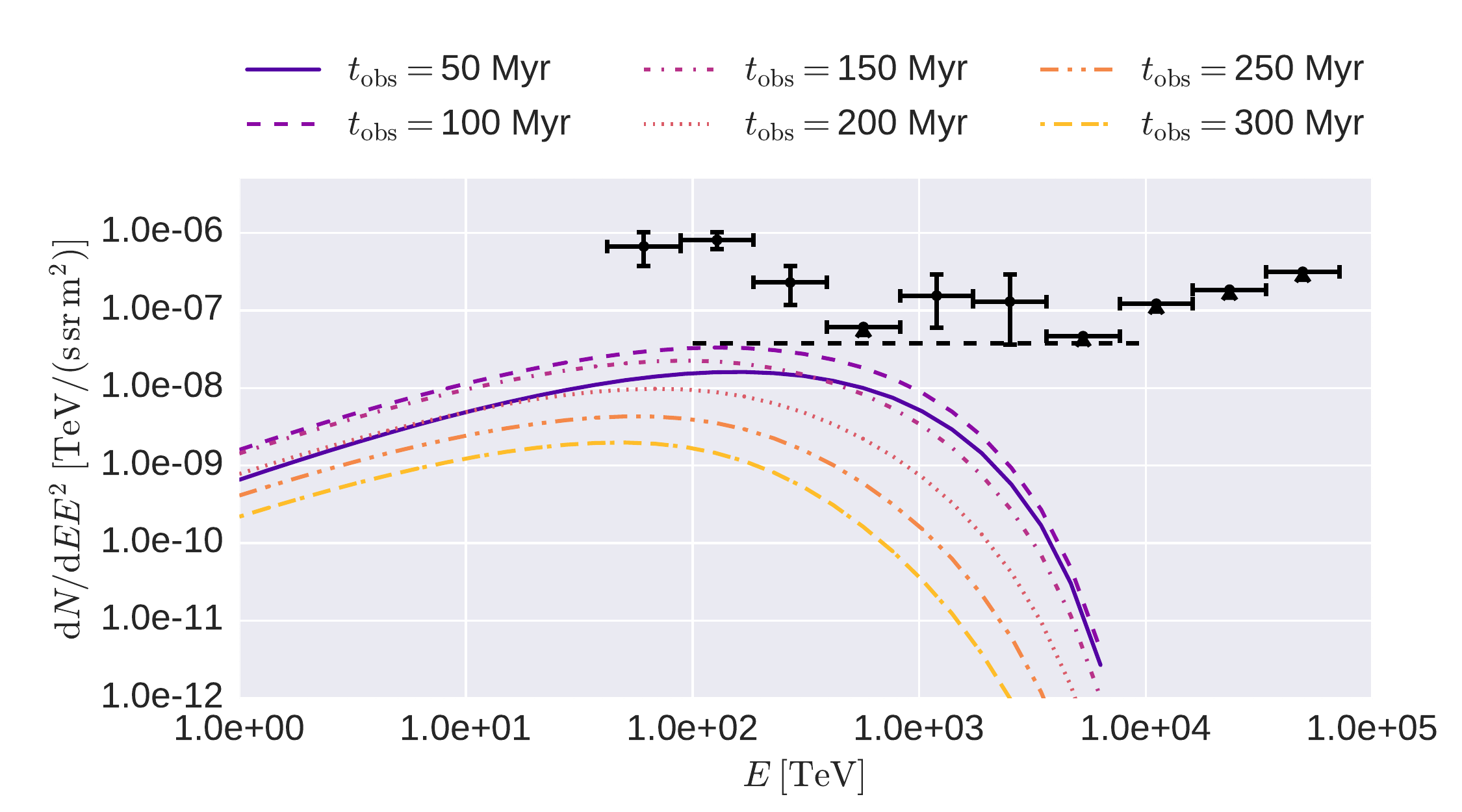}
\end{minipage}
\begin{minipage}{.49\textwidth}
\includegraphics[width=\linewidth]{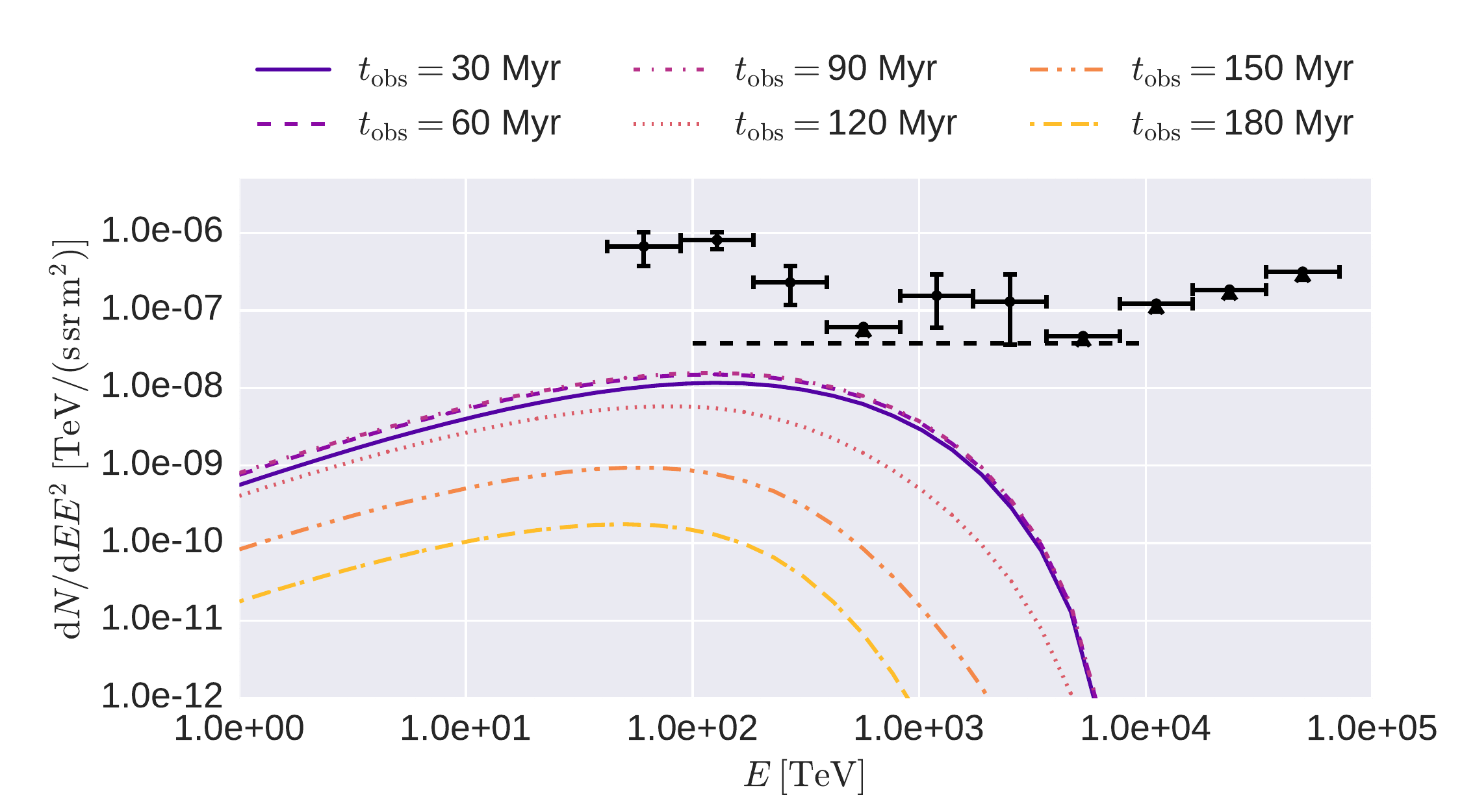}
\end{minipage}
\caption{Time evolution of the neutrino flux produced in proton-proton-interaction of CRs accelerated at the GTS. A diffusion index of $\delta=0.5$ (left) [Tab.\ \ref{tab:Simulations}, Sim.\ 18]
and $\delta=0.6$ (right) [Tab.\ \ref{tab:Simulations}, Sim.\ 19] is used 
in a spherical symmetric model including advection and adiabatic cooling. For comparison the measured neutrino flux by IceCube (black points) \citep{HESE-6-years} (We multiplied the provided single-flavor data points by a factor of three, assuming a 1-1-1 flavor ratio) and the 1-year 
diffuse flux limit from km3Net are shown \citep{KM3NeT}. (Note the different time scales.)}
\label{fig:Neutrinos_0506}
\end{figure*}

\centering
\bibliography{PropagationPaper}

\end{document}